\documentclass[letterpaper,useAMS,natbib]{mn2e}   

\usepackage{epsfig}
\usepackage{graphicx}
\usepackage{times}
\usepackage{amsmath}  
\usepackage{amssymb}
\usepackage{tabularx}  
\usepackage{multirow}
\newif\ifAMStwofonts
\AMStwofontstrue

\def\vhel{\ifmmode{V_{{\rm HEL}}}\else{$V_{{\rm HEL}}$}\fi}
\def\vsys{\ifmmode{V_{\rm sys}}\else{$V_{\rm sys}$}\fi}
\def\kms{\ifmmode{~{\rm km\,s}^{-1}}\else{~km~s$^{-1}$}\fi}
 \def\vlsr{\ifmmode{v_{\rm lsr}}\else{$v_{\rm lsr}$}\fi}

\def\ltsim{\ifmmode\stackrel{<}{_{\sim}}\else$\stackrel{<}{_{\sim}}$\fi}
\def\gtsim{\ifmmode\stackrel{>}{_{\sim}}\else$\stackrel{>}{_{\sim}}$\fi}

\newcommand{\be}{\begin{equation}}
\newcommand{\ee}{\end{equation}}
\newcommand{\ba}{\begin{eqnarray}}
\newcommand{\ea}{\end{eqnarray}}
\newcommand{\brr}{\begin{array}}
\newcommand{\err}{\end{array}}
\newcommand{\bc}{\begin{center}}
\newcommand{\ec}{\end{center}}

\newcommand{\mincir}{\raise
  -2.truept\hbox{\rlap{\hbox{$\sim$}}\raise5.truept \hbox{$<$}\ }}
\newcommand{\magcir}{\raise
  -2.truept\hbox{\rlap{\hbox{$\sim$}}\raise5.truept \hbox{$>$}\ }}
\newcommand{\siml}{\raise
  -2.truept\hbox{\rlap{\hbox{$\sim$}}\raise5.truept \hbox{$<$}\ }}
\newcommand{\simg}{\raise
  -2.truept\hbox{\rlap{\hbox{$\sim$}}\raise5.truept \hbox{$>$}\ }}

\newcommand{\aap}{A\&A}
\newcommand{\apj}{ApJ}
\newcommand{\aj}{AJ}
\newcommand{\apjl}{ApJ}
\newcommand{\apjs}{ApJS}
\newcommand{\mnras}{MNRAS}

\title[Galactic foreground contributions to the \emph{WMAP}5 maps] {Galactic foreground  contributions to the \emph{WMAP}5 maps } 
\author[Macellari et al.]{N. Macellari,$\!^{1}$~E. Pierpaoli,$\!^1$~C. Dickinson,$\!^2$~\&~J. E. Vaillancourt$^3$\\  
$^1$ University of Southern California, Los Angeles, CA 90089-0484, U.S.A. \\
(macellar@usc.edu, pierpaol@usc.edu)\\  
$^2$ Jodrell Bank Centre for Astrophysics, Alan Turing Building, School of Physics \& Astronomy, University of Manchester, Oxford Road, Manchester, M13 9PL, U.K.\\ (Clive.Dickinson@manchester.ac.uk)\\
$^3$ Division of Physics, Mathematics, \& Astronomy, California Institute of Technology, 1200 E. California Blvd., Pasadena, CA 91125, U.S.A. 
\\ Current address: Stratopsheric Observatory for Infrared Astronomy, Universities Space Research Association, NASA Ames Research Center, Moffett Field, CA \\(jvaillancourt@sofia.usra.edu)}

\begin{document}

\date{Accepted ???. Received ???; in original form ???}

\maketitle

\begin{abstract}
  We compute the cross correlation of the intensity and polarisation
  from the 5-year \emph{WMAP} data in different sky-regions with
  respect to template maps for synchrotron, dust, and free-free
  emission.  We derive the frequency dependence and polarisation
  fraction for all three components in 48 different sky regions of
  HEALPix ($N_\mathrm{side} = 2$) pixelisation.  The anomalous
  emission associated with dust is clearly detected in intensity over
  the entire sky at the K (23\,GHz) and Ka (33\,GHz) \emph{WMAP} bands, and is found to be
  the dominant foreground at low Galactic latitude, between $b =
  -40^{\circ}$ and $b = +10^{\circ}$.  The synchrotron spectral index
  obtained from the K and Ka \emph{WMAP} bands from an all-sky
  analysis is $\beta=-3.32 \pm 0.12$ for intensity and $\beta=-3.01
  \pm 0.03$ for the polarised intensity.  

  The polarisation fraction of the synchrotron is constant in
  frequency and increases with latitude from $\approx 5\%$ near the
  Galactic plane up to $\approx 40\%$ in some regions at high
  latitude; the average value for $|b|<20^{\circ}$ is $8.6 \pm 1.7\, \textrm{(stat)} \pm 0.5\, \textrm{(sys)}\,\%$ while for $|b|>20^{\circ}$ it is $19.3\pm
  0.8\, \textrm{(stat)} \pm 0.5\, \textrm{(sys)}\,\%$. Anomalous dust and free-free emission appear to be
    relatively unpolarised. Monte carlo simulations showed that there
    were biases of the method due to cross-talk between the
    components, at up to $\approx 5\,\%$ in any
    given pixel, and $\approx 1.5\,\%$ on average, when the true
    polarisation fraction is low (a few per cent or less). Nevertheless, the average
    polarisation fraction of dust-correlated emission at K-band is $3.2 \pm 0.9\, \textrm{(stat)} \pm 1.5\, \textrm{(sys)}\,\%$ or less than $5\,\%$ at 95\,\% confidence. When comparing real data with simulations, 8 regions show a detected polarisation above the 99th percentile of the distribution from simulations with no input foreground polarisation, 6 of which are detected at above $2\sigma$ and display polarisation fractions between $2.6\,\%$ and $7.2\,\%$, except for one anomalous region, which has $32\pm12\,\%$.
The dust polarisation
    values are consistent with the expectation
    from spinning-dust emission, but polarised dust emission from
    magnetic-dipole radiation cannot be ruled out. Free-free emission
    was found to be unpolarised with an upper limit of $3.4\,\%$ at 95\,\% confidence.
    \end{abstract}


\section{Introduction}
\label{sec:introduction}

During the last few years, great advances have been made in measuring
\marginpar{test} the Cosmic Microwave Background (CMB) anisotropies.
More advances are expected from the \emph{Planck} mission that will
provide all-sky temperature and polarisation CMB measurements with 
high accuracy and unprecedented frequency coverage.  As CMB
experiments reach higher sensitivity, it becomes necessary to better
understand and remove the contribution from competing Galactic and
extragalactic astrophysical signals in the same frequency bands.  It
is well known that the emission of these foregrounds rivals the
temperature signal over a significant fraction of the sky and
dominates the polarisation signal over most of the sky (Kogut et al.\
2007).  As the temperature power spectrum has already been measured
with great precision (e.g., \emph{WMAP}5, Nolta et al.\ 2009; ACBAR,
Reichardt et al.\ 2009; Boomerang, Jones et al.\ 2006; CBI, Sievers et
al.\ 2009; VSA, Dickinson et al.\ 2004; QUaD, Pryke et al.\ 2009),
most observations are now concentrating on polarisation measurements
(e.g., BICEP, Chiang et al.\ 2009; QUaD, Brown et al.\ 2009).  For the
study of polarisation anisotropies however, foregrounds may be the
limiting factor, as the primordial signal is much weaker than
intensity while the foregrounds are not well characterised at the
relevant frequencies at this time.  For this reason, suitable
component separation techniques have been developed and are being
applied to current data (Kogut et al.\ 2007; Eriksen et al.\ 2008;
Gold et al.\ 2009; Dunkley et al.\ 2009; Delabrouille \&
Cardoso~2009), in preparation for \emph{Planck} data, and possible
future missions (Leach et al.\ 2008; Betoule et al.\ 2009).  In some
instances such techniques rely on prior information acquired on the
emission of a specific foreground, for which a prior characterisation
is necessary.  Moreover, the study of foregrounds in the microwave
band is of astrophysical interest on its own, as it can be informative
about physical phenomena occurring in our Galaxy and in others.

Foregrounds are either Galactic or extragalactic in nature.  In this
paper we are interested in the Galactic emission, which dominates over
extragalactic emission at intermediate to large scales ($\gtrsim
1^{\circ}$).  The Galactic emission has three main components: dust,
synchrotron, and free-free, of which only dust and synchrotron are
expected to show significant polarisation.  In contrast with
extragalactic emission, we do not expect the Galactic foreground to be
isotropic.  For this reason, Galactic foregrounds need a
characterisation that is both scale dependent and position dependent.

The most common methods for extracting foreground information from CMB
data involve fitting a given theoretical model to the data.  These
fitting procedures typically assume that a specific foreground follows
a power-law in frequency and is characterised by a given polarisation
fraction and a (fixed) angle.  For instance, this is the approach of
the \emph{WMAP} teams (Kogut et al.\ (2007) for the \emph{WMAP} 3-year
and Gold et al.\ (2009) for the \emph{WMAP} 5-year data).

By so doing, however, the diverse behaviour of different foregrounds
in different regions of the sky may be neglected.  For instance, a
spinning-dust contribution may not have a power-law frequency
dependence, and may make different contributions in different areas of
the sky.  A variety of dust grains may populate different areas of our
galaxy, and have different contributions to the intensity and
polarisation signal at the observed frequencies.  The synchrotron
emission clearly depends on the energy spectrum of electrons
in a given region, while the polarisation signal may be affected by
depolarisation effects which are more severe in denser or less uniform
regions.

In this paper, we aim to characterise the contribution of the three
Galactic components in the microwave band in temperature and
polarisation by studying the \emph{WMAP5} data (Nolta et al.\ 2009;
Hinshaw et al.\ 2009), including their frequency and spatial
dependencies.  To achieve this goal we develop a foreground analysis
that assumes no frequency model and only relies on a morphological
characterisation of foregrounds as derived at non-CMB frequencies.
Specifically, we use a cross-correlation (C-C) analysis with standard
foreground templates.  Such an analysis method has been widely used by
other authors studying CMB intensity data (e.g., de Olivera-Costa et
al.\ 1999, 2002; Banday et al.\ 2003; Bennett et al.\ 2003; Davies et
al.\ 2006), but has not been applied to polarisation data.  An
advantage of using such a method is the possibility of detecting
particular emission processes associated with one of the three
foreground templates even if its frequency and/or polarisation
characteristics are unknown. For example, this may be true in the case
of a spinning-dust component.

A possible limitation of the C-C method is that it allows
characterisation of only foreground components whose spatial
distribution is well traced by the templates.  These templates were
derived at very different frequencies than the CMB measurements, and
may not fully represent the morphology of a particular component in
the range of frequencies at hand.  Most analyses of \emph{WMAP5}
polarisation data are based on modelling the various emission
components in the $Q$ and $U$ maps; here we will perform our analysis
in the polarisation intensity $P$.  This choice avoids the problem of
modelling the Galactic magnetic field to compute the polarisation
angle, or assuming a specific polarisation angle as a function of
position for each component.  Under the hypothesis that all components
have the same polarisation angle (even if possibly a different
polarisation intensity), the method used in analysing the temperature
maps can be readily extended to the polarisation maps.  Although a
more general approach may be sought, as both dust grains' orientation
and synchrotron emission are dictated by the direction of the magnetic
field, this assumption may not be unreasonable.

Section~\ref{sec:wmap} describes the \emph{WMAP} data used in this
work and the choices made about spatial resolution and masking, while
Section~\ref{sec:templates} gives an introduction to the various
sources of foreground emission and the spatial templates that will be
used to trace each component.  The cross-correlation analysis
technique and the computation of uncertainties are derived in
Section~\ref{sec:cc}.  The results of the analysis in both temperature
and polarisation are presented in Section~\ref{sec:results}, along
with separate discussions for each of the 3 emission components. The
main conclusions are summarised in Section~\ref{sec:conclusions}.

\section{\emph{WMAP} data}
\label{sec:wmap}

We use the \emph{WMAP} 5-year data (Hinshaw et al.\ 2009), consisting
of 5 full-sky maps at frequencies of 22.8\,GHz (K-band), 33.0\,GHz
(Ka-band), 40.7\,GHz (Q-band), 60.8\,GHz (V-band), and 93.5\,GHz
(W-band), as provided by the LAMBDA
website\footnote{http://lambda.gsfc.nasa.gov/}.  These maps are
provided in HEALPix\footnote{http://www.eso.org/science/healpix/
  (G\'{o}rski et al.~2005)} format at a resolution of
$N_{\mathrm{side}}=512$ (number of pixels on the sphere = $12\times
N_{\mathrm{side}}^2$).  The maps are then downgraded to a HEALPix
resolution of $N_{\mathrm{side}}=32$, using the ud\_grade\footnote{Simple averaging is used, rather than a noise-weighted average.} HEALPix routine, to give a total of 12,288 pixels.
We chose this resolution as a compromise between obtaining a reasonable signal-to-noise ratio
 and having small enough pixels to keep the information about foreground variations needed in the C-C analysis; with
$N_{\mathrm{side}}=16$ the resolution is becoming too coarse. Larger pixels are also likely to invalidate the assumption of a constant polarisation fraction. Parallel transport of polarisation vectors on the sphere is not considered in the averaging since the effect is at the $10^{-4}$ level or lower.

We convert the data from thermodynamic temperature to brightness
(antenna) temperature using the conversion:
\begin{equation}
 T_A = \frac {x^2 \mathrm{e}^x}{\left(\mathrm{e}^x-1\right)^2} \; \; T_{\mathrm{CMB}}
\end{equation}
where $x \equiv (h\nu)/(kT_0)$ and
$T_{0}$ is the CMB temperature
of $2\fdg 725$\,K (Mather et al.\ 1999); this corresponds to a
correction of $1\%$ at K-band and $25\%$ at W-band.

We apply the \emph{WMAP} team mask KQ85 (Gold et al.\ 2009) to the
\emph{WMAP} maps, in order to avoid the Galactic plane and bright
point sources.  The mask is provided in a $N_{\mathrm{side}}=512$
HEALPix resolution, so we downgrade it to our working resolution
($N_{\mathrm{side}}=32$) and consider in the analysis only the coarse
pixels that are completely outside of the $N_{\mathrm{side}}=512$
masked region.  This procedure reduces to 8099 the number of data
pixels (66\% of the sky) used in the analysis.

In the polarisation analysis we compute the total polarisation as
$P=\sqrt{Q^2+U^2}$.  The mask and the resolution chosen are the same
as the intensity analysis.  This allows direct comparison of the C-C
coefficients fit for the temperature and polarisation data and
calculation of the fractional polarisation at the same sky positions.


\section{Galactic Components Emission and Templates}
\label{sec:templates}

\subsection{Synchrotron}

Synchrotron emission arises from accelerating cosmic ray electrons
spiralling in the Galactic magnetic field.  The emission depends on
the energy spectrum of the electrons and on the intensity of the
magnetic field, which results in significant spatial variations in the
spectral index on the sky. The synchrotron spectrum is approximated by
a power law $T\sim \nu^\beta$, typically with $\beta \approx -2.7$ at
radio wavelengths, steepening to $\beta \approx -3.0$ at \emph{WMAP}
frequencies and with typical spatial variations of $\pm
0.2$. Moreover, it is known that a pure synchrotron spectrum steepens
with frequency due to the effects of spectral ageing, although in
reality, a flattening of the spectrum can occur due to the presence of
multiple components (e.g., Kogut et al.\ 2007).

Most of the information we have for synchrotron emission comes from
low frequency radio surveys, where synchrotron dominates the sky. In
particular, large-area radio surveys at 408\,MHz (Haslam et al.\ 1981,
1982), 1420\,MHz (Reich \& Reich 1986) and 2.3\,GHz (Jonas et al.\
1998) give us a clear picture of the synchrotron component and these
have been used extensively for foreground subtraction and previous
cross-correlation studies. The drawbacks of using such templates are
i) there is a large extrapolation in frequency from $\sim 1$\,GHz to
\emph{WMAP} frequencies that is likely to result in some distortion of
the morphology, ii) baseline (offset) issues result in striping in the
maps, and iii) discrete extragalactic sources contaminate the diffuse
emission. Several efforts have been made to reduce the effects of
striping and source contamination (e.g., Davies et al.\ 1996; Platania
et al.\ 2003) including the NCSA version of the Haslam et al.\
408\,MHz map, available at the LAMBDA website. We use this as our
template for synchrotron emission.

Synchrotron emission is, by nature, highly polarised. For a power-law
distribution of electron energies $N(E) \propto E^{-p}$ propagating in
a uniform magnetic field, the resulting emission is polarised with
fractional linear polarisation $f=(p+1)/(p+7/3)$
aligned perpendicular to the magnetic field (Rybicki \& Lightman
1979). The frequency dependence of synchrotron emission is also
related to the electron energy distribution, $T(\nu) \propto
\nu^{\beta}$ with spectral index $\beta=-(p+3)/2$.
For $\beta \approx -3$, the synchrotron emission can have a maximum
fractional polarisation as high as $f_{s}=0.75$. However,
line-of-sight and beam averaging effects will tend to reduce this,
with typical values at high latitude of $\sim 10$--40\%.


\subsection{Dust}
\label{sec:dust}

The Galactic foreground at far-infrared and submillimetre wavelengths
($\nu \gtrsim 100$~GHz) is dominated by thermal emission from warm ($T
\sim 10$ -- 100 K) interstellar dust grains.  The intensity spectrum
peaks in the range 100 -- 200 $\micron$ and is well modelled by an
emissivity-modified gray-body of the form $\nu^\beta B_\nu(T)$, where
$B_\nu(T)$ is the Planck function at frequency $\nu$ and temperature
$T$.  Using data at 100 and 240 $\micron$, Finkbeiner, Davis \&
Schlegel\ (1999) modelled the all-sky emission with two components
having mean temperatures of 9.4\,K and 16\,K and spectral indices of
$\beta = 1.7$ and $2.7$, respectively.  The predictions at longer
wavelengths are in good agreement with observations at 353\,GHz by
\emph{Archeops} (Ponthieu et al.\ 2005) and at 94\,GHz by \emph{WMAP}
(Bennett et al.\ 2003).  We use the predictions at 94\,GHz, which are
available on the LAMBDA website, as our template for dust emission.

Polarisation from thermally emitting dust is due to aspherical grains
whose spin-axes have become aligned locally with interstellar magnetic
fields (e.g., see reviews by Lazarian 2003, 2007).  The most efficient
direction for emission and absorption/extinction is the long grain
axis (perpendicular to the spin axis).  The observed polarisation
direction is then perpendicular to the aligning magnetic field in the
case of grain emission, but parallel to the field in the case of
background-starlight extinction. The expected size of the polarisation
depends on a number of unknown factors including the efficiency of the
alignment mechanism and the inclination of the magnetic field to the
line-of-sight.  Typical background-starlight polarisation has values
in the $\sim 1$--4 \% range (e.g., Heiles 2000; Fosalba et al.\ 2002),
while polarised emission in dense Galactic clouds have been observed
at 0.5 -- 10 \% (e.g., Dotson et al.\ 2000, 2010; Matthews et
al. 2009).  \emph{WMAP} polarisation observations at 94\,GHz (Page et
al.\ 2007; Kogut et al.\ 2007) and \emph{Archeops} at 353\,GHz
(Ponthieu et al.\ 2005) measure dust polarisations of $\sim 1$\% in
the Galactic plane, increasing to a few percent at higher latitudes.
Studies of the frequency dependence of polarised emission have found
that the spectrum has a polarisation minimum at $\lambda \sim
350$\,\micron, increasing to longer wavelengths (e.g., Hildebrand et
al.\ 1999).  However, these studies are limited to $\lambda = 60$ --
1300 $\micron$ and to bright/dense Galactic clouds.  Extrapolating
this spectrum to the conditions expected to prevail in the
high-latitude diffuse ISM, one expects a more featureless spectrum for
wavelengths a few times longer than about 1\,mm (Hildebrand \& Kirby
2004).

Other possible sources of polarised dust emission at microwave
frequencies are electric-dipole emission from ``spinnng dust'' grains,
as well as magnetic-dipole emission from vibrating ``magnetic dust''
grains (e.g., Lazarian \& Finkbeiner 2003; Draine \& Lazarian
1998a,b,1999; Lazarian \& Draine 2000; Ali-Ha{\"i}moud, Y., Hirata,
C.~M., \& Dickinson, C.\ 2009).  Evidence for the existence of
spinning-dust exists in the strong correlations observed between the
spatial distribution of thermal dust emission and that of anomalous
emission at frequencies of 20 -- 60 GHz.  This anomalous emission
cannot be accounted for by models using only the standard components
of free-free, synchrotron, and/or thermal dust emission.  This
dust-correlated anomalous emission is observed on both large-scales
(e.g., Kogut et al.\ 1996; Finkbeiner 2004; Finkbeiner, Langston, \&
Minter 2004; Davies et al.\ 2006; Hildebrandt et al.\ 2007;
Miville-Desch\^{e}nes et al.\ 2008) as well as in pointed observations
of specific Galactic dust clouds (e.g., Finkbeiner 2004; Watson et
al.\ 2005; Casassus et al.\ 2006; Dickinson et al.\ 2009a, 2010; Scaife et
al.\ 2009; Planck collaboration\ 2011).

The polarisation amplitude of spinning-dust emission is likely to be
small.  Lazarian \& Draine (2000), modelling the alignment of very
small dust grains in the ISM, predict that spinning dust will be
polarised by no more than $\sim$7\% at 2\,GHz, falling to $\lesssim
0.5\%$ above frequencies of 30\,GHz. Empirically, UV observations of
background-starlight polarisation (caused by the same very small
grains) indicate that these grains are not as well aligned with the
local magnetic field as are larger grains (e.g., Martin 2007).
Magnetic-dust is expected to be polarised at much higher levels, as
high as 40\% (Draine \& Lazarian 1999). The only polarisation
observations of the anomalous emission to date have found low
levels. Battistelli et al.\ (2006) report $3.4^{+1.5}_{-1.9}$\%
towards the Perseus molecular cloud at 11\,GHz while Mason et al.\
(2009) report an upper limit of 3.5\% (99.7\% confidence level)
towards the dark cloud Lynds 1622 at 9\,GHz. L\'{o}pez-Caraballo et al.\, (2011) used \emph{WMAP} data to constrain the polarization towards the Perseus molecular cloud to be less than 1.0, 1.8 and 2.7\,\% at 23, 33 and 41\,GHz, respectively. These observations are consistent with polarisation from spinning-dust, but
cannot rule out the existence of vibrating magnetic-dust.


\subsection{Free-free }
\label{sec:ha_template}

Free-free (bremsstrahlung) emission is due to electron-electron
scattering from warm ($T_{e}\approx 10^4$\,K) ionised gas in the
interstellar medium. In the optically thin regime, and at radio
frequencies, the brightness temperature is given by
\[T_b=8.235\times 10^{-2}a T_e^{-0.35}\nu_{\mathrm
  GHz}^{-2.1}(1+0.08)({{\rm EM})_ {{{\rm cm}^{-6}}{\rm pc}}}
\]
where $T_e$ is the electron temperature, EM is the emission measure,
and $a$ is a factor close to unity.  The electron temperature in the
warm ionised medium ranges from $\approx 3000$ to $\sim 20,000$ K but
is $\approx 8000$\,K at the Solar galacocentric distance. The spectrum
follows a well defined power-law with a spectral index of $-2.1$ with
only a small variation in temperature. At \emph{WMAP} frequencies, the
effective free-free spectral index is $\approx -2.14$ for $T_e\approx
8000$\,K.

The well-determined spectrum can be used to separate free-free from
other diffuse radio components, most notably, synchrotron (which is
almost always steeper).  However, due to the complexity of the
components, and limited frequency range of \emph{WMAP}, a cleaner
separation can be achieved by using an appropriate spatial
template. The free-free brightness temperature is proportional to
${\rm EM} \equiv \int n_e^2 dl$, i.e., the integrated column density
squared along the line of sight. Fortunately, this is also true of
recombination lines such as the bright optical line, H$\alpha$. Maps
of H$\alpha$ therefore give an almost one-to-one tracer for radio
free-free emission (Dickinson et al.\ 2003) and they are now available
for the full-sky (Finkbeiner 2003). There is some dependancy on $T_e$
and the local conditions, but these are relatively small. The main
difficulty arises at low Galactic latitudes (typically
$|b|<5^{\circ}$) where dust absorption can reduce the H$\alpha$
intensity by a large factor, thus rendering it useless for predicting
radio free-free brightness.  However, outside the Kp2 mask, almost all
lines of sight have a small to negligible maximum correction factor;
typical values are $\ll1$\,mag.  The correction factor can be
estimated from the IRAS $100\,\mu$m redenning maps (Schlegel et al.\
1998).  We therefore use the Finkbeiner (2003) composite H$\alpha$ map
as our standard free-free template, but also compare a similar map
produced by Dickinson et al.\ (2003).

Free-free emission is intrinsically unpolarised because the scattering
directions are random.  However, a secondary polarisation signature
can occur at the edges of bright free-free features (i.e., H\;{\sc ii} regions)
from Thomson scattering (Rybicki \& Lightman 1979).  This could cause
significant polarisation ($\sim 10\%$) in the Galactic plane,
particularly when observing at high angular resolution. However, at
high Galactic latitudes, and with a relatively low resolution, we
expect the residual polarisation to be $<1\%$.


\section{ The Cross-Correlation Analysis}
\label{sec:cc}

\subsection{The Method}

The C-C method (described in Davies et al.\ 2006) assumes that the
total intensity in every pixel is well modelled by a finite sum of
separate emission mechanisms such that
\begin{align}
\label{seconda}
{\bf I_\mathrm{tot}} (\nu, \hat n)
& =\sum_{i=1}^{n} \; T_i (\nu, \hat n) \\
 &  =\sum_{i=1}^{n} \; \theta_i(\nu, \hat n) \; T_i(\nu _{0i}, \hat n),
\label{secondc}
\end{align}
where ${\bf I_\mathrm{tot}} (\nu,\hat n)$ is the total emission at
frequency $\nu$ in the direction (or map pixel) $\hat n$,
$T_i(\nu_{0i}, \hat n)$ is the emission from the $i^\mathrm{th}$
template at the template frequency $\nu_{0i}$ (hereafter simply
$T_i(\hat n)$), and $\theta_i(\nu,\hat n)$ is the spectral shape of
the emission template, normalised in every pixel such that
$\theta_i(\nu_{0i}) = 1$.  The C-C technique is applied to sky-regions
containing different pixels and assumes that the proportionality
coefficients between the total intensity and each foreground component
are constant within each region.  That is, the $\theta_i(\nu,\hat n)$
are independent of position within each region, although they can vary
from region-to-region:
\begin{equation}
\label{seconde}
{\bf I _\mathrm{tot}} (\nu, \hat n)=\sum_{i=1}^{n} \; \theta_i(\nu) \; T_i(\hat n).
\end{equation}

For the emission mechanisms studied in this work we have $n=3$
templates, although the C-C method can generally be extended to any
number $n$. For any given \emph{WMAP} frequency $\nu$, an estimate of
the parameters $\theta_i(\nu)$ can be computed by minimising $\chi^2$,
as explained in Davies et al.\ (2006).  In the component formalism of
Equation~({\ref{seconde}}) we have
\begin{equation}
\label{CCformula2}
\theta_i (\nu)=  \sum_{j=1}^{3} (A^{-1})_{ij} \cdot \left [ \sum_{\hat n}^R(T_j(\hat n)^T \cdot {\bf M_{SN}^{-1}} (\nu,\hat n)\cdot {\bf I }(\nu, \hat n)) \right ]
\end{equation}
where the sum over $\hat n$ is the sum over all pixels in sky-\-region
$R$, $A_{ij}$ is the $3 \times 3$ template matrix $\sum_{\hat
  n}^R\left[T_i(\hat n)^T \cdot {\bf M_{SN}^{-1}}(\nu,\hat n)\cdot
  T_j(\hat n) \right]$, and ${\bf M_{SN}}(\nu,\hat n)$ is the
covariance matrix of the expected noise of the \emph{WMAP} data and
CMB signal (described in the next section).  The objective is to
invert the covariance matrix and compute the coefficients $\theta_i
(\nu)$ for each of the 3 Galactic foreground components.

Within an area $R$ of the sky, the C-C analysis relies on the
morphological characteristics of a given component. The mean value of
the templates' emission within $R$ (which can be thought of as a
constant offset in the template's map) does not provide information
for the purpose of recovering the signal relative to each individual
component, as it has no morphological structure.  Moreover, keeping
these mean offsets in the templates fixed will bias the results of the C-C\@.
Since all templates, as well as all five \emph{WMAP}
maps, may have such uniform offsets they can each be
represented as a single uniform term in the sum of Equation~(\ref{secondc}).  
That is, each template and dataset have the form
\begin{equation}
T_i(\hat n) = T_i^\prime(\hat n) - \delta_i^\prime, 
\label{eq-off1}
\end{equation}
where $T^\prime$ is an existing template map or \emph{WMAP} data, and
$T_i$ is its true value in which all constant-offsets have been
removed.  Equation (\ref{seconde}) is then re-written as
\begin{equation}
{\bf I_\mathrm{tot}^\prime}(\nu,\hat n) = \sum_{i=1}^{n} \; \theta_i(\nu) \; T_i^\prime(\hat n) + \delta_\mathrm{tot}(\nu)
\label{eq-off2}
\end{equation}
where
\begin{equation}
\delta_\mathrm{tot}(\nu) = \delta^\prime_\mathrm{tot}(\nu) - \sum_{i=1}^n\;\theta_i(\nu)\;\delta_i^\prime.
\label{eq-off3}
\end{equation}
If each analysis is limited to a single sky region covered by a single
set of parameters $\theta_i(\nu)$ then each of the individual terms on
the right-handside of Equation (\ref{eq-off3}) are monopole terms. By
extension, the total $\delta_\mathrm{tot}(\nu)$ must also be a
constant monopole.  Therefore, we can define a new template
$T_\mathrm{off}(\nu,\hat n)\equiv T_\mathrm{off} \equiv 1$ at all
frequencies and all sky positions. Equation (\ref{eq-off2}) can then
be re-written
\begin{equation}
{\bf I_\mathrm{tot}^\prime}(\nu,\hat n)= \sum_{i=1}^{n+1} \; \theta_i(\nu) \; T_i^\prime(\hat n)
\end{equation}
where $\theta_{(n+1)}(\nu) = \delta_\mathrm{tot}(\nu)$ and
$T_{(n+1)}^\prime(\hat n) = T_\mathrm{off}$. The C-C analysis then
proceeds just as in the case of a zero-offset but simply with an
additional template term, and the value of $\delta_\mathrm{tot}(\nu)$
is returned along with the other $n$ fit parameters, $\theta_i(\nu)$.

The introduction of an extra fitting parameter (the amplitude of the
monopole-term) increases the uncertainties returned in the other
parameters of the C-C analysis.  This effect is the largest at high
Galactic latitudes where there is little variation in the foreground
emission.  That is, one should expect the correlation of the data with
the monopole-template to increase with increasing latitude even if a
true constant-offset does not exist in the data maps.

When these offsets are included in the C-C the reduced-$\chi^2$ is
significantly better than without the offset. At K-band it decreases
by about 24\% for the temperature analysis
and by about $8\%$ in the analysis of the polarised emission.
These offsets are found to exhibit a distinct synchrotron-like
frequency spectrum, whereas one would naively expect little
correlation with frequency. This suggests that the Haslam et al.\
408\,MHz synchrotron template is dominating the fit offset terms.

\subsection{Uncertainties}
\label{sec:errors}

Writing the C-C error equation following the component formalism, the
uncertainties on the fitted C-C coefficients for the $i^\mathrm{th}$
template are
\begin{equation}
\label{CCerr2}
\delta \theta_i (\nu)= \sqrt{\sum_{\hat n}^R[(T^T(\hat n) \cdot {\bf M_{SN}}^{-1}(\nu,\hat n)\cdot T(\hat n))^{-1}]_{ii}}.
\end{equation}
In Equation~(\ref{CCerr2}), the covariance matrix $\bf M_{SN}=\bf
M_{S} + \bf M_{N}$, where $\bf M_{N}$ and $\bf M_{S}$ are the
covariances of the instrumental noise and the CMB, respectively.  The
CMB signal covariance only has a significant contribution to the
temperature analysis, where it dominates the instrumental noise.

We compute the signal covariance matrix in the usual way, taking the
CMB power spectrum $\emph C_l$ from the \emph{WMAP} best-fit
cold-dark-matter $(\Lambda CDM)$ power-law spectrum model (Nolta et
al.\ 2009).  In practice, we take into account spatial correlations
within $30^{\circ}$ of a given pixel, effectively neglecting
correlation values that are at most $0.5\%$ that of the pixel
auto-correlation.  The noise covariance in intensity is determined
from the uncorrelated pixel noise as specified for each pixel in the
\emph{WMAP} data; this is an $N \times N$ diagonal matrix with the
inverse of the observation number in each pixel on the diagonal
($\sigma _I = \sigma _0 / \sqrt{N_{obs}}$).

We use the same strategy as in Eq.\ (\ref{CCerr2}) to compute the errors in the polarisation coefficients defined in Section~\ref{sec:FracPol} (Eq.\ \ref{rho}). In the polarisation analysis the covariance matrix is dominated by the
noise, which 
exceeds the contribution from the polarisation power
spectrum $C_l^{EE}$ by two orders of magnitude.\footnote{On the
  diagonal elements, $\bf{M_S}$ $= 0.45 \,\mu$K$^2$, while $\bf M_{N}$
  $= 55 \,\mu$K$^2$.} $\bf M_{N}$ is, to a good approximation, a
diagonal matrix as the pixel-to-pixel noise correlation is at most
$1\%$ of the diagonal term. We verified this by inspecting the $4
\times N \times N$ complete noise matrix available at the LAMBDA
website, even though this is given at $N_{\mathrm{side}}=16$ and our
work is done using $N_{\mathrm{side}}=32$ for the reasons introduced
in Section~\ref{sec:wmap}.  
While it is true that we have no information on exact pixel-to-pixel
correlations at $N_{\mathrm{side}}=32$, the $N_{\mathrm{side}}=16$
noise matrix suggests there are no large-scale noise correlations for
the polarisation analysis.

The polarisation C-C results are, therefore, computationally easier to
determine than the temperature results, because we only need to
compute $N$-dimensional arrays as opposed to the $N \times N$ needed
in the temperature case.  Therefore, we compute the noise for the
total polarisation in each pixel by combining noise in the $Q$ and $U$
maps at each frequency and properly taking into account $Q$-$U$
correlations.

In every pixel $\hat n$ we have a matrix
\begin{equation*}
N(\hat n)=
\begin{pmatrix}
QQ(\hat n) & QU(\hat n)  \\
UQ(\hat n) & UU(\hat n)   \\ 
\end{pmatrix}
\end{equation*}
where $QQ(\hat n) $ and $UU(\hat n) $ are noises and $QU(\hat n) $ is
the $\hat n$ pixel $Q$-$U$ noise correlation, all in terms of
observation number.  Uncertainties on $Q$ and $U$ are given by the
inverse of $N(\hat n) $
\begin{center}
\begin{eqnarray*}
\sigma _q(\hat n)  & = & \sigma _0 \sqrt{[(N(\hat n) )^{-1}]_{11}} \\
\sigma_u(\hat n) & = & \sigma _0 \sqrt{[(N(\hat n) )^{-1}]_{22}}
\end{eqnarray*}
\end{center}
Propagating these noise terms to compute the uncertainty on
$P=\sqrt{Q^2+U^2}$ we obtain:
\begin{equation}
\sigma_p(\hat n)  = \sqrt{\left( \frac{Q}{P} \sigma_q(\hat n)  \right)^2+\left( \frac{U}{P} \sigma _u(\hat n)  \right)^2}.
\end{equation}
The $\sigma_p(\hat n) ^2$ are the elements of the (diagonal) noise matrix ${\bf M_N}$ for the total polarisation $P$. The uncertainty on the polarised correlation coefficients is:

\begin{equation}
\label{CCpolerr}
\delta \rho_i (\nu)= \sqrt{\sum_{\hat n}^R[(T^T(\hat n) \cdot {\bf M_{N}}^{-1}(\nu,\hat n)\cdot T(\hat n))^{-1}]_{ii}}.
\end{equation}
 
When choosing the area over which to perform the C-C analysis, one should
consider the trade-off between the uncertainties and the accuracy of the modelling.
While it is more likely that in a small area a single fitting
coefficient is sufficient to represent a given foreground, the uncertainties
calculated with Eq.~(\ref{CCerr2}) and (\ref{CCpolerr}) decrease with the square root of the
number of pixels considered.  In the following, our choice to consider
relatively large areas is a trade-off between these two competing aspects.

\subsection{Polarisation analysis and fractional polarisation}
\label{sec:FracPol}

When dealing with the cross-correlation of the polarised emission some
preliminary comments are necessary.  Since Equation~(\ref{seconde})
holds for any measured intensity and polarisation value, if templates
from all Stokes parameters were available at some set of reference
frequencies, i.e., $Q_i(\nu_{0i},\hat n)$ and $U_i(\nu_{0i},\hat n)$,
then Equation~(\ref{seconde}) could be solved separately to determine
the coefficients $\theta_i(\nu)$ in polarisation.

That is, Equation (\ref{seconde}), written for $Q$ and $U$ in each
pixel $\hat n$ becomes
\begin{align}
\label{capQ}
Q_\mathrm{tot}(\nu,\hat n) & =\sum_{i=1}^{n} \; Q_i(\nu,\hat n) \\
U_\mathrm{tot}(\nu,\hat n) & =\sum_{i=1}^{n} \; U_i (\nu,\hat n).
\label{capU}
\end{align}
Inserting the definitions of the Stokes parameters in terms of the
fractional linear polarisation $f$ and the polarisation angle
$\gamma$, $Q=f I \cos(2\gamma)$ and $U=f I \sin(2\gamma)$,
Equations~(\ref{capQ}) and (\ref{capU}) become
\begin{align}
Q_\mathrm{tot}(\nu,\hat n) & =  \sum_{i=1}^n \; q_i(\nu,\hat n) \;  T_i(\nu ,\hat n) \\
U_\mathrm{tot}(\nu,\hat n) & =  \sum_{i=1}^n \; u_i(\nu,\hat n) \;  T_i(\nu ,\hat n)
\end{align}
where the correlation coefficients are
\begin{align}
\label{q-u}
q_i(\nu,\hat n) & = f_i(\nu,\hat n) \, \theta_i(\nu) \, \cos[2\gamma_i(\nu ,\hat n)] \quad \mathrm{and}\\
u_i(\nu,\hat n) & = f_i(\nu,\hat n)\, \theta_i(\nu) \, \sin[2\gamma_i(\nu ,\hat n)].
\label{q-u2}
\end{align}
To apply C-C as we did in the temperature analysis, we assume again
that the coefficients $q_i(\nu,\hat n)$ and $u_i(\nu,\hat n) $ are
constant over the analysed region of the sky.

We could still apply this procedure if we had complete information
about the polarisation angle for every foreground.  For example, one
could use simulated Galactic magnetic-field maps to estimate the
position angle of a thermal-dust polarisation-template and the K-band
polarisation data as a synchrotron angle template as has been done by
other authors (e.g., Dunkley et al.\ 2009; Page et al.\ 2007).

In this work we wish to determine if useful knowledge can be extracted
from the \emph{WMAP} polarisation data without making any assumptions
about the dust or synchrotron polarisation angles.  It is possible to
do so by considering the total polarisation
\begin{align}
P_\mathrm{tot}^2(\nu,\hat n)  & \equiv f^2(\nu,\hat n) T^2(\nu,\hat n) =  Q^2(\nu,\hat n) + U^2(\nu,\hat n) & & \\
          & =  \left(\sum_{i=1}^n q_i(\nu,\hat n)\,T_i(\hat
            n)\right)^2 + \left(\sum_{i=1}^n u_i(\nu,\hat n)\,T_i(\hat
            n)\right)^2.
\label{eq-p18}
\end{align}

Equation~\ref{eq-p18} can be re-written by replacing $q_i$ and $u_i$ using Equations
(\ref{q-u}) and (\ref{q-u2}) so that
\begin{multline}
P_\mathrm{tot}^2(\nu,\hat n) = \sum_{i=1}^n  f_i^2 \theta_i ^2 T_i(\hat n) ^2 + \\
+   \sum_{i=1}^n  \sum_{\substack{j=1 \\ j\neq i}}^n  f_if_j \theta_i\theta_jT_i(\hat n)T_j (\hat n) 
 \cos\left[2\left(\gamma_i(\hat n) - \gamma_j(\hat n)\right)\right].
\label{eq-p20}
\end{multline}
(Note that we have dropped the frequency-dependence in Equation~[\ref{eq-p20}]
for simplicity only.)
If we further assume that the polarisation angles are equal for all
emission components ($\gamma_i (\hat n)=\gamma_j (\hat n)$) then
\begin{align}
\label{PPP}
P_\mathrm{tot}(\nu,\hat n) = \sum_{i=1}^n  \rho_i(\nu) \,  T_i(\hat n)
\end{align}
where
\begin{equation}
\label{rho}
\rho_i(\nu) \equiv f_i(\nu) \, \theta_i(\nu).
\end{equation}

Equation (\ref{PPP}) is analogous to Equation (\ref{seconde}), but
with the polarised-intensity correlation coefficients $\rho_i(\nu)$
rather than $\theta_i(\nu)$.  Therefore, $\rho_i(\nu)$ can be computed
by analogy with the C-C used to calculate the $\theta_i(\nu)$ given by
Equation~(\ref{CCformula2}).  The fractional polarisation coefficient
$f_i(\nu)$ can be extracted for each component at each frequency from
these two correlation coefficients:
\begin{equation}
\label{frac}
f_i (\nu)=\frac{\rho _i (\nu)}{\theta_i (\nu)},
\end{equation}
with an uncertainty of:
\begin{equation}
\label{frac_error}
\delta f_i (\nu)= f_i(\nu) \times \sqrt{\left(\frac{\delta \rho_i}{\rho_i}\right)^2 + \left(\frac{\delta \theta_i}{\theta_i}\right)^2}
\end{equation}

The assumption of an equal polarisation angles for all emission
components is reasonable because the direction of all diffuse emission
types is determined by the direction of the magnetic field.  This
assumption leads to a total polarisation (summed over all emission
components) with angles equal to the individual components, so that it
is easily falsifiable by any measurement in which the total position
angle changes with frequency.  While such angle rotations with
frequency are observed in the \emph{WMAP} data they are typically
smaller than $\sim 20^{\circ}$ at a resolution of $4^{\circ}$.
Moreover, the nearly-constant angle assumption is supported by Kogut
et al.\ (2007) who found that the fits based on K-band
(synchrotron-dominated) polarisation angles or the
starlight-polarisation angles gave similar results, with a difference
in the goodness-of-fit of about $1\%$.  The advantage of our approach
is that no models are required to estimate the polarisation angles.

The fractional polarisation $f_i(\nu)$, as given in
Equation~(\ref{frac}), is the ratio of \emph{correlation
  coefficients}.  One should keep in mind that negative values of the
correlation coefficients $\rho_i(\nu)$, and therefore $f_i(\nu)$, are
in fact physically meaningful.  Anti-correlations between the
polarisation data and a total intensity template will occur in regions
where the polarised emission drops with increasing total intensity.
For example, this behaviour can be expected in regions where the
polarisation drops due to increased line-of-sight averaging towards
regions of increasing column density.  This point provides another
reason to consider small areas of the sky where the polarisation
fraction is more likely to be constant.

\subsection{Polarisation noise bias}

Given the $Q$ and $U$ values in each pixel, we calculate the polarised
intensity using
\begin{equation}
P(\hat n) =  \Big(\langle Q\rangle ^2 (\hat n) \; + \; \langle U\rangle ^2 (\hat n) \Big)^{1/2},
\end{equation}
and its associated uncertainty, where the direction $\hat n$ denotes each
$N_{\mathrm{side}}=32$ ($\approx 2^{\circ} \times 2^{\circ}$) piece of sky.
The correlation analysis is then carried out on the $P(\hat n)$ in each
section of the sky defined in Section~\ref{sec:results}.

The polarisation is a complex quantity which can be described by
either its real and imaginary parts ($Q$ and $U$), or its amplitude
and phase ($P$ and $\gamma$).  If one assumes gaussian statistics for
$Q$ and $U$, then the non-linear transformation into $P$ and $\gamma$
results in non-gaussian statistics for these quantities.  The key
point for our analysis is that $P$ has an asymmetric probability
distribution (e.g., Vaillancourt 2006; Simmons \& Stewart 1985)
resulting in a positive bias.  We have corrected for this bias using
the formula:
\begin{align}
  P_\mathrm{true} = 0\qquad \qquad & \quad\mbox{for}\quad  P/\sigma_p < \sqrt{2} \\
  P_\mathrm{true} \approx \sqrt{P^2 - \sigma_p^2} & \quad\mbox{for}\quad  P/\sigma_p > \sqrt{2}
\end{align}
where $\sigma _P$ is the polarisation uncertainty defined in
Section~\ref{sec:errors}.  While the last expression is not exactly
precise one can limit the effects of the bias by only using data above
some polarisation signal-to-noise level.\footnote{No bias-correcting
  algorithm is exact. For more information see Simmons \& Stewart
  (1985), Vaillancourt (2006), and references therein.} Such a bias can be estimated using monte carlo simulations as described in Section~\ref{sec:sim}.

\section{Results}
\label{sec:results}

\begin{figure}
\begin{center}
\includegraphics[width=0.5\textwidth,angle=0]{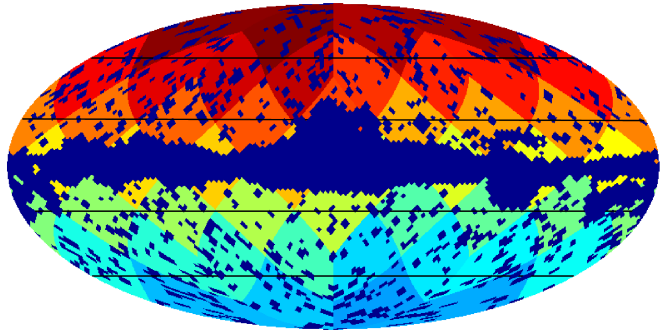}
\caption{All-sky map with the 48 HEALPix $N_\mathrm{side} = 2$ regions
  in different colours with the KQ85 mask applied (dark
  blue). Horizontal black lines represent the isolatitude slices
  (latitude analysis) with $b=-90^{\circ} / -50^{\circ}, \:
  -50^{\circ} / -20^{\circ},\: -20^{\circ} / \: 20^{\circ}, 20^{\circ}
  / \:50^{\circ}, 50^{\circ} / \: 90^{\circ}$. \label{PixReg}}
\end{center}
\end{figure}

We used C-C to compute, at each \emph{WMAP} band, the emission
coefficients of the three Galactic foreground components plus the
offset monopole template.  The analysis presented in this work was
done in three different spatial configurations: on the whole sky
outside of the KQ85 Galactic mask (hereafter the ``all-sky''
analysis), in different isolatitude slices (hereafter the ``latitude''
analysis) with $b=-90^{\circ} / -50^{\circ}, \: -50^{\circ} /
-20^{\circ},\: -20^{\circ} / \: 20^{\circ}, 20^{\circ} / \:50^{\circ},
50^{\circ} / \: 90^{\circ}$, and in 48 HEALPix $N_{\mathrm{side}}=2$
regions (hereafter the ``pixel'' analysis).  Selecting such partitions
allows us to investigate how the foreground properties depend on
direction and latitude (see also Kogut et al.\ 2007;
Miville-Desch\^{e}nes et al.\ 2008), while increasing the likelihood
that the area we are analysing has a uniform fractional polarisation
(see Section~\ref{sec:FracPol}). The best pixel area size among those
tried is $N_{\mathrm{side}}=2$; 
it is small enough that the assumption of constant polarisation
fraction is more likely to be satisfied, and large enough to contain
spatial variations needed for the C-C method.
Moreover, these areas match with previous works (e.g. Dunkley et
al. 2009) 
allowing us to compare our results.
Fig.~\ref{PixReg} shows the sky-locations of pixels used for the
pixel-analysis and the latitude slices used for the latitude-analysis
(see Table~\ref{tab:FracPol} for the mean latitudes and longitudes of
each pixel).

From the C-C coefficients we derive results on the relative level of
each foreground in a given region, its frequency dependence in
temperature and polarisation, and its polarisation fraction.  We note
here that it is not always possible to extract all this information at
all frequencies and for all regions due to limited signal-to-noise
ratios. Therefore, we concentrate on results
where there is $2\sigma$ statistical significance.

\subsection{Tests on a simulated sky}
\label{sec:sim}

In order to test the analysis method described in the preceding
sections, we ran the CCA code on 500 noise realisations of a simulated sky with a given
foreground parametrisation. We created sky maps in both temperature and polarisation, based on the three template temperature maps from Section~\ref{sec:templates} and scaled them using the average coefficients from Davies et al.\ (2006) at each frequency.  We used a constant polarisation
fraction of $20\%$ for synchrotron, $3.5\%$ for dust\footnote{The 3.5\% polarisation represents the polarisation for the total dust-correlated component at K-band, which will be dominated by anomalous microwave emission, and virtually no contribution from thermal dust.}, and $0\%$ for
free-free and a polarisation angle for both synchrotron and dust based on the \emph{WMAP}5 K-band map. We also made a set of simulations with no dust polarisation to compare cases without and without dust polarisation. The CMB model is a Gaussian realisation based on the \emph{WMAP}5 concordance model. We used the LAMBDA $N_{\mathrm{side}}=512$ $I$, $Q$, and
$U$ inverse covariance matrices to produce a $N_{\mathrm{side}}=512$
noise realisation, which was added to the CMB and foreground maps, and then degraded to the resolution used in our analysis, i.e.,
$N_{\mathrm{side}}=32$.

We compared the results for each of the three components for the 500 realisations in each pixel. We focused on the results from K-band, where the data had the most significance. The recovered coefficients were, in general, close to the input values with no large biases. In temperature, the results were all within $3\sigma$ of the input value, except for one region for synchrotron (region 10 at $3.3\sigma$) and one region for dust (region 12 at $3.1\sigma$). The estimated uncertainties corresponded well to the scatter in the coefficients from the 500 realisations.

In polarisation, the results for synchrotron and free-free
  were in good agreement with the input values. The mean recovered
  fractional polarisation for synchrotron was $20.0\pm0.3\,\%$ for the
  case with no dust polarisation and $19.5 \pm0.3\,\%$ with dust
  polarisation. For free-free, we obtained mean values of $-0.02\pm
  0.18\,\%$ and $-0.04\pm0.19\,\%$ for each case, respectively. This
  gives us confidence that the method works as expected and yields
  reasonable results.

However, the values recovered for dust did indicate bias i.e. an additional systematic error. The recovered fractional polarisation was $0.003\pm0.15\,\%$ and $0.38\pm0.15\,\%$ for the simulations with no dust polarisation, and $3.5\,\%$ dust polarisation, respectively. This shows that simply averaging all the polarisation values within a pixel will not yield a reliable estimate if the true value is as low as a few percent. The situation improves when only using regions with significant ($>2\sigma$) detections. We found 9 regions that were significant at the $2\sigma$ level, which resulted in a weighted average of $2.0\pm0.2\,\%$, i.e., an underestimate by $1.5\,\%$ in absolute value. For the case with no dust polarisation, no pixels were detected at greater than the $2\sigma$ level. Similarly, for free-free, we obtained an average value of $1.4\pm0.4\,\%$ when using the 6 regions that were significant, i.e., an overestimate of $1.4\,\%$ in absolute value.

From these tests, we conclude there is a bias in the average fraction polarisation of $\approx 1.5\,\%$, when the true fractional polarisation is a few percent or less. For a single region, we found cases where the bias is up to $\approx 5\,\%$. The reason for this bias appears to lie in cross-talk between the various components. For example, for regions we recover a significant dust polarisation, with no dust polarisation in the simulation, one can see a bias in either the synchrotron or free-free polarisation; this can explain why the average synchrotron polarisation is slightly low, at the $1.7\sigma$ level, in the case where there is dust polarisation. Nevertheless, we can still obtain useful constraints on the fractional polarisation with the current data, keeping in mind the amplitude of possible biases.

Also, we can use the results of these simulations to estimate the significance of results for individual pixels. These will be discussed separately, in the following sections on the synchrotron, free-free and dust results. For the cases where there is significant bias observed in the simulations, the fractional polarisation results will be presented in terms of a statistical (stat) and systematic (sys) error. For example, for the significant dust regions above, the weighted average is $2.0\pm0.2\, \textrm{(stat)} \pm1.5\, \textrm{(sys)}\,\%$.

\subsection{All--sky versus pixel analysis }
\label{sec:results_general}

\begin{figure}
\begin{center}
\includegraphics[width=0.5\textwidth,angle=0]{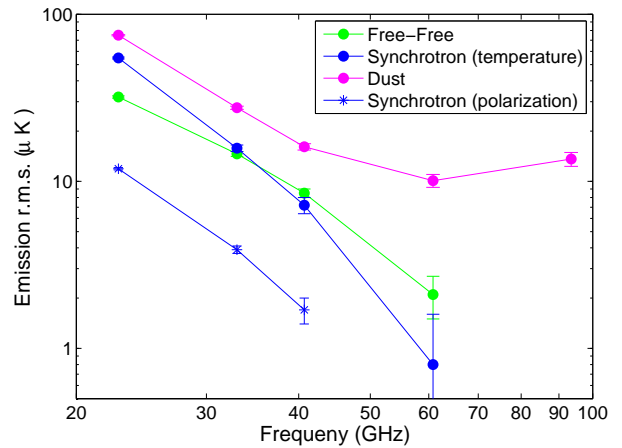}
\caption{R.M.S. emission spectrum of Galactic foreground components
  from the all-sky (KQ85 cut) intensity and polarisation analysis, in
  $\mu$K\@.   Only the polarised synchrotron emission is
  plotted as the dust-correlated and free-free-correlated polarised
  emission are consistent with zero for the all-sky region.
\label{confT}}
\end{center}
\end{figure}

We show in Fig.~\ref{confT} the contribution of each foreground to the
temperature emission as a function of frequency. These estimates,
reported in Table~\ref{tab:allSky}, were obtained by multiplying the
C-C coefficient for a given foreground at a given frequency with the
root-mean-square (r.m.s) of each foreground template outside the KQ85
mask.  The figure clearly shows that the dust-correlated emission is
the dominant component in temperature at all frequencies (see also
Davies et al.\ 2006). The rise from 61-to-94 GHz indicates the high
frequency contribution of thermal dust, while at lower {\it WMAP}
frequencies, there is a significant contribution from anomalous dust.
It is this low frequency dust-correlated emission that has been
tentatively identified with emission from small rapidly-spinning dust
grains (Draine \& Lazarian 1998a,b).  However, the dust contribution
to polarisation is typically subdominant at low frequencies and will
be discussed further in Section~\ref{sec:ddd}.

\begin{table}
  \begin{center}
    \begin{tabular}{cccccc}
      \hline
         & {\bf K-band} & {\bf Ka-band} & {\bf Q-band}  & {\bf V-band} & {\bf W-band}\\
      \hline
      & \multicolumn{5}{c} 
      {\textbf{Temperature r.m.s. emission [$\mu K$]}} \\ 
      \hline
      S  &  54.8  $^{\pm  0.7}$&   15.8 $^{\pm  0.7}$ &   7.2  $^{\pm  0.8}$ &   0.8   $^{\pm  1}$   &  -0.8 $^{\pm  1.3}$  \\      
      D             &   75.0  $^{\pm  0.6}$&   27.6  $^{\pm  0.6}$ &  16.1  $^{\pm  0.7}$ &    10.1 $^{\pm  0.9}$ &  13.6 $^{\pm  1.3}$ \\
      F-F       & 32.0   $^{\pm  0.5}$&  14.6  $^{\pm  0.5}$  &  8.5   $^{\pm  0.5}$ &  2.0   $^{\pm  0.6}$    &  -0.2  $^{\pm  1.1}$  \\
      \hline
     & \multicolumn{5}{c} 
      {\textbf{Polarisation r.m.s emission  [$\mu K$]}} \\   
      \hline
      S   &  11.9  $^{\pm  0.2}$ &    3.9  $^{\pm  0.2}$  &  1.7  $^{\pm  0.3}$ &  0.1   $^{\pm  0.4}$  &    0.1  $^{\pm  0.4}$ \\
      D              &  -0.2  $^{\pm  0.1}$ &  -0.1  $^{\pm  0.2}$ &   0.0  $^{\pm  0.3}$ &   0.1    $^{\pm  0.3}$  &    0.0  $^{\pm  0.4}$ \\
      F-F    & -0.2  $^{\pm  0.1}$ &  -0.3  $^{\pm  0.2}$ &   -0.2  $^{\pm  0.2}$ &   0.0  $^{\pm  0.4}$ &   0.0  $^{\pm  0.6}$ \\
      \hline
    \end{tabular}\caption{Emission r.m.s.\ values for the all-sky
      analysis in both temperature and polarisation for synchrotron (S), 
      dust (D), and free-free (F-F) in the five \emph{WMAP} bands. These values are plotted in  Fig.~\ref{confT}.  \label{tab:allSky}} 
  \end{center}
\end{table}

\begin{figure*}
\begin{center}
\includegraphics[width=1\textwidth,angle=0]{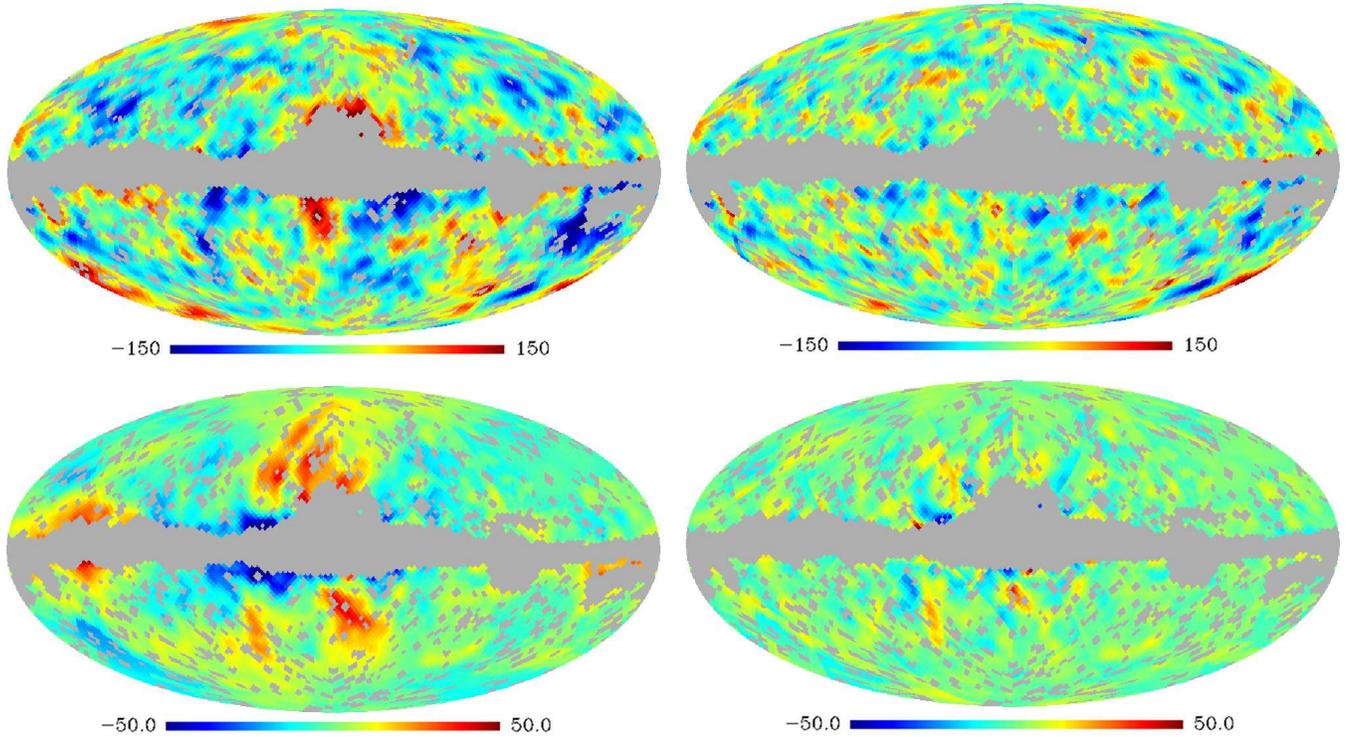}
\caption{Residual maps (on mollweide projection) for the \emph{WMAP}
  K-band temperature ({\it top}) and polarised intensity ({\it
    bottom}) analyses.  The results using the all-sky (KQ85 cut)
  region are reported ({\it left}) together with the results using the
  48 HEALPix $N_\mathrm{side} = 2$ regions ({\it right}). The units of
  the color-scale are $\mu$K. \label{residuals}}
\end{center}
\end{figure*}

\begin{table}
\begin{center}
  \begin{tabular}{ccc}
    \hline
    & {\bf K-band } & {\bf Ka-band} \\
    \hline
    All-Sky Temperature &  0.94 &   0.68   \\
    All-Sky Polarisation  & 5.37   &   0.73   \\
    \hline
    Latitude Temperature &  0.97  &   0.74   \\
    Latitude Polarisation  & 3.72   &   0.56   \\
    \hline
    Pixels Temperature &  0.65  &   0.48   \\
    Pixels Polarisation  & 1.65   &  0.30   \\
    \hline
  \end{tabular}\caption{Reduced-$\chi ^2$ for the three analysis types implemented in the paper. 
The values are the average over the different regions (when more than one is used). \label{tab:Chisquared}} 
\end{center}
\end{table}

Given the C-C coefficients and foreground templates, and the
assumption that the templates trace all the foreground emission, we
can generate a model of the total foreground emission in the
\emph{WMAP} data.  Comparing this model to the actual \emph{WMAP} data
yields a set of residuals which we use to estimate the extent to which
a single fit-coefficient (in the case of the all-sky analysis)
represents the foregrounds across the whole sky.  Results for the
K-band, where the foregrounds are the brightest, are reported in
Fig.~\ref{residuals}(\emph{left}) for both temperature and
polarisation.  In the all-sky temperature residuals, the largest
residuals occur near the Galactic plane, particularly near the
Galactic centre.  This feature has been interpreted as the Galactic
haze, possibly associated with annihilating particles (Finkbeiner et
al.\ 2004; Cumberbatch et al.\ 2009).  However, this feature is not
evident in the pixel-analysis where we compute separate C-C
coefficients in 48 different regions
(Fig.~\ref{residuals}(\emph{right})). This is a consequence of a
varying synchrotron spectral index, which is somewhat flatter than
average near the Galactic centre, due to a diversified population of
electrons in the Galaxy.

Bright residuals in the all-sky analysis are also concentrated on the
well-known Galactic spurs such as the North-Polar Spur (NPS) above and
below the Galactic centre region. These spurs are known to have a
slightly steeper synchrotron spectral index than average and,
therefore, are not well suited for the all-sky polarisation
analysis. Furthermore, these large coherent features are expected to
have significantly different fractional polarisations. The residual
map for the 48 pixel-analysis (Fig.~\ref{residuals}) shows less
extended structure, with more localised residuals near the mask
edge. This indicates that a local analysis is more suited to handle
the complex foreground variations. In summary, Fig.~\ref{residuals}
shows that it is suboptimal to model foregrounds without considering that
they are spatially varying in the sky.

To have a quantitative understanding of the goodness-of-fit we compute
the reduced-$\chi^2$ for each type of analysis, both in temperature
and polarisation. These are reported for the \emph{WMAP} K- and
Ka-bands in Table~\ref{tab:Chisquared}. Again we see that neither the
all-sky nor the latitude analyses are as accurate as the
pixel-analysis, especially in the K-band polarisation.  A limitation of the pixel-analysis is that, in any
given pixel, there may not be sufficient data points to yield a
signal-to-noise ratio larger than 1. This is a particular problem for
high frequencies (mostly V- and W-bands) in polarisation.  In most
cases we will be limited to studying the \emph{WMAP} foregrounds at K-band only.  The 48 emission coefficients in K-band, for each
foreground component in intensity and polarisation, are reported in
Table~\ref{tab:FracPol}.

\begin{table*}
\begin{center}
  \begin{tabular}{cccccccccccc}
    \hline
    {\bf Pixel} &  {\bf Pixels}  &   $\langle$ {\bf Lat} $\rangle$  &$\langle$ {\bf Long} $\rangle$ & {\bf Synch} K & {\bf Dust} K & {\bf Dust} W & {\bf Synch \%} & {\bf Dust \%} & {\bf F-F \%} & $\chi^2_T$ & $\chi^2_P$ \\
    \hline
$1$ & 207 & 65.7 & 46.2 & 38.9 $^{\pm 3.8}$ & 19.6 $^{\pm 4.2}$ & 11 $^{\pm 5.3}$        & {\bf 40.6 $^{\pm 4.7}$} &   $<$ 11.0                     &   ---                & 0.45  & 0.41     \\  
$2$ & 214 & 66.7 & 134.8 & ---                    &  ---                     &   ---                   &   ---                              &   ---                     &   ---                     & 0.31  & 0.81     \\  
 $3$ & 214 & 66.8 & 224.5 & ---                    & 9.5 $^{\pm 3.2}$ &   ---                        &   ---                            &   ---                    &   ---                     & 0.49  & 0.73    \\  
$4$ & 202 & 67 & 318 & 30.6 $^{\pm 3.4}$         & 9.5 $^{\pm 3.4}$ &  ---                       & {\bf 23.8 $^{\pm 3.6}$ } &   ---                     & $<$ 11.6              & 0.84  & 0.41         \\
\hline  
$5$ & 189 & 42.3 & 21.3 & 49.5 $^{\pm 3.3}$     & 56.8 $^{\pm 4.3}$ &  ---                        & {\bf 26.1 $^{\pm 2.4}$} & {\bf 7.2 $^{\pm 2.3}$}  &   ---               & 0.45  & 3.01         \\  
$6$ & 195 & 43 & 69.7 &  ---                          & 19.9 $^{\pm 3.2}$ & 6.2 $^{\pm 3.5}$     &   ---                             & {\bf 10.6 $^{\pm 4.9}$ } &   ---                & 0.49  & 2.11          \\  
$7$ & 209 & 42 & 110.3 & 9.9 $^{\pm 2.6}$        & 32.7 $^{\pm 3.4}$ & 7.4 $^{\pm 3.7}$     &  {\bf 17.8 $^{\pm 7.5}$} &   $<$ 5.0                   &   ---               & 0.55  & 0.75         \\  
$8$ & 195 & 41.9 & 159.8 & 10.2 $^{\pm 4.5}$   & 30.8 $^{\pm 4.3}$ & 8.4 $^{\pm 4.8}$     & 42.4 $^{\pm 21.6}$          &   $<$ 7.6                    &  ---                  & 0.49  & 1.42        \\  
$9$ & 211 & 41.7 & 200.1 & 8.7 $^{\pm 3.7}$     &  ---                     &  ---                   & 16.7 $^{\pm 14.4}$          &  ---                    &   ---                       & 0.56  & 0.22   \\  
$10$ & 232 & 41.6 & 250.8 & 12.9 $^{\pm 3.4}$ & 12.8 $^{\pm 2.8}$ &  ---                    & {\bf 18.3 $^{\pm 7.6}$} &   $<$ 12.8                    & $<$ 12.4        & 0.60  & 0.31             \\  
$11$ & 205 & 42.1 & 289.8 & 16.8 $^{\pm 2.9}$ & 23.8 $^{\pm 3.4}$ &   ---                        & {\bf 21.3 $^{\pm 6.2}$} &  ---                     & $<$ 11.6            & 0.47  & 1.05       \\  
$12$ & 214 & 42.4 & 339.6 & 25.7 $^{\pm 4.1}$ & 43.5 $^{\pm 4.2}$    &  ---                         & {\bf 22.5 $^{\pm 5.6}$} &   $<$ 5.8              &   ---               & 0.37  & 2.16           \\ 
\hline 
$13$ & 44 & 30.9 & 4.1 & 21.1 $^{\pm 4.5}$       & {\bf  69.6 $^{\pm 4.9}$ } & 14.1 $^{\pm 5.2}$ & {\bf 44.3 $^{\pm 12.8}$} & $<$ 3.6         & $<$ 6.4             & 1.32  & 2.32             \\  
$14$ & 200 & 21.1 & 44.9 &  {\bf 104.8 $^{\pm 4.4}$} & 54.7 $^{\pm 4.3}$ &  ---                    & {\bf 23.7 $^{\pm 1.4}$} &  ---                    &   ---                & 0.98  & 2.02          \\  
$15$ & 158 & 21.4 & 88.6 &  ---                     &  {\bf 110.5 $^{\pm 3.4}$} & 17.2 $^{\pm 3.4}$ &  ---                    & $<$ 1.6         & $<$  5.0                      & 0.63  & 1.87     \\  
$16$ & 167 & 22.7 & 135 &  {\bf 81.9 $^{\pm 5.3}$} & {\bf  76.4 $^{\pm 4.2}$} &  ---             & {\bf 19.8 $^{\pm 1.9}$} &   ---                    & 7.6 $^{\pm 4.8}$     & 0.61  & 0.28  \\  
$17$ & 211 & 20.1 & 180.5 & 22.9 $^{\pm 3.9}$  &  {\bf 80 $^{\pm 3.4}$}    & 20.9 $^{\pm 4.0}$    &{\bf 29.6 $^{\pm 6.9}$} & 2.6 $^{\pm 1.3}$  &  ---                   & 2.25  & 1.39      \\  
$18$ & 209 & 20.3 & 225.1 &  ---                    & {\bf  62.1 $^{\pm 4.3}$} & {\bf 26.8 $^{\pm 4.4}$} &   ---                   & $<$ 2.4           &  ---                      & 1.10  & 0.43    \\  
$19$ & 172 & 22.6 & 270.4 & 12.2 $^{\pm 4.4}$  & 22.9 $^{\pm 3.7}$           &   ---                              & {\bf 34.9 $^{\pm 16.2}$} &   $<$  9.8                    & $<$ 3.8       & 1.26  & 1.22         \\  
$20$ & 162 & 22.9 & 315 &  {\bf 97.3 $^{\pm 3.5}$} & 45.2 $^{\pm 4.6}$     &   ---                      & {\bf 14.8 $^{\pm 1.1}$} & $<$ 7.2              &   ---                  & 0.52  & 2.10       \\  
\hline
$21$ & 28 & -8.1 & 23.5 &  {\bf 59.7 $^{\pm 5.9}$} & 54.3 $^{\pm 7.2}$       &   ---                       &   ---                    & 8.4 $^{\pm 5.9}$  &  ---                      & 0.81  & 6.08   \\  
$22$ & 86 & 0 & 66.1 & 48.3 $^{\pm 6.7}$            &  {\bf 61.2 $^{\pm 5.2}$} &   ---                    & 5.7 $^{\pm 3.4}$ &$<$ 5.0                   & $<$ 6.0                      & 0.50  & 5.14  \\  
$23$ & 50 & -5.9 & 110.4 & 33.1 $^{\pm 5.4}$     &  {\bf 84.1 $^{\pm 6.4}$} &  ---                   & {\bf 10.8 $^{\pm 5.1}$} &{\bf 6.0 $^{\pm 1.8}$ } & $<$ 7.0           & 0.78  & 1.33              \\  
$24$ & 53 & 6.3 & 158.1 & 12.8 $^{\pm 6.2}$       &  {\bf 102.5 $^{\pm 6.1}$} & 29.4 $^{\pm 7.2}$ & 43.9 $^{\pm 26.9}$ & 2.3 $^{\pm 2}$  &      ---                      & 0.57  & 1.30   \\   
$25$ & 88 & 5.6 & 203.6 & ---                          &  {\bf 100.9 $^{\pm 5.3}$} &{\bf 22.7 $^{\pm 6.0}$} &  ---                    & 2.8 $^{\pm 1.8}$  & 8.7 $^{\pm 5.1}$     & 1.78  & 0.84  \\  
$26$ & 64 & 3 & 244.4 & 11.7 $^{\pm 7.6}$          & 54.3 $^{\pm 5.5}$               &   ---                      &   ---                   & $<$ 5.8           & $<$ 3.2                      & 1.15  & 0.46     \\  
$27$ & 74 & -0.8 & 291.3 &  {\bf 58.2 $^{\pm 4.7}$ }& {\bf  60.1 $^{\pm 4.3}$} &  ---                 &{\bf 9.1 $^{\pm 2.2}$} & 2.7 $^{\pm 1.7}$  & $<$ 12.6            & 0.90  & 1.11       \\  
$28$ & 16 & -12.2 & 336.7 &  {\bf 88 $^{\pm 17.2}$} & 35.8 $^{\pm 16.2}$       &   ---                   &   ---                      &   ---                     &   ---                   & 0.88  & 2.80      \\  
\hline
$29$ & 180 & -22.3 & 1.5 &  {\bf 62.5 $^{\pm 4.9}$} &  {\bf 64.9 $^{\pm 3.5}$} & 13.8 $^{\pm 4.1}$ & {\bf 19.6 $^{\pm 2.6}$ }&   ---                    &  ---               & 0.47  & 0.49          \\  
$30$ & 214 & -21 & 45.2 & 39.7 $^{\pm 5.6}$      &  {\bf 74.4 $^{\pm 5.6}$ }  & 15.8 $^{\pm 5.6}$     &   6.8 $^{\pm 3.6}$      &$<$ 4.0                  &   ---                  & 1.00  & 2.20       \\  
$31$ & 177 & -21.8 & 89.4 &  ---                     & {\bf  88.1 $^{\pm 3.6}$}  & {\bf 28.3 $^{\pm 4.4}$} &  ---                    & $<$ 2.0                 &   ---                  & 0.49  & 0.81       \\  
$32$ & 195 & -19.9 & 135.2 & 34.2 $^{\pm 4.7}$ & {\bf  80 $^{\pm 4.1}$ }     & 12.6 $^{\pm 4.6}$            & {\bf 36.9 $^{\pm 6}$ }   & 1.5 $^{\pm 1.4}$  &    $<$ 6.2          & 0.99  & 1.17            \\  
$33$ & 102 & -24.9 & 178 & 17.8 $^{\pm 3.5}$    &  {\bf 107.1 $^{\pm 3.8}$} & {\bf 30.2 $^{\pm 4.0}$} & {\bf 52.9 $^{\pm 12.5}$} & $<$ 2.4               &  $<$ 6.0       & 1.13  & 0.82         \\  
$34$ & 150 & -20.5 & 226.9 &{\bf 72.6 $^{\pm 6.4}$} &{\bf 104.8 $^{\pm 5.2}$}& {\bf 22.7 $^{\pm 7.4}$ }& {\bf 7.1 $^{\pm 2.3}$} &{\bf 3.2 $^{\pm 1.5}$ } & $<$ 10.2    & 0.45  & 0.85  \\  
$35$ & 145 & -22.2 & 272.1 &  ---                   & 51.1 $^{\pm 5.2}$                 &   ---                               &  ---                           & {\bf  4 $^{\pm 1.8}$}  &          $<$ 2.2    & 0.50  & 0.74  \\  
$36$ & 181 & -21 & 314 & 39 $^{\pm 4.6}$          &  {\bf 79.3 $^{\pm 3.5}$}    & 14.3 $^{\pm 4.4}$               & {\bf 26.4 $^{\pm 4.2}$} & $<$ 2.2                  & ---     & 0.49  & 1.23  \\  
\hline
$37$ & 207 & -41.8 & 19.3 & 24.7 $^{\pm 4.5}$   & 18.7 $^{\pm 6}$             &   ---                    & {\bf 21.6 $^{\pm 6.6}$} &   ---                     &   ---                    & 0.39  & 0.61      \\  
$38$ & 191 & -42.7 & 69.6 & ---                      & 25.3 $^{\pm 3}$           &   ---                    &   ---                   & $<$ 6.6                  &   ---                     & 1.05  & 2.12     \\  
$39$ & 206 & -41.9 & 109.9 & 10.8 $^{\pm 3.7}$ & 16.4 $^{\pm 2.4}$        &   ---                            &   ---                    &$<$ 10                  &   ---                    & 0.47  & 0.60      \\  
$40$ & 202 & -42 & 159.8 & ---                       & {\bf  90.5 $^{\pm 3.2}$} &   ---                    &  ---                    & $<$ 2.6               & ---                           & 0.98  & 1.09    \\  
$41$ & 213 & -42.8 & 200 & ---                       & 47.8 $^{\pm 2.6}$       & 7.2 $^{\pm 2.8}$             &  ---                    &$<$ 3.8                   & $<$ 8.4                    & 1.27  & 0.80    \\  
$42$ & 201 & -42.2 & 251.1 & ---                     & 13.8 $^{\pm 4.4}$       &   ---                           &  ---                   &   ---                   &   ---                      & 0.59  & 0.96   \\  
$43$ & 217 & -42.4 & 290.5 & ---                     &  {\bf 66 $^{\pm 4}$ }& {\bf 15.7 $^{\pm 4.4}$}      &   ---                   & $<$ 2.2                 & $<$ 10.6         & 0.33  & 0.70             \\  
$44$ & 215 & -41.6 & 339.7 & 33.8 $^{\pm 4.2}$  & 13.1 $^{\pm 4.3}$        &   ---                         & {\bf 38.5 $^{\pm 5.6}$} &   ---                     &   ---             & 0.56 & 1.30             \\  
\hline
$45$ & 203 & -66.4 & 45.7 & 11.8 $^{\pm 2.8}$    & 11.5 $^{\pm 3.2}$ &   ---                         & {\bf 25.2 $^{\pm 9.1}$ } & {\bf 31.7 $^{\pm 11.7}$}  &   ---           & 0.40  & 0.32               \\  
$46$ & 202 & -66.3 & 134.4 & 20.5 $^{\pm 2.9}$  & ---                     &   ---                        &  ---                    &  ---                    &   ---                      & 0.55  & 1.60    \\  
$47$ & 206 & -66.7 & 224.2 & 10.3 $^{\pm 3.6}$  & 12.7 $^{\pm 2.5}$ &  ---                           & 13.8 $^{\pm 8.8}$ & $<$ 9.8                   &   ---                  & 0.91  & 0.72        \\  
$48$ & 193 & -65.7 & 313.6 &  ---                     & ---                    &   ---                         &   ---                    &  ---                     &   ---                   & 0.37  & 0.63       \\  

    \hline
  \end{tabular}
  \caption{Results of the pixel analysis.  Here we report the pixel
    region number, the number of
    data points, the average latitude and the average longitude of
    each region, the synchrotron-correlated temperature emission at
    K-band in $\mu$K, the dust-correlated temperature emission at
    K-band and at W-band in $\mu$K, and fractional polarisation
    percentage of synchrotron, anomalous dust and free-free at
    K-band. Bold faces are used to emphasise a component with a
    temperature emission higher than $60\,\mu$K at K-band and $15\,\mu$K at W-band, and are use to indicate regions where a particular
    component is found to be more than $2\sigma$ significantly
    polarised.  We indicate $2\sigma$ upper limits when in that region
    the fractional polarisation is compatible with zero. No data for fractional polarisations
    indicates upper limits reaching $100\%$.  In column  5, 6 and 7 we display detections that are significant at the $1\sigma$ level. Note that the systematic uncertainties are not included here and can be significant when the polarisation fractions are low, i.e. less than a few per cent (see Section~\ref{sec:sim}).
    The $\chi ^2$ values from temperature and polarisation analysis at
    K-band are also reported. \label{tab:FracPol}}
\end{center}
\end{table*}

\subsection{Synchrotron}
\label{sec:sss}

Synchrotron is the dominant contribution to the polarised K- and
Ka-band emission, especially in the northern hemisphere, as reported
in Table~\ref{tab:FracPol},
because of the strong emission associated with the north polar spur.
On the other hand, synchrotron is comparable to anomalous dust
emission in temperature, and again its emission is more intense in the
northen part of the sky.

From the all-sky analysis, using K- and Ka- band and the corrected
Haslam et al.\ map described before, we find $\beta_s=-3.32 \pm 0.12$
in intensity and $\beta_s=-3.01 \pm 0.03$ in polarisation, where the
errors come from the uncertainties on the C-C coefficients.  The
smaller polarisation uncertainties occur here because synchrotron is
the dominant polarised foreground at low frequencies, as opposed to
the intensity analysis where the dust-correlated and
free-free-correlated emission are comparable to the synchrotron.
These synchrotron spectral indexes are in good agreement with previous
results, i.e. $\beta_s \approx -3$ (e.g. Miville-Desch\^{e}nes et al.\
2008).

From the pixel analysis we find that the synchrotron emission is
polarised from $\approx 5\%$ in some regions at low Galactic latitude
up to $\approx 40\%$ at high latitudes, as reported in
Fig.~\ref{polPixel} (\emph{top}).  In Fig.~\ref{FracPolLatitude} the
same quantity is plotted as a function of the mean Galactic latitude
of each region. The decline of the polarised emission at low latitudes
is well known in the literature (e.g., Wolleben et al.\ 2006) and is
interpreted as a depolarisation effect near the Galactic plane ($|b|
\lesssim 20^{\circ}$) due to integration along the line-of-sight of
emission with different polarisation angles. The average polarisation
fraction in the pixel regions with ($|b| < 20^{\circ}$) is $8.6
\pm 1.7\, \textrm{(stat)} \pm 0.5\, \textrm{(sys)}\,\%$.  Uncertainties at high latitudes are usually larger than
at low latitude because the templates have less variations in their
structure at high latitudes. This renders the C-C analysis
increasingly degenerate, as explained in
Section~\ref{sec:errors}. Nevertheless, the synchrotron fractional
polarisation at high Galactic latitude is in the range $\sim 10-40\%$
with an average value of $19.3 \pm 0.8\, \textrm{(stat)} \pm 0.5\, \textrm{(sys)}\,\%$ and standard deviation of
$\pm11.0\%$ ($|b|>20^{\circ}$). As discussed in Section~\ref{sec:sim}, the synchrotron values are the most robust of the three components, and we believe the errors from the C-C method are representative, even after averaging.

Our estimate of the synchrotron fractional polarisation is consistent
with other work at the $1\sigma$ level.  Kogut et al.\ (2007) computed
the fractional synchrotron polarisation by dividing the polarised
emission (estimated using a pixel-by-pixel frequency fit), by a
synchrotron intensity map (based on a MEM analysis; Hinshaw et al.\
2007). They find a typical fractional synchrotron polarisation of
5--25 \%. However, since this intensity map was computed ignoring a
possible contribution from anomalous dust, the synchrotron intensity
may be overestimated, resulting in an underestimate of the fractional
polarisation.  On the other hand, Miville-Desch\^{e}nes et al.\ (2008)
included a spinning-dust contribution to the total K-band intensity,
and obtained a higher synchrotron fractional polarisation, up to 40\%.
The fact that our results are compatible with these implies that in
most of the pixel regions the assumptions described in
Section~\ref{sec:FracPol} are reasonable.

We have computed the synchrotron spectral index in every pixel with a
S/N ratio $>2$ again using only the K- and Ka-bands; see
Table~\ref{tab:SincSIndex}.  The problem in this case is that the
pixels with S/N ratio $> 2$ are few, and we conclude that the C-C
technique is not appropriate for a detailed spectral-index
analysis. Nonetheless, the spectral indices in polarisation are
consistent with other analyses (e.g., Dunkley et al.\ 2009) who find
$\beta_{s} \simeq -3$; we find an average value is $\beta_{s}=-3.24\pm
0.20$.

In temperature, the results are not compatible with polarisation, with
an average spectral index of $\beta_{s}=-2.02\pm 0.20$. This is
clearly too flat and must be an artifact of the analysis. We checked
that these results, especially in the 5 low $\beta_s$ regions, are not
due to a high correlation between synchrotron and dust. Without
sufficient differences in spatial morphology between synchrotron and
dust, confusion can occur between the two components in the C-C.  As
we expected, the dust and synchrotron templates are correlated with an
all-sky (outside the KQ85 mask) correlation coefficient of $r=0.48$.
In the regions where high synchrotron spectral indices are observed (4, 5, 16, 27 and 32)
the correlation between dust and synchrotron is higher ($r=0.71$).
This suggests again that, in temperature, the C-C in some pixel
regions is not able to correctly separate the synchrotron, free-free
and dust components. In fact, most of the areas where the spectral
index is biased by this effect are the ones that have problems for
synchrotron in our simulations, e.g. regions $16$, $27$,
$32$. Moreover, when fitting over larger regions, the spectral index moves to more typical
values of $\beta_{s}\approx -3$.  The same difficulty with C-C analysis was found in
previous work, e.g. Davies et al.~(2006), where the introduction of the
0.4\,GHz point of Haslam et al.\ was needed to obtain meaningful
spectral index values. However, we note that this problem is most significant at the higher frequency channels above 33\,GHz while K-band the coefficients are not strongly affected. Indeed, the spectral index computed from the weighted mean K-band coefficient alone, which corresponds to the mean spectral index between 408\,MHz and 23\,GHz, is $\beta=-2.97\pm0.01$.

\begin{table}
\begin{center}
  \begin{tabular}{ccc}
    \hline
    Pixel & $\beta_s$ Temperature  &  $\beta_s$ Polarisation  \\
    \hline
   1  &   $-2.88$   $^{\pm  0.92    }$  &   $-3.01$  $^{\pm   0.56  }$   \\
    4   & $-1.79$   $^{\pm  0.65  }$   & $-3.58$  $^{\pm   0.98  }$   \\
    5  & $-1.85$    $^{\pm 0.41   }$  &  $-3.39$  $^{\pm   0.64  }$     \\
    14 &  $-2.02$    $^{\pm 0.30  }$  &  $-3.44$ $^{\pm    0.41  }$   \\
    15  &    ---                        &  $-3.22$   $^{\pm  0.95  }$   \\
    16  & $-1.35$  $^{\pm  0.34  }$  & $-3.50$  $^{\pm   0.89  }$     \\
    20 &  $-2.28$   $^{\pm  0.25  }$  &  $-4.13$  $^{\pm   0.94  }$    \\
    21 &  $-2.30$    $^{\pm 0.73  }$ &    ---                       \\
    22 &  ---                          &    ---                        \\
    27  & $-1.48$    $^{\pm 0.45  }$  &    ---                     \\
    28 &  ---                          &    ---                        \\
    29 &  ---                           &     $-2.81$   $^{\pm  0.99   }$   \\
    32 &  $-1.73$  $^{\pm 0.92  }$ &  $-2.53$ $^{\pm    0.71  }$   \\
    44 &  ---                           &    $-3.36$ $^{\pm    0.82  }$    \\
    \hline
  \end{tabular}
  \caption{Synchrotron spectral index values in different
    pixel regions from the intensity and polarisation analyses at K- and Ka-bands.  
    Only the regions with a significant result are reported ($\sigma
    _{\beta _S} <1.0$).
    \label{tab:SincSIndex}}
\end{center}
\end{table}

\subsection{Dust}
\label{sec:ddd}

\begin{figure}
\begin{center}
\includegraphics[width=0.5\textwidth,angle=0]{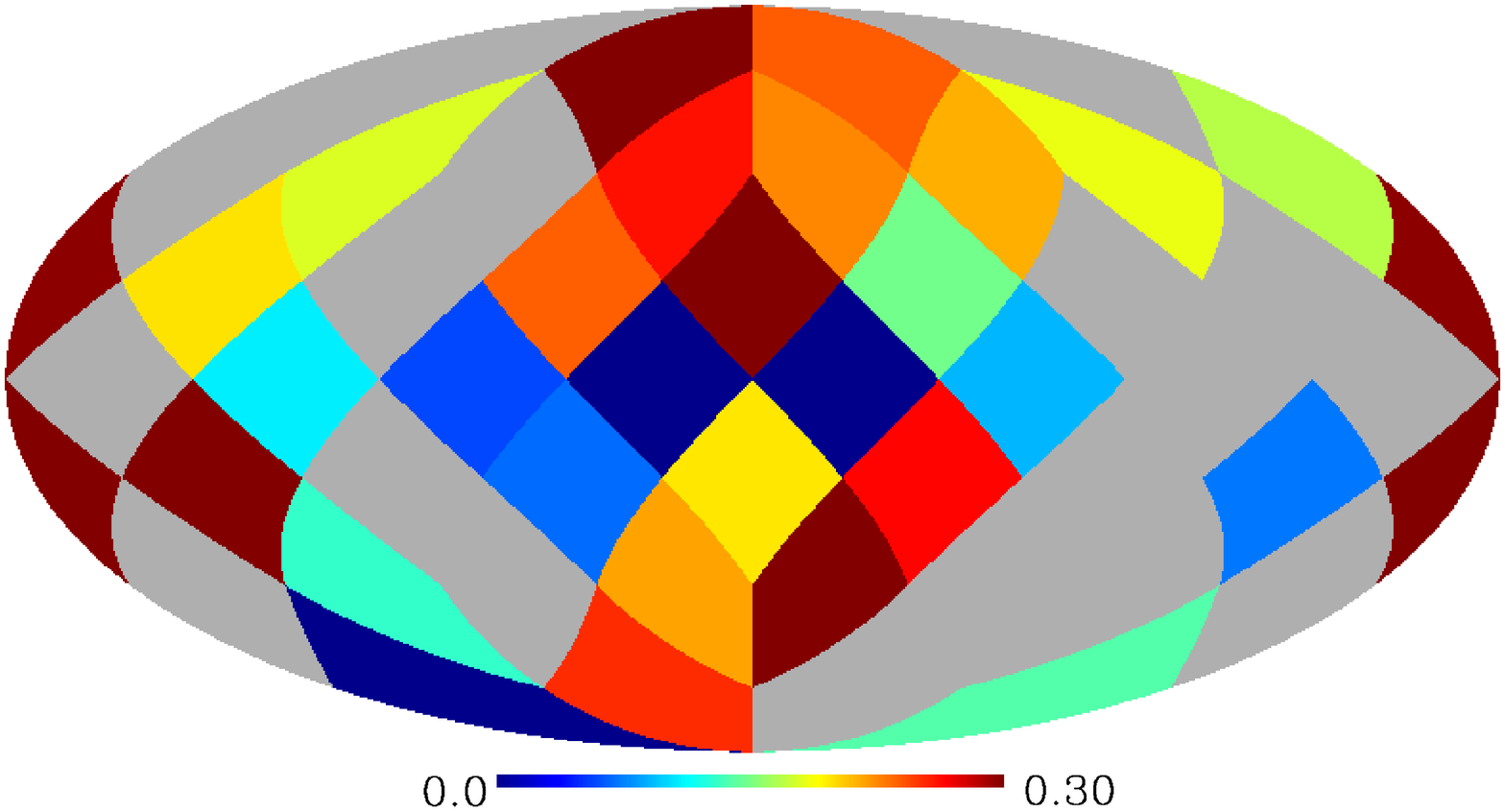}
 \includegraphics[width=0.5\textwidth,angle=0]{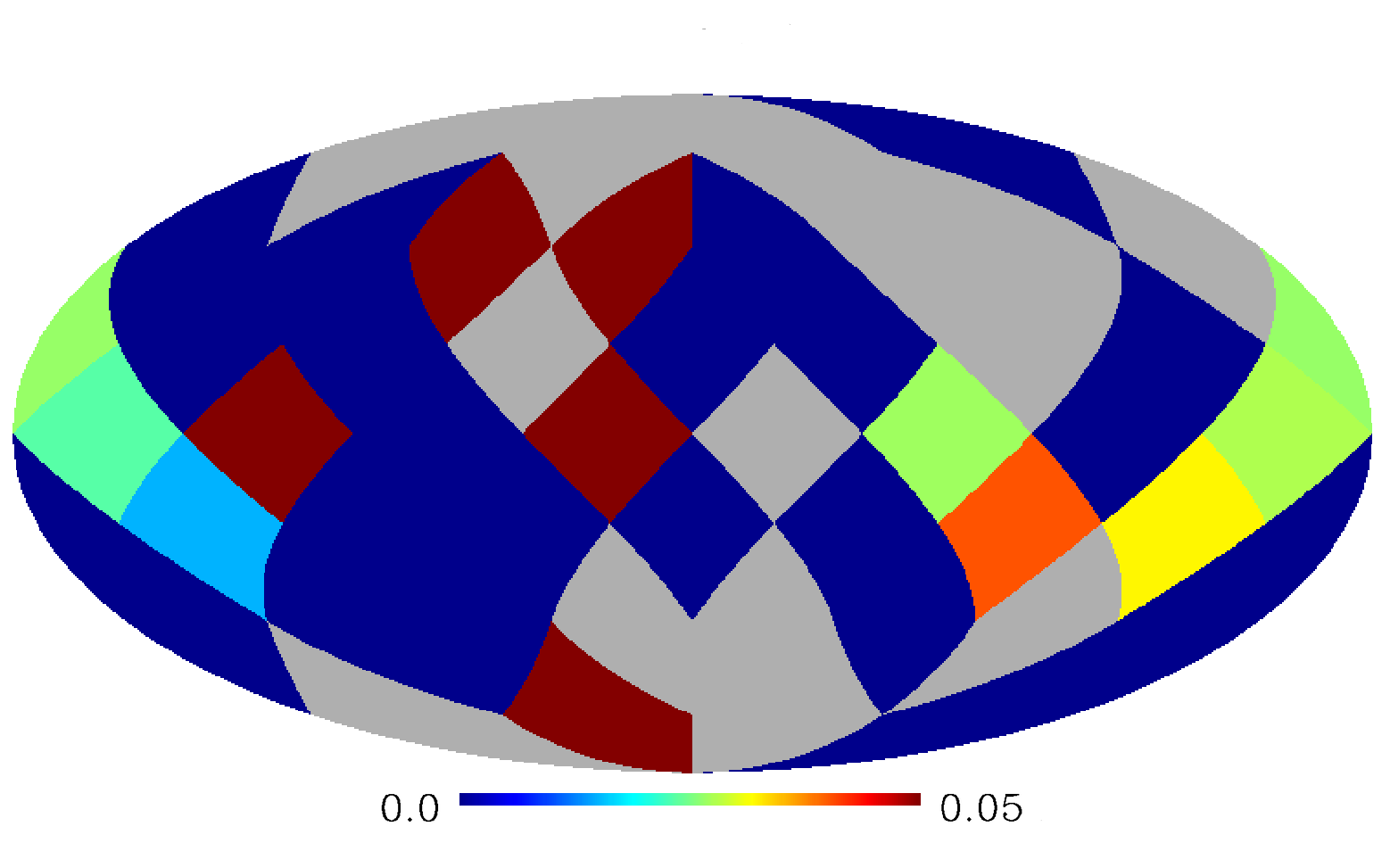}
\includegraphics[width=0.5\textwidth,angle=0]{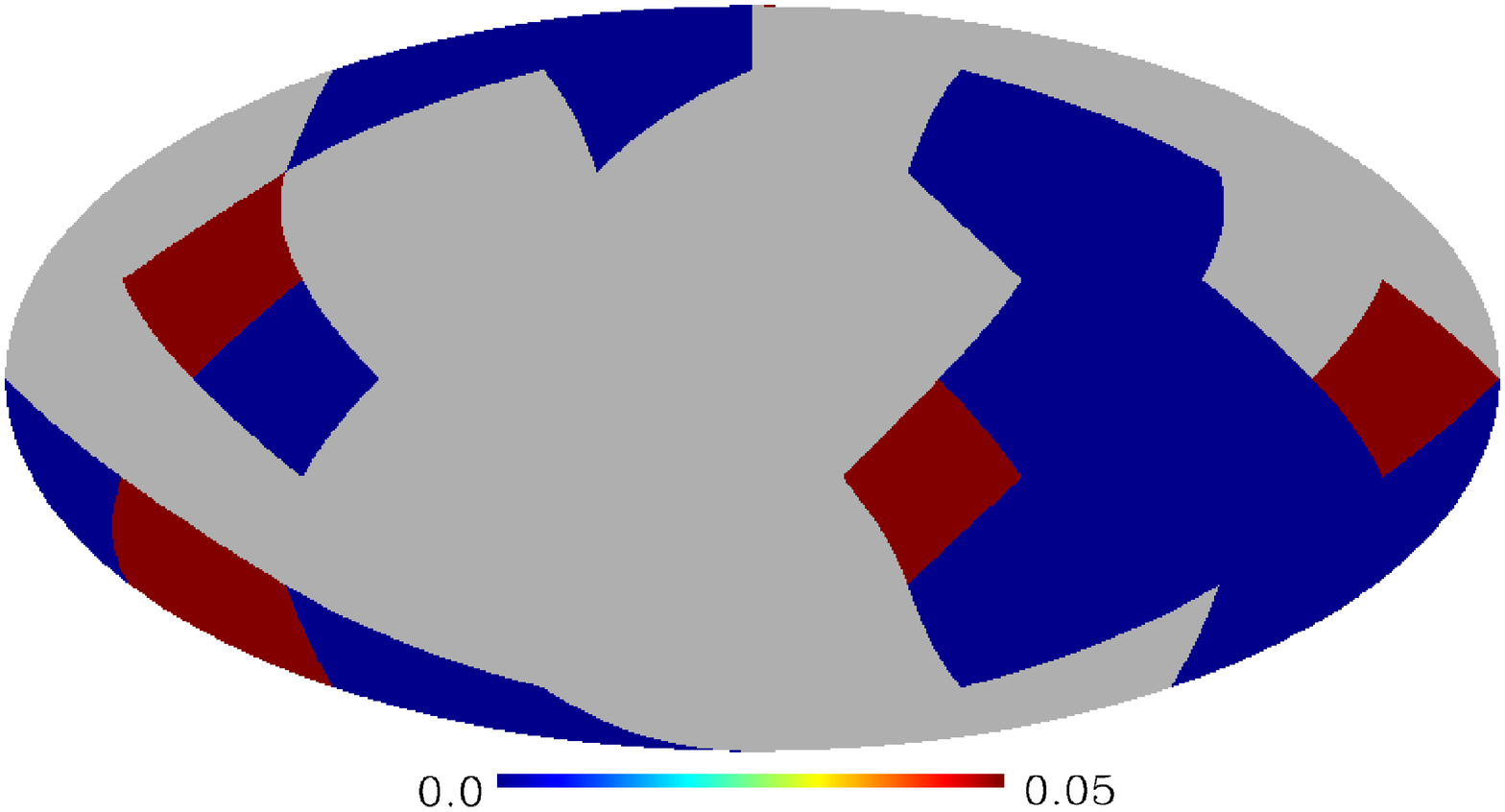}
\caption{Synchrotron-correlated (\emph{top}), dust-correlated
  (\emph{middle}), and free-free-correlated (\emph{bottom}) fractional
  polarisations in the 48 regions of the K-band pixel-analysis. The
  regions compatible with a zero polarisation fraction within
  $1\sigma$ are plotted in blue. Regions with $2\sigma_p > 30\%$, or
  with negative fractional polarisation, are plotted in
  grey.\label{polPixel}}
\end{center}
\end{figure}

\begin{figure}
\begin{center}
\includegraphics[width=0.5\textwidth,angle=0]{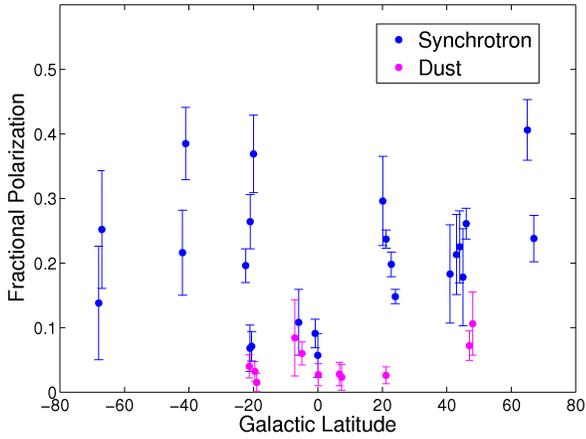}
  \caption{Synchrotron- and dust-correlated fractional polarisation of \emph{WMAP} 
K-band as a function of mean Galactic latitude in the 48 regions of the pixel-analysis. 
We plot only regions where $f_P > \sigma_{f_P}$.
\label{FracPolLatitude}}
\end{center}
\end{figure}

\begin{figure}
\begin{center}
\includegraphics[width=0.5\textwidth,angle=0]{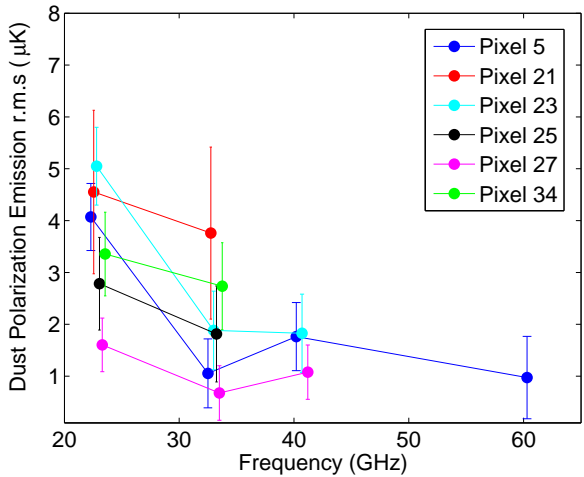}
 \caption{Dust-correlated polarised emission r.m.s. in $\mu$K as a
   function of frequency in the pixel-analysis. We report only the
   six pixel regions where there is a $2\sigma$ significant emission
   in the K-band.  \label{PolDustFreq}}
\end{center}
\end{figure}

It is clear from the all-sky analysis of Fig.~\ref{confT} that the
total unpolarised signal across all 5 \emph{WMAP} bands is dominated
by dust-correlated emission. The rise in dust emission from 61 to
94~GHz is consistent with the expectation of thermally emitting dust.
The drop with frequency from 22 to 61~GHz has been termed
``anomalous'' dust-correlated emission and has been attributed to such
sources as spinning-dust emission (e.g., Lazarian \& Draine 1998a,b)
and dust-correlated synchrotron emission (Bennett et al.\ 2003).  The
results in Fig.~\ref{confT}, and K--Ka spectral index for anomalous
dust of $\beta_d \approx -2.5$, is consistent with the results of
Davies et al.\ (2006).

The results for emission presented in Table~\ref{tab:FracPol} show
that dust emission also dominates the unpolarised K-band
emission for most of the sky in the pixel-analysis.  
We find that the
dust-correlated emission is dominant at low frequencies in
many pixels even where the thermal dust emission is not present, (e.g., pixels 16, 22, 23, 27, and 40).  

Fig.~\ref{polPixel} shows the polarisation fraction of each foreground
at K-band, computed using Equation~(\ref{frac}). The same values are
reported in Table~\ref{tab:FracPol}. As expected, synchrotron emission
is highly polarised, while dust and free-free almost unpolarised.
We compared the values found in various pixels with the
  distributions obtained from simulations. As a general comment on the
  simulations we can say that a) the range of the recovered dust
  polarisation fraction, assuming the null hypothesis, is typically
  5--10\,\% and varies between a few percent and 25\% (for regions very
  distant from the Galactic plane, e.g., 2, 3, 47, 48); b) the
  distributions from simulations with and without dust polarisation are, in
  general, very similar, although higher mean polarisation values
  are not always associated with the case corresponding to the
  polarised dust (this may indicate cross-talk with other components);
  c) the variance in the distributions is typically similar to the error
  bars computed by the CCA on real data.

When we compare the detected values of the polarisation
  fraction in the K-band with the distribution we find that in some
  pixels they exceed the tail of the simulation distribution. The
  pixels for which the simulation-recovered values are always below
  the (real) detected dust polarisation are 5, 17, 21, 23, 25, 27, 34,
  35 (see examples in Fig.~\ref{fig:dust_distrib}){\footnote{For comparison, the number of pixels in which the fractional polarisation is always below the distribution is two.}. In some regions
  (5, 6, 17, 21, 23, 24, 25, 27, 32, 34, 35, 45) the detections are at
  more than the $2\sigma$ level. Some of these regions also show a (small) 
  increase in synchrotron polarisation with respect to simulation
  results (e.g., 5, 17, 32), with an accompanying anti-correlation
  with the dust values. Indeed, in some cases, the recovered dust
  polarisation fraction becomes slightly negative, such as in region 5
  (see Fig.~\ref{fig:dust_distrib}), which is evidence of cross-talk
  between components. Six regions (5, 17, 23, 34, 35, and 45) out of the 48 exceed the simulated distributions at the 99th percentile and are individually significant at $>2\sigma$. They showed fractional polarisations in the range $2.6$--$7.2\,\%$ except for the anomalous pixel 45, which appears to have a high polarisation fraction of $32\pm12\%$, detected at a
  significance level of $2.7\sigma$. This region does not contain any
  strong foreground features in total-intensity or in
  polarisation. Additionally, the simulation recovers the correct
  polarisation but with a relatively large uncertainty. The detected
  value in the real data is well outside the distribution, as shown in
  Fig.~\ref{fig:dust_r45}. This region should be investigated further
  with higher sensitivity data (e.g., {\it WMAP} 9-year, {\it
    Planck}).

\begin{figure*}
\begin{center}
\includegraphics[width=0.33\textwidth,angle=0]{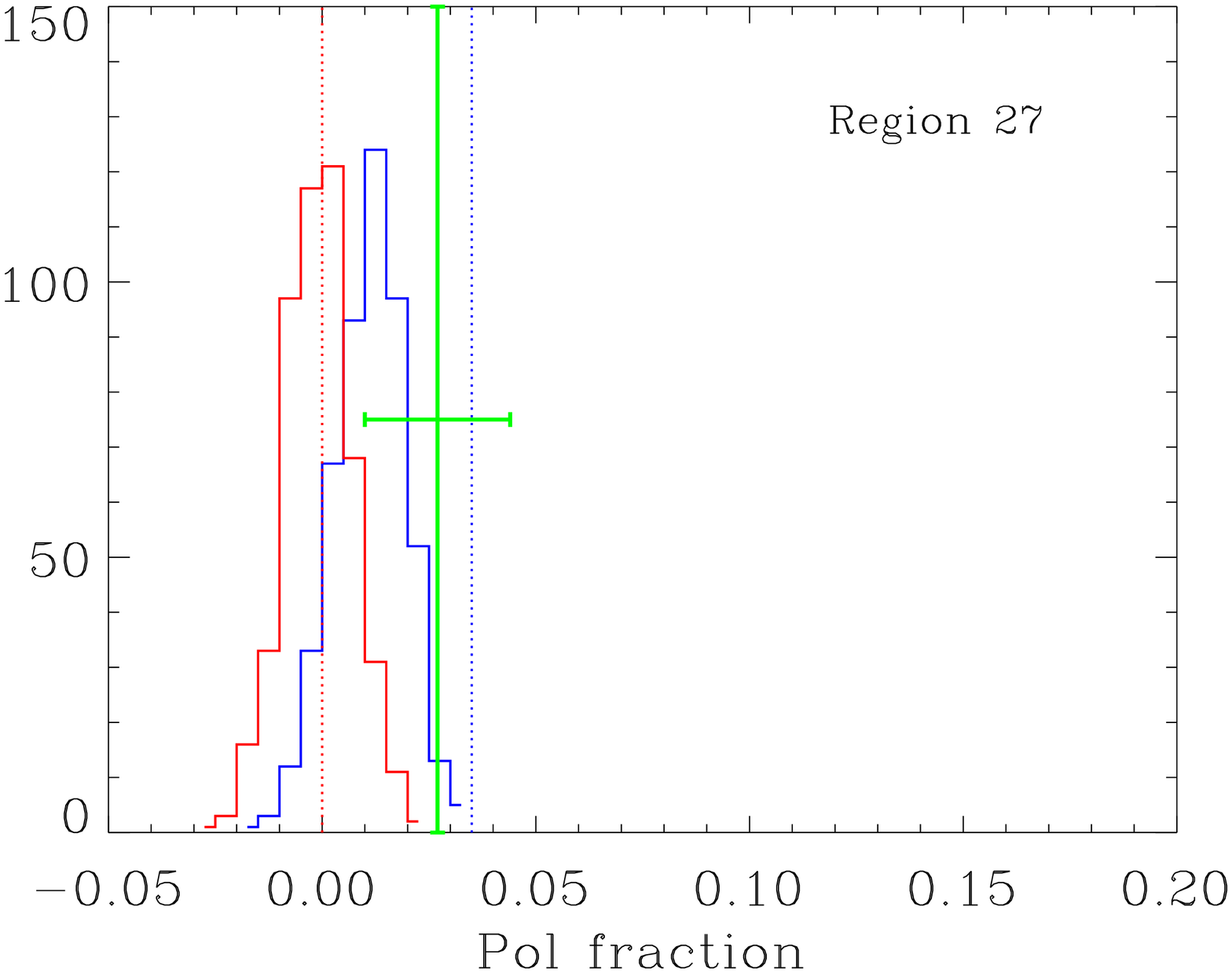}
\includegraphics[width=0.33\textwidth,angle=0]{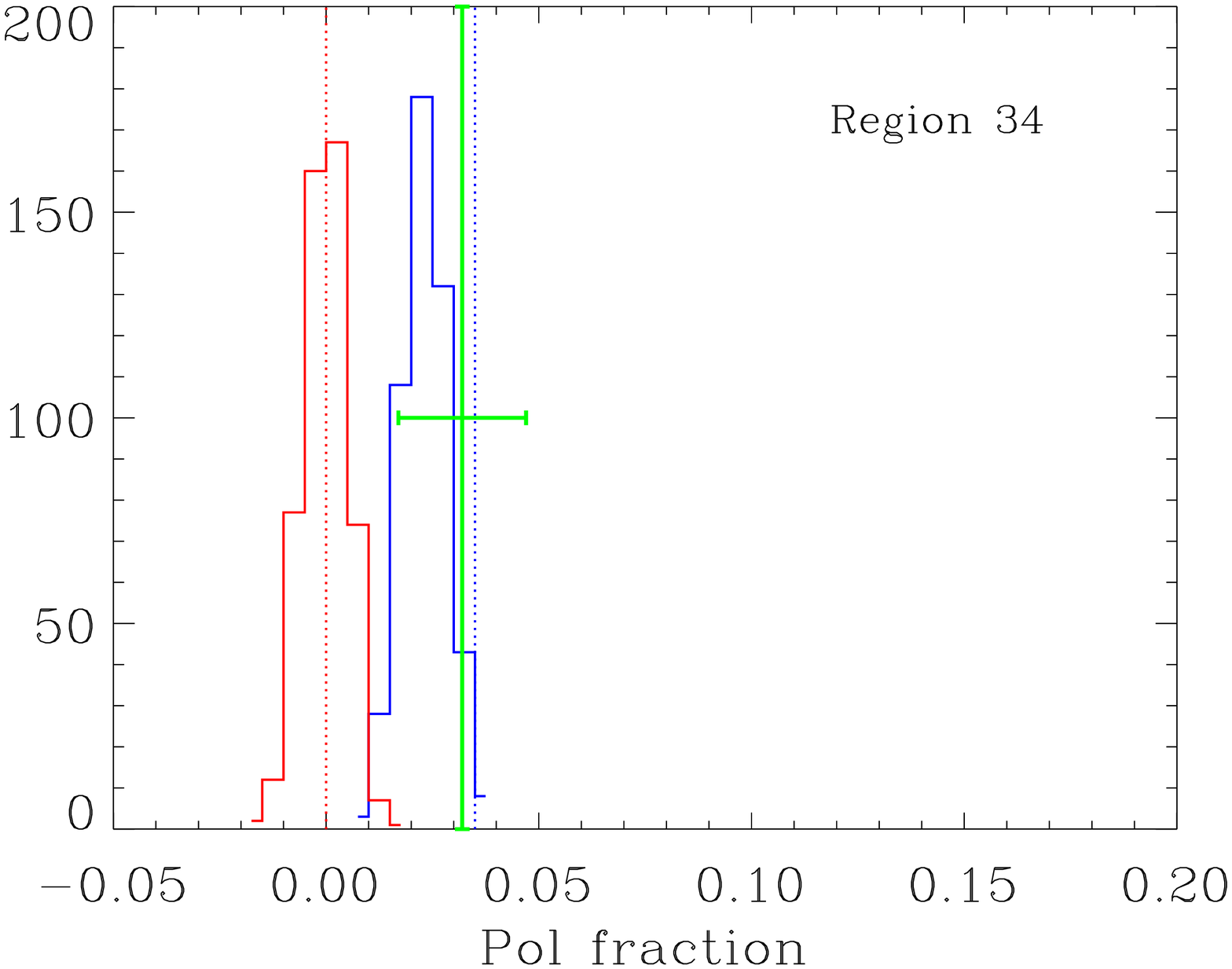}
\includegraphics[width=0.33\textwidth,angle=0]{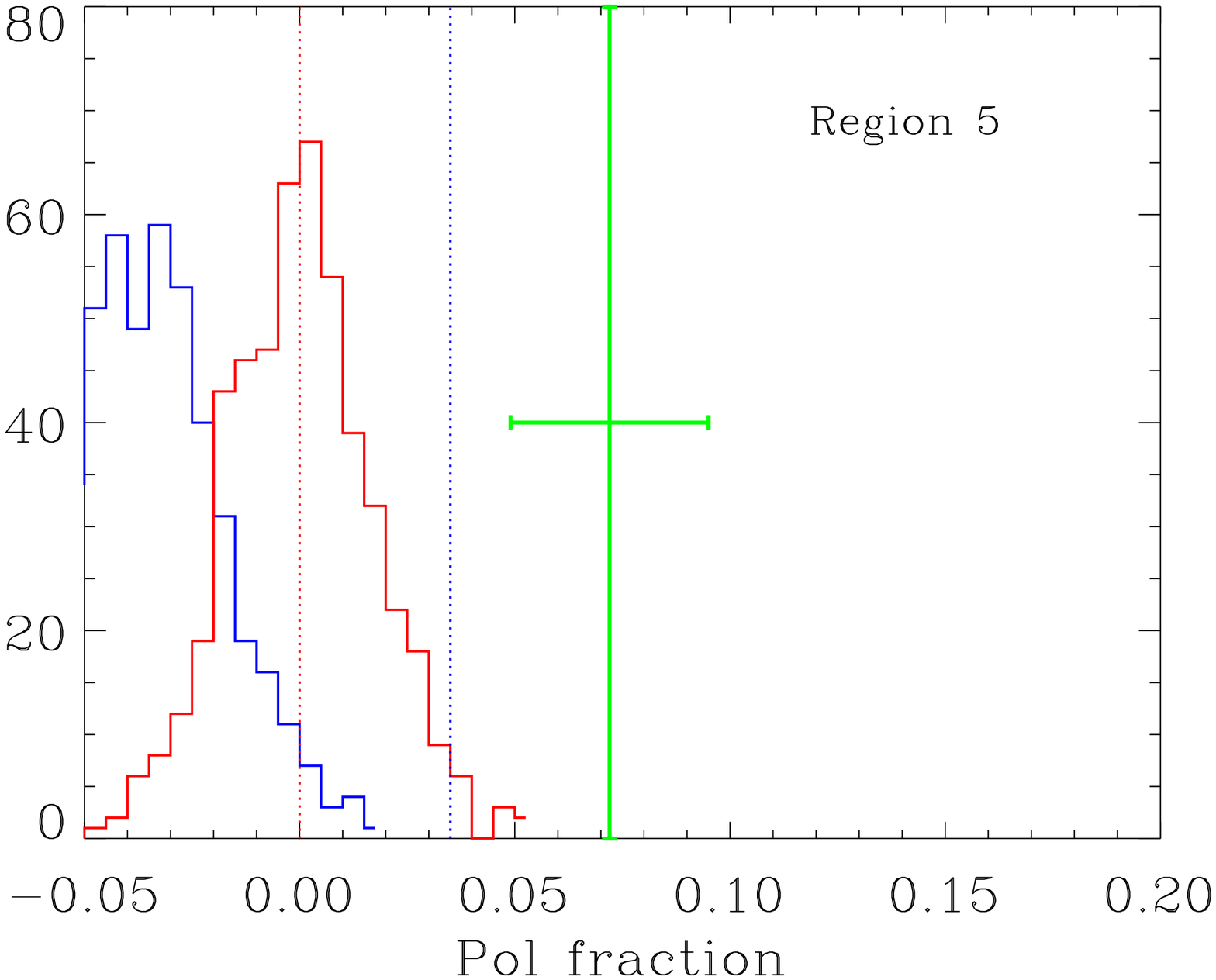}
\caption{Histograms of the K-band dust polarisation fraction
    for three regions, from 500 realisations, for the case with no
    dust polarisation (red line) and with $3.5\,\%$ dust polarisation
    (blue line). The input values are shown as vertical dotted
    lines. The green line and error-bar represent the best-fitting
    value and its $1\sigma$ uncertainty from the real
    data. \label{fig:dust_distrib}}
\end{center}
\end{figure*}

\begin{figure}
\begin{center}
\includegraphics[width=0.5\textwidth,angle=0]{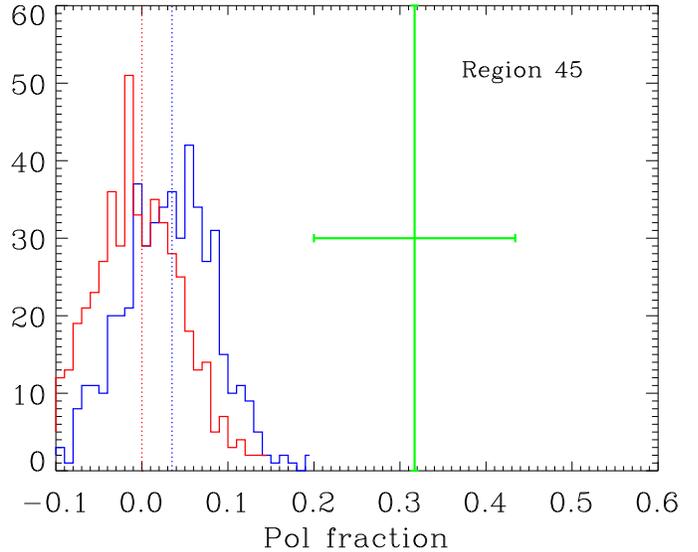}
\caption{Histogram of the K-band dust polarisation fraction
    for region 45, from 500 realisations, for the case with no dust
    polarisation (red line) and with $3.5\,\%$ dust polarisation (blue
    line). The input values are shown as vertical dotted lines. The
    green line and error-bar represent the best-fitting value and its $1\sigma$
    uncertainty from the real data. \label{fig:dust_r45}}
\end{center}
\end{figure}

Although we cannot rely heavily on individual pixels, we can
  still continue, knowing that there can be biases of up to $\approx
  5\,\%$ in any given region, and $\approx 1.5\,\%$ when averaging
  significant ($>2\sigma$) pixels, as discussed in
  Section~\ref{sec:sim}. Furthermore, we can choose pixels that in the
  simulations were able to recover the correct dust amplitude within
  $1\sigma$ of the true value (regions 6, 12, 25, 33, 34, 35, 39, and
  45). Although these may not be completely representative of the real
  data, they should be more reliable than simply averaging all or just
  significant pixels. For these, we find a weighted average of $3.2\pm0.9\, \textrm{(stat)} \pm 1.5\, \textrm{(sys)}\,\%$. We do not claim this as a real detection, since
  this is only a $\approx 2\sigma$ detection when considering the
  overall bias seen in the simulations described in
  Section~\ref{sec:sim}. However, it is clear that the average K-band dust
  polarisation fraction is less than about $5\,\%$ at 95\,\% confidence.

A comparison of the dust and
synchrotron values at K-band (Fig.~\ref{FracPolLatitude}; as a
function of Galactic latitude) shows that the K-band synchrotron
polarisation increases from $f_s \sim 5\%$ to $\sim 40 \%$, while the
dust remains fairly constant within the error-bars at $f_d \sim 2\% -
10\%$ in the pixels where significant dust polarisation emission is
found.  Over most of the sky dust is found to be unpolarised, with a
$2 \sigma$ upper limit on fractional polarisation of about $5\%$ in
seventeen pixel regions and about $10\%$ in four other regions.

Given the limitations of our results, particularly at frequencies
above 33\,GHz, we cannot make a detailed analysis of the frequency
dependence of the dust polarisation. Fig.~\ref{PolDustFreq} reports
the r.m.s. of the dust polarisation emission. The fractional
polarisation for these same data are plotted in
Fig.~\ref{FreqFracPolD} (pixel 21 has been removed from this plot as
the temperature-analysis yields negative C-C coefficients).  Here we
report just K-, Ka,- and Q-bands, as the higher frequency data have
signal-to-noise ratios less than $1\sigma$. While the fractional
polarisation data do exhibit a slight frequency dependence the
uncertainties are quite large. Due to these large uncertainties we
provide only a broad discussion, rather than a detailed analysis, of
these results in the remainder of this section.

In the all-sky analysis (Fig.~\ref{confT}) the total emission from the
anomalous-dust component appears to dominate the thermal dust
component for frequencies below 61\,GHz (V-band). We find similar
results in the temperature data of the five pixels of
Fig.~\ref{FreqFracPolD} (not shown).  Lazarian \& Draine (2000)
predicted that spinning-dust polarisation should be no greater than
2\% at 20\,GHz and become almost completely unpolarised above 40\,GHz.
If the anomalous component is due to spinning-dust then the
dust-correlated polarisation should be consistent with
zero-polarisation in the frequency range 40--61~GHz (Q--V bands), with
the possibility of increasing polarisation towards lower
frequencies. Within the error-bars, this is at least consistent with
the behaviour observed in Fig.~\ref{FreqFracPolD}.  Alternatively,
Draine \& Lazarian (1999) predict polarisations as high as 40\% in the
1--200~GHz range for emission from vibrating magnetic-grains (although
the actual polarisation is not expected to be so high given some level
of beam depolarisation and depolarisation due to line-of-sight
integration). Depending on the exact number of magnetic-domains and
the grains' aspect ratios, the magnetic-grains' polarisation may drop
from its maximal value to zero almost anywhere within the same
frequency range; the observed polarisation as a function of frequency
could have almost any behaviour (i.e., rising, falling, flat, or all
three). Given the large uncertainties, we can only note here that the
data plotted in Fig.~\ref{FreqFracPolD} is not inconsistent with this
behaviour.

\begin{figure}
\begin{center}
\includegraphics[width=0.5\textwidth,angle=0]{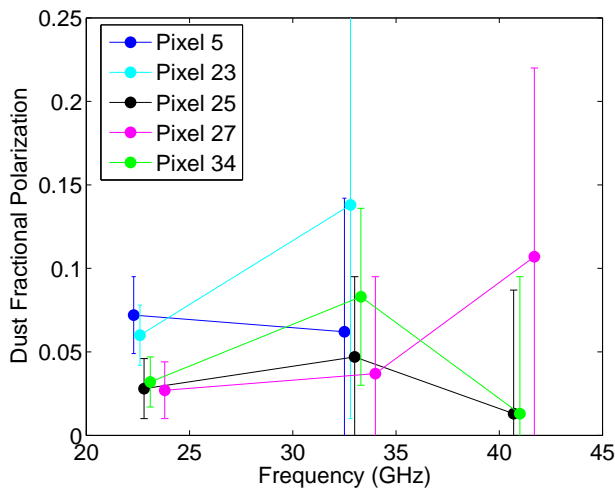}
\caption{Dust-correlated polarisation fraction as a function of
  frequency in the pixel-analysis. We report only the five pixel
  regions where there is a $>1\sigma$ result in the K-band that also
  have a non-zero value at Ka and Q bands. \label{FreqFracPolD}}
\end{center}
\end{figure}

\subsection{Free Free}
\label{sec:fff}

A simple fit to a power-law spectrum for the free-free emission in the
all-sky analysis yields a free-free spectral index of $\beta_f=-2.15
\pm 0.22$.  In the latitude analysis and in the all-sky analysis we
find free-free emission is unpolarised ($\lesssim 1\%$) over the sky.
This result is confirmed in the pixel-analysis, where over most of the
sky free-free is unpolarised at the $1\sigma$ level, as reported in
Table~\ref{tab:FracPol}. 

The upper limits in the distributions
  of the polarisation fraction from simulations in K-band typically
  range between 5 and 15 percent, apart from the regions around the
  Galactic poles that are essentially unconstrained. We find that in
  region 16 the detected free-free polarisation is higher than all
  simulation results but still compatible with zero at the $2\sigma$
  level when C-C errors are considered. In 16 of the 48 pixel regions
  we were able to put $2 \sigma$ upper limits on fractional
  polarisations ranging from $10\%$ (in seven regions) to about $5\%$
  (in nine regions).  These upper limits are more stringent where the
  uncertainties are small, i.e., where the free-free emission is
  higher, such as near the Galactic plane at longitudes between
  $200^\circ$ and $300^\circ$. Some pixels (e.g., 14, 16, 17, 24, 31,
  33, 40, and 43) do show significant cross-talk in the
  simulations. By comparing the simulations with the real data, we can
  isolate regions that we believe are not strongly affected by
  cross-talk and where the C-C errors are comparable or larger to the
  scatter in the simulations (therefore no significant bias). These are 4, 10, 11, 13,15, 19, 22, 23,
  26, 27, 32, 34, and 41.  Examples are shown in
  Fig.~\ref{fig:free_distrib}. The weighted average of these more
  reliable pixels is $0.0\pm1.7\,\%$, corresponding to an upper limit
  of $3.4\,\%$ (95\,\% confidence). The standard deviation in all 48 pixels is
  $\pm 5.5\%$. We note that, unlike for dust, no regions exceed the simulation distributions at the 99th percentile and are individually significant (at $>2\sigma$).

\begin{figure*}
\begin{center}
\includegraphics[width=0.33\textwidth,angle=0]{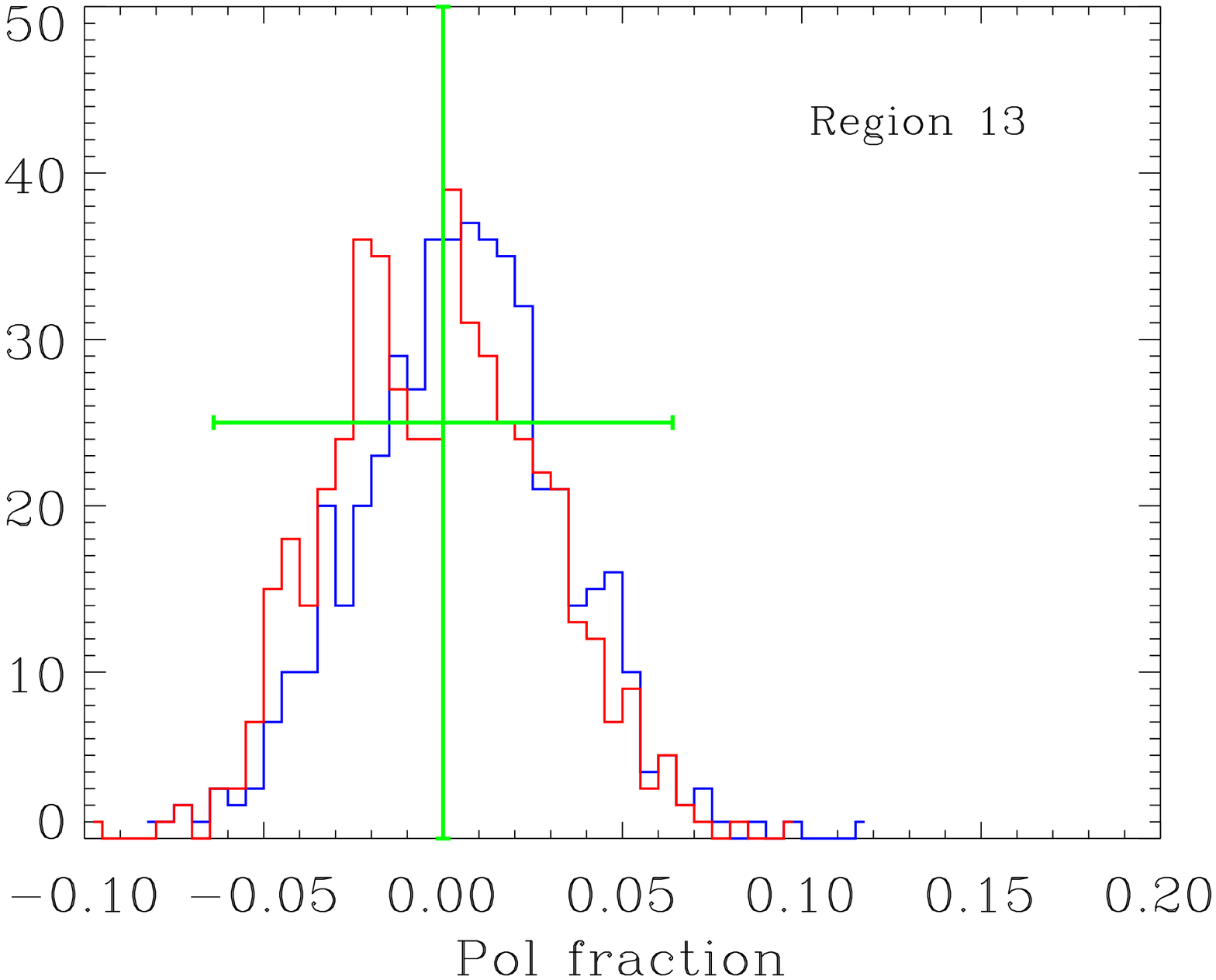}
\includegraphics[width=0.33\textwidth,angle=0]{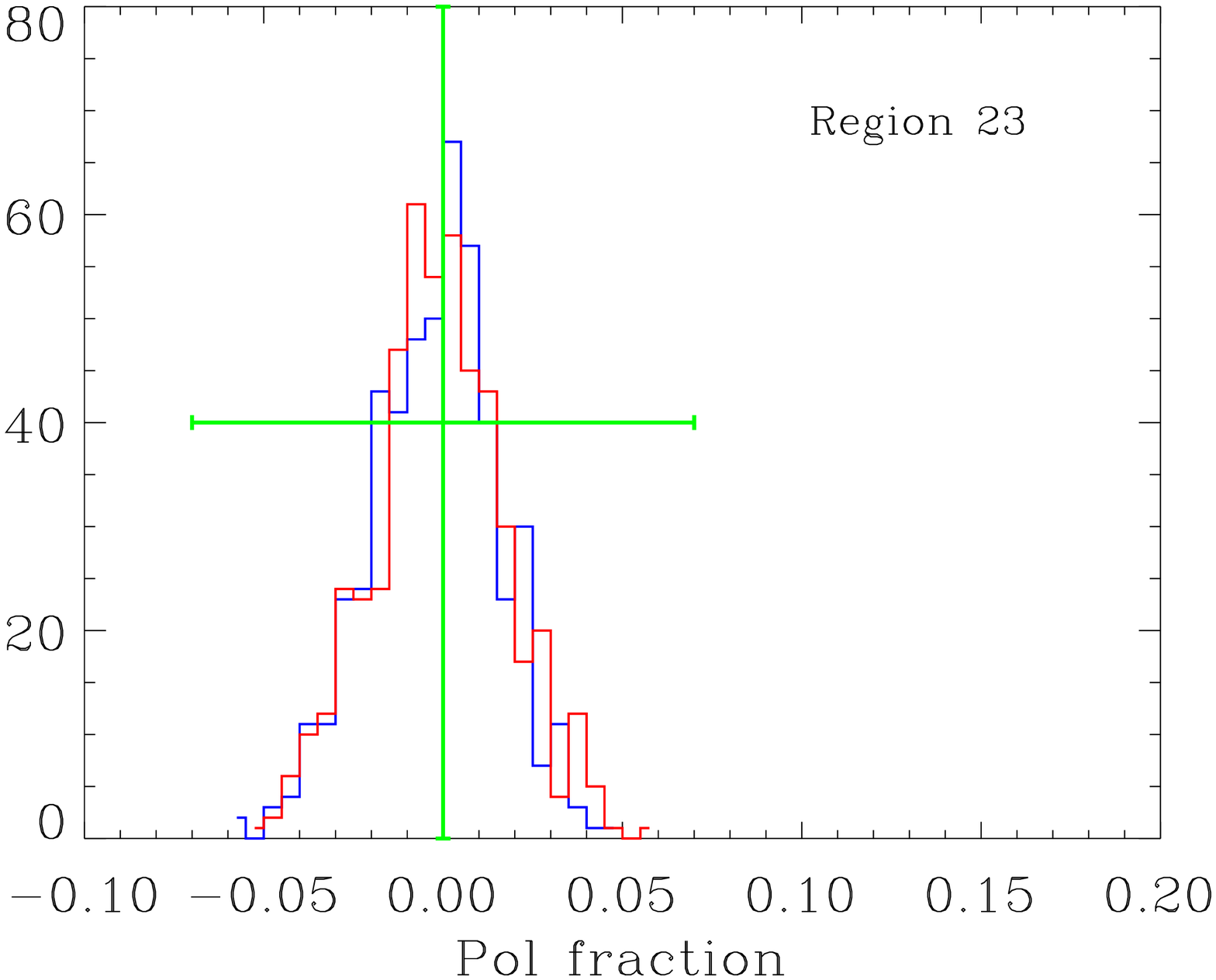}
\includegraphics[width=0.33\textwidth,angle=0]{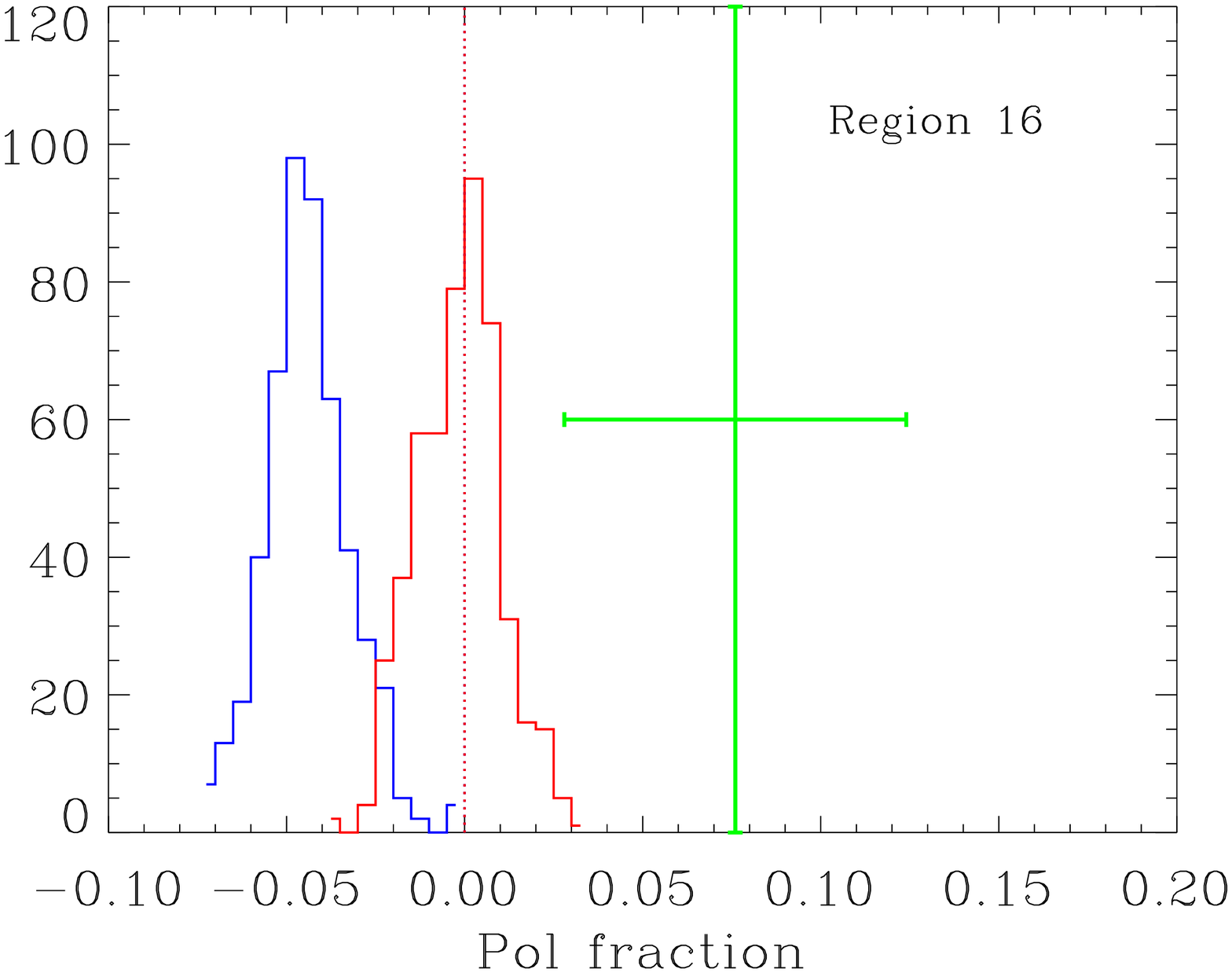}
\caption{Histograms of the free-free polarisation fraction at
    K-band for three regions, from 500 realisations at K-band, for the
    case with no dust polarisation (red line) and with $3.5\,\%$ dust
    polarisation (blue line). The input values are shown as vertical
    dotted lines. The green line and error-bar represents the best-fitting
    value and its $1\sigma$ uncertainty from the real
    data. \label{fig:free_distrib}}
\end{center}
\end{figure*}

\section{Conclusions}
\label{sec:conclusions}

We computed the cross-correlation of the intensity and polarisation
\emph{WMAP} data in different sky regions using templates for
synchrotron, dust, and free-free emission, plus a monopole-offset
template.  This technique is reliable because it takes into account
the correlations between foregrounds, and because it requires no
assumptions about the foregrounds' frequency behaviours.  However,
this study relies on the assumption of equal polarisation angles for
the three foregrounds and on the hypothesis of constant fractional
polarisation fraction in each of the the analysed sky-regions.  We
derive the frequency dependence and polarisation fraction for all
three components in 48 different sky regions (delineated in HEALPix
format at $N_\mathrm{side} = 2$ resolution).  

This resolution is chosen as a compromise between (a) the assumption
of uniform polarisation fraction in any given sky-region and (b) the need
to average a large number of data points in order to minimise uncertainties.

The anomalous low-frequency emission correlated with thermal-dust is
clearly detected in intensity over the entire sky at the \emph{WMAP}
K- and Ka- bands.  It is also found to be the dominant foreground at
low Galactic latitude in the pixel analysis, between $b = -40^{\circ}$
and $b = 10^{\circ}$.  The synchrotron spectral-index obtained from
the K- and Ka-bands is consistent with previous results, although
small differences are found between the polarisation and temperature
results in the pixel-analysis.  On the other hand, the values from the
all-sky analysis ($\beta=-3.32 \pm 0.12$ for intensity and
$\beta=-3.01 \pm 0.03$ from polarisation) are compatible with an
average value of $\beta\simeq -3$.  

The polarisation fraction of the
synchrotron is constant in frequency and increases with latitude from
$5\%$ near the Galactic plane to about $40\%$ {\bf} in some regions at
high latitude; this is consistent with a depolarisation effect due to
integrating through high-column densities in the plane. The average
synchrotron fractional polarisation at low latitude ($|b|<20^{\circ}$)
is $8.6\pm 1.7\, \textrm{(stat)} \pm 0.5\, \textrm{(sys)}\,\%$, while at high latitude ($|b|>20^{\circ}$) is
$19.3\pm 0.8\,\textrm{(stat)} \pm 0.5\, \textrm{(sys)}\,\%$ with a standard deviation of $\pm11.0\%$. The robustness of these values was verified using monte carlo simulations.

Previous work has usually assumed that the anomalous dust-correlated
emission in K-band is unpolarised.  We find this assumption to
  be reasonable with most of the sky having polarisation fractions
  less than $10\,\%$. However, monte carlo simulations revealed that
  some pixels are likely to be affected by cross-talk, at up to the
  $\approx 5\,\%$ level. Taking an average of the significant
  ($>2\sigma$) regions led to a bias of $\approx 1.5\,\%$ in the
  simulations. The average polarisation fraction, for the more
  reliable pixels, was $3.2\pm0.9\, \textrm{(stat)} \pm 1.5\, \textrm{(sys)}\,\%$. Given the average bias seen in
  the simulations ($1.5\,\%$), we do not claim a strong detection,
  while it is clear that the polarisation fraction is less than
  $5\,\%$ at 95\,\% confidence. We found several pixels that appeared to have a significant
  detection of dust polarisation at K-band while being robust in the
  simulations. Eight out of 48 regions exceeded the 99th percentile of the distribution from simulations with no input foreground polarisation, 6 of which are detected at above $2\sigma$ and display polarisation fractions between $2.6\,\%$ and $7.2\,\%$, except for region 45 at $(l,b)\sim (46\degr,-66\degr)$, which was found
  to be polarised at $31.7\pm11.7\,\%$ ($2.7\sigma$). 

Free-free emission is found to be unpolarised over the
entire sky with an upper limit on the fractional
polarisation of about $5\%$ in nine regions, and about $10\%$ in seven
regions. Unlike for the dust, no regions were found to have a polarisation fraction that exceeded the simulation distributions. Guided by simulations, we find an upper limit of $3.4\,\%$ (95\,\% confidence).

\section*{Acknowledgments}

This research was carried out at the Jet Propulsion Laboratory,
California Institute of Technology, under a contract with the National
Aeronautics and Space Administration and funded through the Director's
Research and Development Fund Program.  EP is an NSF-ADVANCE fellow
(AST 06-49899) also supported by NASA grant NNX07AH59G and Planck
subcontract 1290790.  NM and EP were supported by JPL SURP award
1314616 for this work, and would like to thank Caltech for hospitality
during this period.  EP wishes to thank the Aspen Center for Physics
where part of this work was carried out.  CD acknowledges an STFC
Advanced Fellowship and ERC grant under the FP7.  JEV is supported by
NSF AST 05-40882 and 08-38261 through the Caltech Submillimeter
Observatory.  The authors would like to thank Mark Halpern and Anthony
Banday for useful conversations.  We acknowledge the use of the Legacy
Archive for Microwave Background Data Analysis (LAMBDA). Support for
LAMBDA is provided by the NASA Office of Space Science. Some of the
results in this paper have been derived using the HEALPix (G\'{o}rski
et al.,~2005) package.

\bibliographystyle{mn2e}


\appendix

\section{Complete tables of Cross-Correlation Coefficients}
\label{sec:appendix}
Here we report the C-C coeffcients for dust, synchrotron and free-free resulting from our analysis.

\begin{table}
 \begin{tabular}{cccccc}
    \hline
     {\bf Pixel} & {\bf K-band } & {\bf Ka-band}  & {\bf Q-band } & {\bf V-band}  & {\bf W-band } \\
    \hline
    & \multicolumn{5}{c} 
      {\textbf{Dust Temperature C-C coefficients}} \\
\\\hline

1 & 12.8 $^{\pm 2.7}$ & 9.2 $^{\pm 2.9}$ & 8.4 $^{\pm 3.2}$ & 7.6 $^{\pm 2.8}$ & 7.2 $^{\pm 3.8}$  \\  
2 & -2.9 $^{\pm 5.2}$ & -10.4 $^{\pm 5.6}$ & -10.7 $^{\pm 6}$ & -11.5 $^{\pm 5.9}$ & -10 $^{\pm 6.8}$  \\  
3 & 8.2 $^{\pm 2.8}$ & 2.7 $^{\pm 2.8}$ & 0.7 $^{\pm 2.8}$ & 0.8 $^{\pm 3.2}$ & 0.3 $^{\pm 3.4}$  \\  
4 & 6.9 $^{\pm 2.5}$ & 1.6 $^{\pm 2.5}$ & -0.4 $^{\pm 2.6}$ & -0.6 $^{\pm 2.7}$ & 0.6 $^{\pm 3.4}$  \\  
5 & 8.2 $^{\pm 0.6}$ & 2.4 $^{\pm 0.7}$ & 1.1 $^{\pm 0.7}$ & 0.7 $^{\pm 0.7}$ & 1 $^{\pm 0.7}$  \\  
6 & 15.4 $^{\pm 2.5}$ & 8.5 $^{\pm 2.7}$ & 6.5 $^{\pm 2.6}$ & 5.7 $^{\pm 2.7}$ & 4.8 $^{\pm 3}$  \\  
7 & 5 $^{\pm 0.5}$ & 1.7 $^{\pm 0.5}$ & 1 $^{\pm 0.5}$ & 0.8 $^{\pm 0.7}$ & 1.1 $^{\pm 0.6}$  \\  
8 & 8.4 $^{\pm 1.2}$ & 4.1 $^{\pm 1.3}$ & 2.9 $^{\pm 1.4}$ & 2.4 $^{\pm 1.3}$ & 2.3 $^{\pm 1.4}$  \\  
9 & 3.1 $^{\pm 3}$ & -0.6 $^{\pm 3.1}$ & -1.9 $^{\pm 3.2}$ & -2.2 $^{\pm 3.5}$ & -1.6 $^{\pm 3.4}$  \\  
10 & 7.8 $^{\pm 1.7}$ & 2.8 $^{\pm 1.8}$ & 1.4 $^{\pm 2}$ & 1.1 $^{\pm 2.1}$ & 0.7 $^{\pm 2}$  \\  
11 & 8.3 $^{\pm 1.2}$ & 3 $^{\pm 1.2}$ & 2 $^{\pm 1.3}$ & 1.7 $^{\pm 1.3}$ & 1.7 $^{\pm 1.3}$  \\  
12 & 9.5 $^{\pm 0.9}$ & 1.9 $^{\pm 1}$ & 0.2 $^{\pm 0.9}$ & -1 $^{\pm 1.1}$ & -0.5 $^{\pm 1.3}$  \\  
13 & 8.3 $^{\pm 0.6}$ & 3.2 $^{\pm 0.6}$ & 2.2 $^{\pm 0.7}$ & 1.5 $^{\pm 0.7}$ & 1.7 $^{\pm 0.7}$  \\  
14 & 5.3 $^{\pm 0.4}$ & 1.5 $^{\pm 0.5}$ & 0.6 $^{\pm 0.5}$ & 0.3 $^{\pm 0.5}$ & 0.7 $^{\pm 0.5}$  \\  
15 & 6.8 $^{\pm 0.2}$ & 2.3 $^{\pm 0.2}$ & 1.3 $^{\pm 0.2}$ & 0.8 $^{\pm 0.3}$ & 1.1 $^{\pm 0.2}$  \\  
16 & 4.8 $^{\pm 0.3}$ & 0.7 $^{\pm 0.3}$ & -0.2 $^{\pm 0.3}$ & -0.6 $^{\pm 0.3}$ & -0.2 $^{\pm 0.4}$  \\  
17 & 7.3 $^{\pm 0.3}$ & 3.3 $^{\pm 0.3}$ & 2.2 $^{\pm 0.3}$ & 1.6 $^{\pm 0.3}$ & 1.9 $^{\pm 0.4}$  \\  
18 & 9.5 $^{\pm 0.7}$ & 5.5 $^{\pm 0.7}$ & 4.7 $^{\pm 0.7}$ & 4 $^{\pm 0.7}$ & 4.1 $^{\pm 0.7}$  \\  
19 & 3.7 $^{\pm 0.6}$ & -0.1 $^{\pm 0.6}$ & -0.7 $^{\pm 0.6}$ & -0.9 $^{\pm 0.7}$ & -0.1 $^{\pm 0.6}$  \\  
20 & 8 $^{\pm 0.8}$ & 2.1 $^{\pm 0.9}$ & 0.9 $^{\pm 0.9}$ & 0.4 $^{\pm 0.9}$ & 0.7 $^{\pm 1.1}$  \\  
21 & 5.2 $^{\pm 0.7}$ & 0.6 $^{\pm 0.8}$ & -0.4 $^{\pm 0.8}$ & -0.7 $^{\pm 0.7}$ & -0.7 $^{\pm 0.8}$  \\  
22 & 4.3 $^{\pm 0.4}$ & 0.6 $^{\pm 0.4}$ & -0.1 $^{\pm 0.4}$ & -0.4 $^{\pm 0.5}$ & 0.1 $^{\pm 0.4}$  \\  
23 & 5.3 $^{\pm 0.4}$ & 0.9 $^{\pm 0.4}$ & -0.2 $^{\pm 0.4}$ & -0.7 $^{\pm 0.5}$ & -0.3 $^{\pm 0.5}$  \\  
24 & 6.5 $^{\pm 0.4}$ & 2.7 $^{\pm 0.4}$ & 1.8 $^{\pm 0.4}$ & 1.4 $^{\pm 0.4}$ & 1.9 $^{\pm 0.5}$  \\  
25 & 5.9 $^{\pm 0.3}$ & 2.3 $^{\pm 0.3}$ & 1.4 $^{\pm 0.4}$ & 0.9 $^{\pm 0.3}$ & 1.3 $^{\pm 0.4}$  \\  
26 & 5.9 $^{\pm 0.6}$ & 0.9 $^{\pm 0.6}$ & -0.2 $^{\pm 0.7}$ & -0.7 $^{\pm 0.7}$ & -0.2 $^{\pm 0.8}$  \\  
27 & 5.6 $^{\pm 0.4}$ & 1.7 $^{\pm 0.4}$ & 0.9 $^{\pm 0.4}$ & 0.5 $^{\pm 0.5}$ & 1 $^{\pm 0.5}$  \\  
28 & 9 $^{\pm 4.1}$ & 0.9 $^{\pm 4.2}$ & 0.4 $^{\pm 4.8}$ & -1.3 $^{\pm 4.9}$ & 0.2 $^{\pm 5.2}$  \\  
29 & 11.5 $^{\pm 0.6}$ & 4.9 $^{\pm 0.6}$ & 3.5 $^{\pm 0.7}$ & 2.4 $^{\pm 0.8}$ & 2.5 $^{\pm 0.8}$  \\  
30 & 8.3 $^{\pm 0.6}$ & 3.3 $^{\pm 0.6}$ & 2.1 $^{\pm 0.7}$ & 1.4 $^{\pm 0.6}$ & 1.8 $^{\pm 0.6}$  \\  
31 & 11.8 $^{\pm 0.5}$ & 6.5 $^{\pm 0.5}$ & 5 $^{\pm 0.6}$ & 4 $^{\pm 0.6}$ & 3.8 $^{\pm 0.6}$  \\  
32 & 7.7 $^{\pm 0.4}$ & 2.7 $^{\pm 0.4}$ & 1.5 $^{\pm 0.4}$ & 0.8 $^{\pm 0.5}$ & 1.2 $^{\pm 0.5}$  \\  
33 & 6.1 $^{\pm 0.2}$ & 2.8 $^{\pm 0.2}$ & 1.9 $^{\pm 0.2}$ & 1.5 $^{\pm 0.2}$ & 1.7 $^{\pm 0.2}$  \\  
34 & 4.9 $^{\pm 0.2}$ & 1.5 $^{\pm 0.3}$ & 0.9 $^{\pm 0.3}$ & 0.7 $^{\pm 0.3}$ & 1.1 $^{\pm 0.3}$  \\  
35 & 5.9 $^{\pm 0.6}$ & 1.5 $^{\pm 0.6}$ & 0.4 $^{\pm 0.7}$ & -0.1 $^{\pm 0.6}$ & 0.3 $^{\pm 0.8}$  \\  
36 & 7.1 $^{\pm 0.3}$ & 2.8 $^{\pm 0.3}$ & 1.7 $^{\pm 0.4}$ & 1.1 $^{\pm 0.4}$ & 1.3 $^{\pm 0.4}$  \\  
37 & 4.4 $^{\pm 1.4}$ & -0.5 $^{\pm 1.5}$ & -1.3 $^{\pm 1.7}$ & -2.1 $^{\pm 1.4}$ & -1.7 $^{\pm 1.6}$  \\  
38 & 8.3 $^{\pm 1}$ & 3.3 $^{\pm 1}$ & 2.1 $^{\pm 1.1}$ & 1.1 $^{\pm 1.1}$ & 1.4 $^{\pm 1.3}$  \\  
39 & 5.4 $^{\pm 0.8}$ & 0.7 $^{\pm 0.8}$ & 0 $^{\pm 0.9}$ & -0.6 $^{\pm 1}$ & -0.1 $^{\pm 0.9}$  \\  
40 & 7 $^{\pm 0.3}$ & 2.1 $^{\pm 0.3}$ & 0.8 $^{\pm 0.3}$ & 0 $^{\pm 0.3}$ & 0.2 $^{\pm 0.4}$  \\  
41 & 8.7 $^{\pm 0.5}$ & 3.1 $^{\pm 0.5}$ & 1.8 $^{\pm 0.5}$ & 1 $^{\pm 0.5}$ & 1.3 $^{\pm 0.5}$  \\  
42 & 7.3 $^{\pm 2.3}$ & 3.8 $^{\pm 2.5}$ & 2.7 $^{\pm 2.6}$ & 2.4 $^{\pm 2.6}$ & 2.5 $^{\pm 2.4}$  \\  
43 & 9 $^{\pm 0.6}$ & 4.2 $^{\pm 0.6}$ & 2.9 $^{\pm 0.7}$ & 1.9 $^{\pm 0.6}$ & 2.1 $^{\pm 0.6}$  \\  
44 & 6 $^{\pm 2}$ & 0.9 $^{\pm 2}$ & 0.4 $^{\pm 2.3}$ & -0.5 $^{\pm 2.5}$ & 0.3 $^{\pm 2.7}$  \\  
45 & 8.5 $^{\pm 2.3}$ & 4.3 $^{\pm 2.5}$ & 1.9 $^{\pm 2.6}$ & 1.9 $^{\pm 2.7}$ & 1.4 $^{\pm 3.2}$  \\  
46 & 1.2 $^{\pm 1.2}$ & -3.4 $^{\pm 1.3}$ & -4.4 $^{\pm 1.3}$ & -4.4 $^{\pm 1.2}$ & -3.3 $^{\pm 1.5}$  \\  
47 & 16.2 $^{\pm 3.2}$ & 8.8 $^{\pm 3.2}$ & 6.6 $^{\pm 3.6}$ & 6.3 $^{\pm 3.9}$ & 5.9 $^{\pm 3.9}$  \\  
48 & 3.3 $^{\pm 5.5}$ & -1.8 $^{\pm 5.9}$ & -3.1 $^{\pm 6.2}$ & -5.7 $^{\pm 6.6}$ & -5.4 $^{\pm 5.9}$  \\     
    \hline
  \end{tabular}\caption{Full fit temperature coefficients for dust in all 48 pixel regions, with associated 1-$\sigma$ errors (in $\mu K \mu K^{-1}_{FDS}$). \label{tab:TabDustT}} 
\end{table}

\begin{table}
 \begin{tabular}{cccccc}
    \hline
     {\bf Pixel} & {\bf K-band } & {\bf Ka-band}  & {\bf Q-band } & {\bf V-band}  & {\bf W-band } \\
    \hline
   & \multicolumn{5}{c} 
      {\textbf{Dust Polarization C-C coefficients}} \\
      \\\hline
1 & -1.1 $^{\pm 0.7}$ & -0.4 $^{\pm 0.7}$ & 0.3 $^{\pm 0.6}$ & -0.3 $^{\pm 0.8}$ & -0.9 $^{\pm 1}$  \\  
2 & -1.2 $^{\pm 1.5}$ & 0.1 $^{\pm 1.5}$ & -0.3 $^{\pm 1.4}$ & -0.3 $^{\pm 1.7}$ & 0.3 $^{\pm 2.1}$  \\  
3 & 0.4 $^{\pm 0.8}$ & 0.3 $^{\pm 0.8}$ & 0.2 $^{\pm 0.8}$ & -0.6 $^{\pm 1}$ & 0.1 $^{\pm 1.1}$  \\  
4 & -0.9 $^{\pm 0.8}$ & 0.2 $^{\pm 0.8}$ & -0.2 $^{\pm 0.9}$ & -0.4 $^{\pm 1}$ & 0.4 $^{\pm 1.2}$  \\  
5 & 0.6 $^{\pm 0.2}$ & 0.2 $^{\pm 0.2}$ & 0.3 $^{\pm 0.2}$ & 0.1 $^{\pm 0.2}$ & 0 $^{\pm 0.3}$  \\  
6 & 1.6 $^{\pm 0.7}$ & 0 $^{\pm 0.7}$ & 0.4 $^{\pm 0.6}$ & -0.1 $^{\pm 0.8}$ & -0.4 $^{\pm 1}$  \\  
7 & -0.1 $^{\pm 0.1}$ & 0.1 $^{\pm 0.1}$ & 0 $^{\pm 0.1}$ & 0 $^{\pm 0.2}$ & 0 $^{\pm 0.2}$  \\  
8 & -0.5 $^{\pm 0.3}$ & -0.2 $^{\pm 0.3}$ & 0 $^{\pm 0.3}$ & 0.2 $^{\pm 0.4}$ & 0 $^{\pm 0.4}$  \\  
9 & 0 $^{\pm 0.9}$ & 0.1 $^{\pm 0.9}$ & 0.1 $^{\pm 0.8}$ & 0.3 $^{\pm 1}$ & -0.2 $^{\pm 1.2}$  \\  
10 & -0.3 $^{\pm 0.5}$ & -0.2 $^{\pm 0.5}$ & 0.1 $^{\pm 0.5}$ & 0.4 $^{\pm 0.6}$ & 0.8 $^{\pm 0.7}$  \\  
11 & -1.3 $^{\pm 0.4}$ & 0 $^{\pm 0.4}$ & -0.1 $^{\pm 0.4}$ & -0.1 $^{\pm 0.5}$ & -0.3 $^{\pm 0.6}$  \\  
12 & -0.4 $^{\pm 0.3}$ & -0.3 $^{\pm 0.3}$ & 0 $^{\pm 0.3}$ & 0.1 $^{\pm 0.3}$ & 0.2 $^{\pm 0.4}$  \\  
13 & -0.1 $^{\pm 0.2}$ & 0.3 $^{\pm 0.2}$ & 0 $^{\pm 0.2}$ & 0 $^{\pm 0.3}$ & -0.1 $^{\pm 0.3}$  \\  
14 & -1.6 $^{\pm 0.1}$ & -0.4 $^{\pm 0.1}$ & -0.2 $^{\pm 0.1}$ & -0.1 $^{\pm 0.1}$ & 0 $^{\pm 0.1}$  \\  
15 & -0.1 $^{\pm 0}$ & 0 $^{\pm 0}$ & 0 $^{\pm 0}$ & 0 $^{\pm 0}$ & 0 $^{\pm 0.1}$  \\  
16 & -0.3 $^{\pm 0.1}$ & 0 $^{\pm 0.1}$ & 0 $^{\pm 0.1}$ & 0 $^{\pm 0.1}$ & 0 $^{\pm 0.1}$  \\  
17 & 0.2 $^{\pm 0.1}$ & 0 $^{\pm 0.1}$ & 0.1 $^{\pm 0.1}$ & 0 $^{\pm 0.1}$ & -0.1 $^{\pm 0.1}$  \\  
18 & -0.2 $^{\pm 0.2}$ & 0 $^{\pm 0.2}$ & -0.1 $^{\pm 0.2}$ & 0 $^{\pm 0.2}$ & -0.3 $^{\pm 0.3}$  \\  
19 & -0.3 $^{\pm 0.2}$ & 0.1 $^{\pm 0.2}$ & -0.1 $^{\pm 0.2}$ & 0 $^{\pm 0.2}$ & 0.2 $^{\pm 0.2}$  \\  
20 & -0.1 $^{\pm 0.3}$ & -0.2 $^{\pm 0.3}$ & 0.1 $^{\pm 0.3}$ & -0.1 $^{\pm 0.4}$ & 0.2 $^{\pm 0.4}$  \\  
21 & 0.4 $^{\pm 0.3}$ & 0.4 $^{\pm 0.3}$ & 0.1 $^{\pm 0.3}$ & 0.1 $^{\pm 0.4}$ & -0.2 $^{\pm 0.5}$  \\  
22 & 0 $^{\pm 0.1}$ & 0 $^{\pm 0.1}$ & 0 $^{\pm 0.1}$ & 0 $^{\pm 0.1}$ & 0 $^{\pm 0.2}$  \\  
23 & 0.3 $^{\pm 0.1}$ & 0.1 $^{\pm 0.1}$ & 0.1 $^{\pm 0.1}$ & 0 $^{\pm 0.1}$ & -0.1 $^{\pm 0.2}$  \\  
24 & 0.2 $^{\pm 0.1}$ & 0.2 $^{\pm 0.1}$ & 0.1 $^{\pm 0.1}$ & 0.2 $^{\pm 0.2}$ & -0.3 $^{\pm 0.2}$  \\  
25 & 0.2 $^{\pm 0.1}$ & 0.1 $^{\pm 0.1}$ & 0 $^{\pm 0.1}$ & 0 $^{\pm 0.1}$ & 0 $^{\pm 0.2}$  \\  
26 & -0.1 $^{\pm 0.2}$ & -0.1 $^{\pm 0.2}$ & 0 $^{\pm 0.2}$ & 0 $^{\pm 0.2}$ & 0 $^{\pm 0.2}$  \\  
27 & 0.2 $^{\pm 0.1}$ & 0.1 $^{\pm 0.1}$ & 0.1 $^{\pm 0.1}$ & 0 $^{\pm 0.1}$ & 0.1 $^{\pm 0.1}$  \\  
28 & -0.8 $^{\pm 1.8}$ & 0.4 $^{\pm 1.9}$ & -1.6 $^{\pm 1.8}$ & -0.1 $^{\pm 2.2}$ & -1.9 $^{\pm 2.6}$  \\  
29 & -0.8 $^{\pm 0.2}$ & -0.2 $^{\pm 0.2}$ & -0.1 $^{\pm 0.2}$ & -0.1 $^{\pm 0.2}$ & 0 $^{\pm 0.2}$  \\  
30 & -0.2 $^{\pm 0.2}$ & -0.1 $^{\pm 0.2}$ & 0.1 $^{\pm 0.2}$ & 0 $^{\pm 0.2}$ & 0 $^{\pm 0.3}$  \\  
31 & 0 $^{\pm 0.1}$ & 0 $^{\pm 0.1}$ & 0.1 $^{\pm 0.1}$ & 0 $^{\pm 0.1}$ & 0 $^{\pm 0.2}$  \\  
32 & 0.1 $^{\pm 0.1}$ & -0.1 $^{\pm 0.1}$ & 0 $^{\pm 0.1}$ & 0.1 $^{\pm 0.1}$ & 0 $^{\pm 0.2}$  \\  
33 & -0.1 $^{\pm 0.1}$ & 0 $^{\pm 0.1}$ & 0.1 $^{\pm 0.1}$ & 0 $^{\pm 0.1}$ & 0 $^{\pm 0.1}$  \\  
34 & 0.2 $^{\pm 0.1}$ & 0.1 $^{\pm 0.1}$ & 0 $^{\pm 0.1}$ & 0 $^{\pm 0.1}$ & 0 $^{\pm 0.1}$  \\  
35 & 0.2 $^{\pm 0.1}$ & 0 $^{\pm 0.1}$ & 0 $^{\pm 0.1}$ & 0.1 $^{\pm 0.1}$ & -0.1 $^{\pm 0.1}$  \\  
36 & -0.1 $^{\pm 0.1}$ & 0 $^{\pm 0.1}$ & 0.1 $^{\pm 0.1}$ & 0 $^{\pm 0.1}$ & 0 $^{\pm 0.1}$  \\  
37 & -0.7 $^{\pm 0.4}$ & -0.2 $^{\pm 0.4}$ & -0.4 $^{\pm 0.4}$ & 0.1 $^{\pm 0.5}$ & 0.1 $^{\pm 0.6}$  \\  
38 & 0.1 $^{\pm 0.3}$ & -0.1 $^{\pm 0.3}$ & 0.1 $^{\pm 0.3}$ & 0 $^{\pm 0.3}$ & 0.1 $^{\pm 0.4}$  \\  
39 & 0.2 $^{\pm 0.3}$ & -0.1 $^{\pm 0.3}$ & -0.1 $^{\pm 0.3}$ & -0.1 $^{\pm 0.3}$ & 0 $^{\pm 0.4}$  \\  
40 & 0 $^{\pm 0.1}$ & 0 $^{\pm 0.1}$ & 0 $^{\pm 0.1}$ & 0 $^{\pm 0.1}$ & 0 $^{\pm 0.1}$  \\  
41 & 0 $^{\pm 0.2}$ & 0 $^{\pm 0.2}$ & 0.1 $^{\pm 0.2}$ & 0 $^{\pm 0.2}$ & 0.2 $^{\pm 0.2}$  \\  
42 & -1.1 $^{\pm 0.7}$ & -0.2 $^{\pm 0.7}$ & 0.1 $^{\pm 0.6}$ & 0 $^{\pm 0.8}$ & 0.1 $^{\pm 1}$  \\  
43 & 0 $^{\pm 0.1}$ & -0.1 $^{\pm 0.1}$ & 0 $^{\pm 0.1}$ & 0 $^{\pm 0.1}$ & 0 $^{\pm 0.1}$  \\  
44 & -0.9 $^{\pm 0.5}$ & -0.2 $^{\pm 0.5}$ & -0.5 $^{\pm 0.5}$ & 0.2 $^{\pm 0.6}$ & 0.1 $^{\pm 0.8}$  \\  
45 & 2.7 $^{\pm 0.7}$ & 0 $^{\pm 0.7}$ & 0.4 $^{\pm 0.7}$ & 0 $^{\pm 0.9}$ & 0.1 $^{\pm 1}$  \\  
46 & 0.5 $^{\pm 0.3}$ & 0 $^{\pm 0.3}$ & 0.2 $^{\pm 0.3}$ & 0.2 $^{\pm 0.4}$ & 0 $^{\pm 0.5}$  \\  
47 & -0.8 $^{\pm 0.8}$ & -0.2 $^{\pm 0.8}$ & -0.1 $^{\pm 0.7}$ & 0.2 $^{\pm 0.9}$ & -0.5 $^{\pm 1.1}$  \\  
48 & -0.7 $^{\pm 1.4}$ & -0.5 $^{\pm 1.5}$ & -0.9 $^{\pm 1.3}$ & 1 $^{\pm 1.7}$ & -1.3 $^{\pm 2}$  \\

    \hline
  \end{tabular}\caption{Full fit coefficients for polarized dust in all 48 pixel regions, with associated 1-$\sigma$ errors (in $\mu K \mu K^{-1}_{FDS}$). \label{tab:TabDustP}} 
\end{table}

\begin{table}
 \begin{tabular}{cccccc}
    \hline
     {\bf Pixel} & {\bf K-band } & {\bf Ka-band}  & {\bf Q-band } & {\bf V-band}  & {\bf W-band } \\
    \hline
   & \multicolumn{5}{c} 
      {\textbf{Synchrotron Temperature C-C coefficients}} \\
      \\\hline

1 & 5.9 $^{\pm 0.6}$ & 2.1 $^{\pm 0.6}$ & 1.3 $^{\pm 0.7}$ & 0.5 $^{\pm 0.7}$ & 0.4 $^{\pm 0.6}$  \\  
2 & -4.4 $^{\pm 2}$ & -6.9 $^{\pm 2.2}$ & -7 $^{\pm 2.2}$ & -6.8 $^{\pm 2.4}$ & -6.2 $^{\pm 2.2}$  \\  
3 & 4.2 $^{\pm 2.5}$ & 1.4 $^{\pm 2.6}$ & -0.1 $^{\pm 2.8}$ & -0.5 $^{\pm 3.2}$ & -0.3 $^{\pm 3.4}$  \\  
4 & 8.4 $^{\pm 0.9}$ & 4.4 $^{\pm 0.9}$ & 3.4 $^{\pm 0.9}$ & 2.8 $^{\pm 1}$ & 1.9 $^{\pm 1.3}$  \\  
5 & 6 $^{\pm 0.4}$ & 3 $^{\pm 0.4}$ & 2.3 $^{\pm 0.4}$ & 1.9 $^{\pm 0.4}$ & 1.5 $^{\pm 0.5}$  \\  
6 & 0.6 $^{\pm 1.4}$ & -1.2 $^{\pm 1.6}$ & -1.7 $^{\pm 1.5}$ & -2 $^{\pm 1.8}$ & -1.7 $^{\pm 1.9}$  \\  
7 & 4.8 $^{\pm 1.2}$ & 2 $^{\pm 1.3}$ & 1.5 $^{\pm 1.5}$ & 0.6 $^{\pm 1.6}$ & 0.7 $^{\pm 1.6}$  \\  
8 & 2.3 $^{\pm 1}$ & -1.9 $^{\pm 1.1}$ & -2.8 $^{\pm 1.2}$ & -3.1 $^{\pm 1.2}$ & -3 $^{\pm 1}$  \\  
9 & 4.2 $^{\pm 1.8}$ & 0.5 $^{\pm 1.9}$ & -0.1 $^{\pm 2}$ & -0.5 $^{\pm 2}$ & -0.9 $^{\pm 2.4}$  \\  
10 & 6.5 $^{\pm 1.7}$ & 3.4 $^{\pm 1.8}$ & 2.1 $^{\pm 1.8}$ & 1.4 $^{\pm 1.9}$ & 1.2 $^{\pm 1.8}$  \\  
11 & 5.4 $^{\pm 1}$ & 1.5 $^{\pm 1}$ & 0.2 $^{\pm 1}$ & -0.1 $^{\pm 1.2}$ & -0.4 $^{\pm 1}$  \\  
12 & 5.9 $^{\pm 1}$ & 2.9 $^{\pm 1}$ & 2 $^{\pm 1}$ & 1.5 $^{\pm 1.2}$ & 1 $^{\pm 1.3}$  \\  
13 & 5.1 $^{\pm 1.1}$ & 0.2 $^{\pm 1.1}$ & -1.1 $^{\pm 1.2}$ & -1.6 $^{\pm 1.1}$ & -1.5 $^{\pm 1.3}$  \\  
14 & 7.2 $^{\pm 0.3}$ & 3.4 $^{\pm 0.3}$ & 2.5 $^{\pm 0.3}$ & 1.9 $^{\pm 0.4}$ & 1.4 $^{\pm 0.4}$  \\  
15 & -1.6 $^{\pm 0.9}$ & -4.2 $^{\pm 0.9}$ & -4.7 $^{\pm 1}$ & -4.9 $^{\pm 1.1}$ & -4.2 $^{\pm 1.1}$  \\  
16 & 12.9 $^{\pm 0.8}$ & 7.9 $^{\pm 0.9}$ & 6.9 $^{\pm 0.9}$ & 5.8 $^{\pm 0.9}$ & 5.1 $^{\pm 1.2}$  \\  
17 & 5.5 $^{\pm 0.9}$ & 2.3 $^{\pm 1}$ & 1.7 $^{\pm 1.1}$ & 1.4 $^{\pm 1.2}$ & 0.8 $^{\pm 1}$  \\  
18 & 1.2 $^{\pm 1.4}$ & -1.2 $^{\pm 1.4}$ & -1.7 $^{\pm 1.4}$ & -1.8 $^{\pm 1.4}$ & -1.9 $^{\pm 1.6}$  \\  
19 & 3 $^{\pm 1.1}$ & -0.9 $^{\pm 1.1}$ & -1.5 $^{\pm 1.2}$ & -1.8 $^{\pm 1.4}$ & -1.7 $^{\pm 1.2}$  \\  
20 & 9.1 $^{\pm 0.3}$ & 3.9 $^{\pm 0.3}$ & 2.6 $^{\pm 0.3}$ & 1.6 $^{\pm 0.3}$ & 1.1 $^{\pm 0.4}$  \\  
21 & 7 $^{\pm 0.7}$ & 3.1 $^{\pm 0.7}$ & 2.2 $^{\pm 0.8}$ & 1.1 $^{\pm 0.8}$ & 1.1 $^{\pm 1}$  \\  
22 & 8.3 $^{\pm 1.2}$ & 6 $^{\pm 1.2}$ & 5.1 $^{\pm 1.3}$ & 4.3 $^{\pm 1.2}$ & 3.5 $^{\pm 1.4}$  \\  
23 & 9.9 $^{\pm 1.6}$ & 5.3 $^{\pm 1.7}$ & 3.8 $^{\pm 1.8}$ & 3.1 $^{\pm 1.9}$ & 2.4 $^{\pm 2.3}$  \\  
24 & 3.8 $^{\pm 1.8}$ & 1.8 $^{\pm 2}$ & 1.8 $^{\pm 2.2}$ & 1.3 $^{\pm 1.9}$ & 0.3 $^{\pm 2.5}$  \\  
25 & -3.7 $^{\pm 0.9}$ & -7.6 $^{\pm 0.9}$ & -8.6 $^{\pm 0.9}$ & -8.9 $^{\pm 1.1}$ & -7.8 $^{\pm 1.2}$  \\  
26 & 2.6 $^{\pm 1.7}$ & 3.1 $^{\pm 1.9}$ & 3.4 $^{\pm 1.9}$ & 3.4 $^{\pm 2}$ & 3.4 $^{\pm 2.4}$  \\  
27 & 10.4 $^{\pm 0.8}$ & 6.1 $^{\pm 0.9}$ & 5 $^{\pm 0.9}$ & 3.8 $^{\pm 0.9}$ & 2.9 $^{\pm 1}$  \\  
28 & 11.3 $^{\pm 2.2}$ & 6.9 $^{\pm 2.4}$ & 5.1 $^{\pm 2.6}$ & 4.1 $^{\pm 2.6}$ & 3 $^{\pm 2.6}$  \\  
29 & 7.2 $^{\pm 0.6}$ & 1.7 $^{\pm 0.6}$ & 0.3 $^{\pm 0.7}$ & -0.7 $^{\pm 0.7}$ & -0.6 $^{\pm 0.7}$  \\  
30 & 4.2 $^{\pm 0.6}$ & 0.9 $^{\pm 0.6}$ & 0.2 $^{\pm 0.7}$ & -0.2 $^{\pm 0.6}$ & -0.3 $^{\pm 0.7}$  \\  
31 & -0.6 $^{\pm 1.1}$ & -3.5 $^{\pm 1.2}$ & -3.8 $^{\pm 1.3}$ & -4 $^{\pm 1.2}$ & -3.9 $^{\pm 1.3}$  \\  
32 & 7.2 $^{\pm 1}$ & 3.9 $^{\pm 1.1}$ & 3 $^{\pm 1.1}$ & 2.3 $^{\pm 1.3}$ & 1.7 $^{\pm 1.3}$  \\  
33 & 4.6 $^{\pm 0.9}$ & 0.4 $^{\pm 0.9}$ & -0.4 $^{\pm 0.9}$ & -1 $^{\pm 0.9}$ & -1.3 $^{\pm 1}$  \\  
34 & 14.6 $^{\pm 1.3}$ & 10.2 $^{\pm 1.3}$ & 8.7 $^{\pm 1.4}$ & 7.1 $^{\pm 1.6}$ & 6 $^{\pm 1.7}$  \\  
35 & 0 $^{\pm 1.6}$ & -2.7 $^{\pm 1.7}$ & -3.1 $^{\pm 1.6}$ & -3 $^{\pm 2}$ & -2.8 $^{\pm 1.8}$  \\  
36 & 4.4 $^{\pm 0.5}$ & -0.1 $^{\pm 0.5}$ & -1.2 $^{\pm 0.6}$ & -1.9 $^{\pm 0.7}$ & -1.8 $^{\pm 0.6}$  \\  
37 & 7.1 $^{\pm 1.3}$ & 3.1 $^{\pm 1.4}$ & 1.9 $^{\pm 1.3}$ & 2 $^{\pm 1.3}$ & 1.1 $^{\pm 1.4}$  \\  
38 & -1.2 $^{\pm 1.1}$ & -3.4 $^{\pm 1.1}$ & -3.4 $^{\pm 1.2}$ & -3.5 $^{\pm 1.1}$ & -3 $^{\pm 1.1}$  \\  
39 & 5.8 $^{\pm 2}$ & 2 $^{\pm 2.1}$ & 0.7 $^{\pm 2.2}$ & -0.4 $^{\pm 2.3}$ & 0.3 $^{\pm 2.8}$  \\  
40 & -6.6 $^{\pm 1.2}$ & -10.7 $^{\pm 1.3}$ & -11.3 $^{\pm 1.3}$ & -11.3 $^{\pm 1.3}$ & -9.5 $^{\pm 1.5}$  \\  
41 & -3 $^{\pm 1.5}$ & -5 $^{\pm 1.5}$ & -5.8 $^{\pm 1.7}$ & -6.1 $^{\pm 1.6}$ & -5.4 $^{\pm 1.8}$  \\  
42 & 2.4 $^{\pm 2.2}$ & -0.3 $^{\pm 2.3}$ & -1 $^{\pm 2.2}$ & -1.4 $^{\pm 2.7}$ & -1.5 $^{\pm 2.9}$  \\  
43 & -0.1 $^{\pm 1.3}$ & -2.3 $^{\pm 1.4}$ & -2.9 $^{\pm 1.5}$ & -3.2 $^{\pm 1.4}$ & -2.6 $^{\pm 1.8}$  \\  
44 & 7.7 $^{\pm 1}$ & 2.9 $^{\pm 1}$ & 1.3 $^{\pm 1}$ & 0.6 $^{\pm 1.1}$ & 0.4 $^{\pm 1.1}$  \\  
45 & 6.1 $^{\pm 1.5}$ & 3.1 $^{\pm 1.5}$ & 2.6 $^{\pm 1.5}$ & 2.2 $^{\pm 1.8}$ & 2 $^{\pm 1.6}$  \\  
46 & 12.7 $^{\pm 1.8}$ & 8.9 $^{\pm 1.8}$ & 8 $^{\pm 2.1}$ & 7.7 $^{\pm 2}$ & 6.1 $^{\pm 1.9}$  \\  
47 & 5.6 $^{\pm 2}$ & 0.1 $^{\pm 2.1}$ & -0.9 $^{\pm 2.4}$ & -2 $^{\pm 2.2}$ & -1.8 $^{\pm 2.7}$  \\  
48 & -1.4 $^{\pm 1.8}$ & -4 $^{\pm 2}$ & -4.4 $^{\pm 1.9}$ & -4.8 $^{\pm 2.2}$ & -3.9 $^{\pm 2.4}$  \\      
    \hline
  \end{tabular}\caption{Full fit temperature coefficients for Synchrotron in all 48 pixel regions, with associated 1-$\sigma$ errors (in $\mu K  K^{-1}_{408 MHz}$). \label{tab:TabSincT}} 
\end{table}

\begin{table}
 \begin{tabular}{cccccc}
    \hline
     {\bf Pixel} & {\bf K-band } & {\bf Ka-band}  & {\bf Q-band } & {\bf V-band}  & {\bf W-band } \\
    \hline
   & \multicolumn{5}{c} 
      {\textbf{Synchrotron Polarization C-C coefficients}} \\
      \\\hline

1 & 2.4 $^{\pm 0.2}$ & 0.8 $^{\pm 0.2}$ & 0.4 $^{\pm 0.1}$ & 0 $^{\pm 0.2}$ & 0.3 $^{\pm 0.2}$ \\ 
2 & 0.6 $^{\pm 0.4}$ & 0.2 $^{\pm 0.4}$ & 0.2 $^{\pm 0.4}$ & 0 $^{\pm 0.5}$ & -0.4 $^{\pm 0.6}$ \\ 
3 & 1.9 $^{\pm 0.7}$ & 0.4 $^{\pm 0.7}$ & 0.3 $^{\pm 0.7}$ & 0.7 $^{\pm 0.8}$ & 0.2 $^{\pm 1}$ \\ 
4 & 2 $^{\pm 0.2}$ & 0.6 $^{\pm 0.2}$ & 0.2 $^{\pm 0.2}$ & -0.1 $^{\pm 0.3}$ & -0.2 $^{\pm 0.3}$ \\ 
5 & 1.6 $^{\pm 0.1}$ & 0.5 $^{\pm 0.1}$ & 0.3 $^{\pm 0.1}$ & 0.1 $^{\pm 0.1}$ & 0.1 $^{\pm 0.1}$ \\ 
6 & 0.5 $^{\pm 0.3}$ & 0.4 $^{\pm 0.3}$ & 0 $^{\pm 0.3}$ & 0 $^{\pm 0.3}$ & 0.1 $^{\pm 0.4}$ \\ 
7 & 0.9 $^{\pm 0.3}$ & 0.2 $^{\pm 0.3}$ & -0.1 $^{\pm 0.3}$ & 0 $^{\pm 0.3}$ & 0.2 $^{\pm 0.4}$ \\ 
8 & 1 $^{\pm 0.2}$ & 0.2 $^{\pm 0.2}$ & 0 $^{\pm 0.2}$ & -0.1 $^{\pm 0.3}$ & 0.1 $^{\pm 0.3}$ \\ 
9 & 0.7 $^{\pm 0.5}$ & 0.5 $^{\pm 0.5}$ & 0.3 $^{\pm 0.5}$ & 0.1 $^{\pm 0.6}$ & 0.3 $^{\pm 0.8}$ \\ 
10 & 1.2 $^{\pm 0.4}$ & 0 $^{\pm 0.4}$ & 0 $^{\pm 0.4}$ & 0 $^{\pm 0.4}$ & 0 $^{\pm 0.5}$ \\ 
11 & 1.2 $^{\pm 0.3}$ & 0 $^{\pm 0.3}$ & -0.1 $^{\pm 0.3}$ & 0.2 $^{\pm 0.3}$ & 0.2 $^{\pm 0.4}$ \\ 
12 & 1.3 $^{\pm 0.3}$ & 0.6 $^{\pm 0.3}$ & 0.2 $^{\pm 0.3}$ & 0.1 $^{\pm 0.3}$ & 0.1 $^{\pm 0.4}$ \\ 
13 & 2.3 $^{\pm 0.4}$ & 0.7 $^{\pm 0.4}$ & -0.1 $^{\pm 0.4}$ & -0.3 $^{\pm 0.5}$ & 0.3 $^{\pm 0.6}$ \\ 
14 & 1.7 $^{\pm 0.1}$ & 0.5 $^{\pm 0.1}$ & 0.3 $^{\pm 0.1}$ & 0.1 $^{\pm 0.1}$ & 0 $^{\pm 0.1}$ \\ 
15 & 1.8 $^{\pm 0.2}$ & 0.6 $^{\pm 0.2}$ & 0.2 $^{\pm 0.1}$ & 0 $^{\pm 0.2}$ & -0.1 $^{\pm 0.2}$ \\ 
16 & 2.6 $^{\pm 0.2}$ & 0.7 $^{\pm 0.2}$ & 0.3 $^{\pm 0.2}$ & 0.1 $^{\pm 0.2}$ & 0.2 $^{\pm 0.3}$ \\ 
17 & 1.6 $^{\pm 0.3}$ & 0.5 $^{\pm 0.3}$ & -0.1 $^{\pm 0.3}$ & 0 $^{\pm 0.3}$ & 0.1 $^{\pm 0.4}$ \\ 
18 & 1 $^{\pm 0.3}$ & 0 $^{\pm 0.3}$ & 0.4 $^{\pm 0.3}$ & 0 $^{\pm 0.4}$ & 0.3 $^{\pm 0.5}$ \\ 
19 & 1.1 $^{\pm 0.3}$ & 0.2 $^{\pm 0.3}$ & 0.2 $^{\pm 0.3}$ & 0 $^{\pm 0.4}$ & -0.1 $^{\pm 0.4}$ \\ 
20 & 1.3 $^{\pm 0.1}$ & 0.3 $^{\pm 0.1}$ & 0.1 $^{\pm 0.1}$ & 0 $^{\pm 0.1}$ & 0 $^{\pm 0.1}$ \\ 
21 & -0.1 $^{\pm 0.3}$ & 0.1 $^{\pm 0.3}$ & 0 $^{\pm 0.3}$ & 0 $^{\pm 0.4}$ & 0.4 $^{\pm 0.5}$ \\ 
22 & 0.5 $^{\pm 0.3}$ & 0 $^{\pm 0.3}$ & -0.1 $^{\pm 0.2}$ & 0 $^{\pm 0.3}$ & -0.1 $^{\pm 0.4}$ \\ 
23 & 1.1 $^{\pm 0.5}$ & 0.3 $^{\pm 0.5}$ & 0.1 $^{\pm 0.5}$ & 0.3 $^{\pm 0.6}$ & 0.3 $^{\pm 0.8}$ \\ 
24 & 1.7 $^{\pm 0.6}$ & 0.2 $^{\pm 0.6}$ & 0.4 $^{\pm 0.6}$ & -0.6 $^{\pm 0.8}$ & 0.8 $^{\pm 0.9}$ \\ 
25 & 1.5 $^{\pm 0.3}$ & 0.1 $^{\pm 0.3}$ & 0.2 $^{\pm 0.3}$ & 0.3 $^{\pm 0.3}$ & 0 $^{\pm 0.4}$ \\ 
26 & 0.6 $^{\pm 0.4}$ & 0.3 $^{\pm 0.4}$ & 0.1 $^{\pm 0.4}$ & 0.1 $^{\pm 0.5}$ & 0.1 $^{\pm 0.6}$ \\ 
27 & 1 $^{\pm 0.2}$ & 0.2 $^{\pm 0.2}$ & -0.1 $^{\pm 0.2}$ & -0.1 $^{\pm 0.3}$ & 0 $^{\pm 0.3}$ \\ 
28 & 0.5 $^{\pm 1.1}$ & -0.2 $^{\pm 1.1}$ & 1 $^{\pm 1}$ & 0.2 $^{\pm 1.3}$ & 1 $^{\pm 1.5}$ \\ 
29 & 1.4 $^{\pm 0.2}$ & 0.5 $^{\pm 0.2}$ & 0.2 $^{\pm 0.2}$ & 0 $^{\pm 0.2}$ & 0 $^{\pm 0.2}$ \\ 
30 & 0.3 $^{\pm 0.1}$ & 0.1 $^{\pm 0.2}$ & 0 $^{\pm 0.2}$ & 0 $^{\pm 0.2}$ & 0 $^{\pm 0.2}$ \\ 
31 & 1.3 $^{\pm 0.2}$ & 0.3 $^{\pm 0.2}$ & 0.1 $^{\pm 0.2}$ & -0.1 $^{\pm 0.3}$ & 0.1 $^{\pm 0.3}$ \\ 
32 & 2.6 $^{\pm 0.2}$ & 1 $^{\pm 0.2}$ & 0.4 $^{\pm 0.2}$ & -0.1 $^{\pm 0.3}$ & -0.2 $^{\pm 0.3}$ \\ 
33 & 2.5 $^{\pm 0.3}$ & 0.8 $^{\pm 0.3}$ & 0.2 $^{\pm 0.3}$ & 0.2 $^{\pm 0.4}$ & -0.4 $^{\pm 0.5}$ \\ 
34 & 1 $^{\pm 0.3}$ & -0.1 $^{\pm 0.3}$ & 0.1 $^{\pm 0.3}$ & -0.1 $^{\pm 0.4}$ & -0.1 $^{\pm 0.5}$ \\ 
35 & 0.6 $^{\pm 0.3}$ & 0.3 $^{\pm 0.3}$ & 0 $^{\pm 0.3}$ & -0.2 $^{\pm 0.4}$ & 0.5 $^{\pm 0.5}$ \\ 
36 & 1.2 $^{\pm 0.1}$ & 0.4 $^{\pm 0.1}$ & 0.2 $^{\pm 0.1}$ & 0.1 $^{\pm 0.1}$ & -0.1 $^{\pm 0.2}$ \\ 
37 & 1.5 $^{\pm 0.4}$ & 0.3 $^{\pm 0.4}$ & 0.3 $^{\pm 0.4}$ & -0.1 $^{\pm 0.4}$ & -0.3 $^{\pm 0.5}$ \\ 
38 & 0.5 $^{\pm 0.3}$ & 0.6 $^{\pm 0.3}$ & 0 $^{\pm 0.3}$ & -0.1 $^{\pm 0.4}$ & 0 $^{\pm 0.4}$ \\ 
39 & 0.7 $^{\pm 0.6}$ & 0.3 $^{\pm 0.7}$ & 0.1 $^{\pm 0.6}$ & 0.1 $^{\pm 0.8}$ & -0.1 $^{\pm 0.9}$ \\ 
40 & 0.5 $^{\pm 0.3}$ & 0.3 $^{\pm 0.3}$ & 0.1 $^{\pm 0.3}$ & 0 $^{\pm 0.3}$ & 0 $^{\pm 0.4}$ \\ 
41 & 0.3 $^{\pm 0.4}$ & 0.1 $^{\pm 0.4}$ & -0.1 $^{\pm 0.4}$ & 0 $^{\pm 0.5}$ & -0.2 $^{\pm 0.6}$ \\ 
42 & 1.8 $^{\pm 0.5}$ & 0.4 $^{\pm 0.5}$ & 0.3 $^{\pm 0.4}$ & 0.1 $^{\pm 0.6}$ & 0.1 $^{\pm 0.7}$ \\ 
43 & 1 $^{\pm 0.2}$ & 0.5 $^{\pm 0.2}$ & 0.1 $^{\pm 0.2}$ & -0.1 $^{\pm 0.2}$ & -0.1 $^{\pm 0.3}$ \\ 
44 & 3 $^{\pm 0.2}$ & 0.9 $^{\pm 0.2}$ & 0.6 $^{\pm 0.2}$ & -0.1 $^{\pm 0.3}$ & 0 $^{\pm 0.3}$ \\ 
45 & 1.5 $^{\pm 0.4}$ & 0.9 $^{\pm 0.4}$ & 0 $^{\pm 0.4}$ & 0 $^{\pm 0.5}$ & 0.1 $^{\pm 0.6}$ \\ 
46 & 0.1 $^{\pm 0.5}$ & 0 $^{\pm 0.5}$ & 0.2 $^{\pm 0.5}$ & -0.1 $^{\pm 0.6}$ & -0.6 $^{\pm 0.7}$ \\ 
47 & 0.8 $^{\pm 0.4}$ & 0.4 $^{\pm 0.4}$ & 0.7 $^{\pm 0.4}$ & 0.2 $^{\pm 0.5}$ & -0.1 $^{\pm 0.6}$ \\ 
48 & 2 $^{\pm 0.4}$ & 0.7 $^{\pm 0.4}$ & 0.4 $^{\pm 0.3}$ & -0.2 $^{\pm 0.4}$ & 0 $^{\pm 0.5}$ \\

    \hline
  \end{tabular}\caption{Full fit coefficients for polarized synchrotron in all 48 pixel regions, with associated 1-$\sigma$ errors (in $\mu K  K^{-1}_{408 MHz}$). \label{tab:TabSincP}} 
\end{table}

\begin{table}
 \begin{tabular}{cccccc}
    \hline
     {\bf Pixel} & {\bf K-band } & {\bf Ka-band}  & {\bf Q-band } & {\bf V-band}  & {\bf W-band } \\
    \hline
   & \multicolumn{5}{c} 
      {\textbf{Free-Free Temperature C-C coefficients}} \\
      \\\hline

1 & -59.9 $^{\pm 23.9}$ & -44.7 $^{\pm 25.1}$ & -39 $^{\pm 24.4}$ & -32 $^{\pm 29.6}$ & -27.1 $^{\pm 33.4}$  \\  
2 & 67.2 $^{\pm 21.7}$ & 77.6 $^{\pm 22.8}$ & 80.9 $^{\pm 26.1}$ & 79 $^{\pm 27.6}$ & 62.8 $^{\pm 23.5}$  \\  
3 & -44.5 $^{\pm 11.3}$ & -40.6 $^{\pm 11.6}$ & -42.1 $^{\pm 11.5}$ & -41.1 $^{\pm 12.3}$ & -33.3 $^{\pm 11.8}$  \\  
4 & 22.1 $^{\pm 3.5}$ & 12.8 $^{\pm 3.6}$ & 9.6 $^{\pm 4}$ & 5.7 $^{\pm 4.6}$ & 2.5 $^{\pm 4.8}$  \\  
5 & 12.1 $^{\pm 10.1}$ & 8.4 $^{\pm 11.1}$ & 10 $^{\pm 10.5}$ & 4.9 $^{\pm 12.5}$ & 3.4 $^{\pm 11.3}$  \\  
6 & -8.1 $^{\pm 8.5}$ & -10.4 $^{\pm 9.2}$ & -10.2 $^{\pm 9.7}$ & -11.6 $^{\pm 9.7}$ & -9.2 $^{\pm 11.5}$  \\  
7 & 17.3 $^{\pm 7.3}$ & 10.4 $^{\pm 8.1}$ & 7.1 $^{\pm 8.4}$ & 5.7 $^{\pm 8.7}$ & 4.6 $^{\pm 10.3}$  \\  
8 & -10.3 $^{\pm 13.1}$ & -10.7 $^{\pm 14.2}$ & -10 $^{\pm 15.1}$ & -9.9 $^{\pm 14.2}$ & -5.6 $^{\pm 14.1}$  \\  
9 & -3.3 $^{\pm 5.4}$ & -7.4 $^{\pm 5.8}$ & -8.1 $^{\pm 5.7}$ & -9.9 $^{\pm 6.1}$ & -8 $^{\pm 6.7}$  \\  
10 & 19.6 $^{\pm 4.8}$ & 16.6 $^{\pm 5.1}$ & 15.3 $^{\pm 5.3}$ & 11.7 $^{\pm 5.9}$ & 8.7 $^{\pm 5.5}$  \\  
11 & 19.2 $^{\pm 3.3}$ & 16.8 $^{\pm 3.4}$ & 15.8 $^{\pm 3.9}$ & 13.7 $^{\pm 3.9}$ & 12.6 $^{\pm 3.4}$  \\  
12 & 6.4 $^{\pm 4.6}$ & 0.4 $^{\pm 5}$ & -2.3 $^{\pm 5.4}$ & -6.2 $^{\pm 4.9}$ & -5.7 $^{\pm 5.5}$  \\  
13 & 22.6 $^{\pm 2.1}$ & 14.3 $^{\pm 2.2}$ & 11.8 $^{\pm 2.5}$ & 7.8 $^{\pm 2.2}$ & 6.3 $^{\pm 2.5}$  \\  
14 & 0 $^{\pm 3.7}$ & -3.2 $^{\pm 3.8}$ & -4 $^{\pm 4.3}$ & -5.2 $^{\pm 4.2}$ & -5.4 $^{\pm 4}$  \\  
15 & 15.2 $^{\pm 1.4}$ & 10.3 $^{\pm 1.5}$ & 8.6 $^{\pm 1.5}$ & 6.7 $^{\pm 1.5}$ & 5.7 $^{\pm 1.8}$  \\  
16 & 7.2 $^{\pm 1.7}$ & 3.1 $^{\pm 1.7}$ & 1.7 $^{\pm 1.8}$ & -0.3 $^{\pm 1.8}$ & -0.8 $^{\pm 2}$  \\  
17 & -0.4 $^{\pm 1.9}$ & -3.3 $^{\pm 2}$ & -4.4 $^{\pm 2.2}$ & -5.5 $^{\pm 2}$ & -5 $^{\pm 2}$  \\  
18 & 3.1 $^{\pm 3.2}$ & -1.1 $^{\pm 3.5}$ & -3.2 $^{\pm 3.8}$ & -3.6 $^{\pm 3.8}$ & -3.6 $^{\pm 4}$  \\  
19 & 9.8 $^{\pm 0.6}$ & 5.6 $^{\pm 0.6}$ & 4 $^{\pm 0.7}$ & 2.2 $^{\pm 0.7}$ & 1.2 $^{\pm 0.6}$  \\  
20 & -1.8 $^{\pm 1.4}$ & -2.6 $^{\pm 1.5}$ & -3.4 $^{\pm 1.6}$ & -3.9 $^{\pm 1.8}$ & -3.9 $^{\pm 1.8}$  \\  
21 & 18.6 $^{\pm 4.2}$ & 6.7 $^{\pm 4.3}$ & 2.1 $^{\pm 5}$ & 1.2 $^{\pm 4.8}$ & -2 $^{\pm 5.3}$  \\  
22 & 11.7 $^{\pm 1.1}$ & 4.2 $^{\pm 1.2}$ & 1.7 $^{\pm 1.2}$ & -0.5 $^{\pm 1.2}$ & -1.2 $^{\pm 1.6}$  \\  
23 & 13.9 $^{\pm 1.8}$ & 8.9 $^{\pm 1.9}$ & 7.5 $^{\pm 2}$ & 5.4 $^{\pm 2.2}$ & 4.5 $^{\pm 2.6}$  \\  
24 & 7.5 $^{\pm 2.9}$ & 0.2 $^{\pm 3.2}$ & -3 $^{\pm 3.1}$ & -6 $^{\pm 3.4}$ & -5.7 $^{\pm 3.3}$  \\  
25 & 9.8 $^{\pm 1.4}$ & 6.8 $^{\pm 1.4}$ & 5.9 $^{\pm 1.5}$ & 4.8 $^{\pm 1.5}$ & 3.5 $^{\pm 2}$  \\  
26 & 10.3 $^{\pm 0.6}$ & 5.3 $^{\pm 0.6}$ & 3.6 $^{\pm 0.6}$ & 1.8 $^{\pm 0.7}$ & 0.9 $^{\pm 0.7}$  \\  
27 & 3.6 $^{\pm 0.8}$ & 0 $^{\pm 0.8}$ & -1.7 $^{\pm 0.9}$ & -2.9 $^{\pm 0.9}$ & -2.6 $^{\pm 0.9}$  \\  
28 & -14.8 $^{\pm 4}$ & -14.9 $^{\pm 4.1}$ & -14.7 $^{\pm 4.8}$ & -12.9 $^{\pm 4.9}$ & -10.8 $^{\pm 5.3}$  \\  
29 & -1.1 $^{\pm 2.5}$ & -1.9 $^{\pm 2.6}$ & -2 $^{\pm 2.8}$ & -2.6 $^{\pm 2.6}$ & -3.6 $^{\pm 3.2}$  \\  
30 & -1.6 $^{\pm 5.1}$ & -0.7 $^{\pm 5.4}$ & -0.5 $^{\pm 5.9}$ & -2.2 $^{\pm 5.8}$ & -3 $^{\pm 5.9}$  \\  
31 & 8.2 $^{\pm 1.1}$ & 2.6 $^{\pm 1.2}$ & 0.7 $^{\pm 1.3}$ & -1 $^{\pm 1.3}$ & -1.5 $^{\pm 1.4}$  \\  
32 & 12.2 $^{\pm 1.6}$ & 6.2 $^{\pm 1.7}$ & 3.7 $^{\pm 1.8}$ & 1.8 $^{\pm 1.9}$ & 0.9 $^{\pm 1.9}$  \\  
33 & 7.7 $^{\pm 0.6}$ & 4.4 $^{\pm 0.7}$ & 3.3 $^{\pm 0.7}$ & 2 $^{\pm 0.7}$ & 1.2 $^{\pm 0.8}$  \\  
34 & 2.6 $^{\pm 0.5}$ & 0 $^{\pm 0.5}$ & -1.2 $^{\pm 0.6}$ & -2.2 $^{\pm 0.6}$ & -2.4 $^{\pm 0.5}$  \\  
35 & 8 $^{\pm 0.4}$ & 4.6 $^{\pm 0.4}$ & 3.4 $^{\pm 0.5}$ & 1.9 $^{\pm 0.5}$ & 1.3 $^{\pm 0.5}$  \\  
36 & -3.5 $^{\pm 1.6}$ & -4.7 $^{\pm 1.7}$ & -4.7 $^{\pm 1.7}$ & -5 $^{\pm 1.7}$ & -4.4 $^{\pm 2}$  \\  
37 & 14.7 $^{\pm 9.4}$ & 17.7 $^{\pm 9.6}$ & 20.5 $^{\pm 10.3}$ & 16.5 $^{\pm 10.8}$ & 20.5 $^{\pm 10.9}$  \\  
38 & 14.7 $^{\pm 6.2}$ & 13.1 $^{\pm 6.7}$ & 11.3 $^{\pm 6.6}$ & 9.2 $^{\pm 8.1}$ & 6.9 $^{\pm 7.7}$  \\  
39 & 21.1 $^{\pm 7.9}$ & 18.7 $^{\pm 8}$ & 20.3 $^{\pm 8.6}$ & 18.6 $^{\pm 9.8}$ & 11.7 $^{\pm 8.9}$  \\  
40 & 14.1 $^{\pm 3.1}$ & 8.7 $^{\pm 3.3}$ & 5.8 $^{\pm 3.5}$ & 3.4 $^{\pm 3.2}$ & 4 $^{\pm 3.8}$  \\  
41 & 4.3 $^{\pm 0.5}$ & 0.9 $^{\pm 0.6}$ & -0.5 $^{\pm 0.6}$ & -1.8 $^{\pm 0.6}$ & -1.9 $^{\pm 0.7}$  \\  
42 & 3 $^{\pm 2.1}$ & -0.2 $^{\pm 2.3}$ & -1.2 $^{\pm 2.6}$ & -2.5 $^{\pm 2.6}$ & -2.6 $^{\pm 2.9}$  \\  
43 & 21.9 $^{\pm 3.3}$ & 16.2 $^{\pm 3.6}$ & 14.5 $^{\pm 3.7}$ & 11.8 $^{\pm 4.2}$ & 9.9 $^{\pm 4.1}$  \\  
44 & -8.4 $^{\pm 7.7}$ & -9.6 $^{\pm 7.9}$ & -9 $^{\pm 8.2}$ & -8.6 $^{\pm 9.6}$ & -9.3 $^{\pm 10.5}$  \\  
45 & -7.5 $^{\pm 16}$ & -1.8 $^{\pm 16.5}$ & 2.8 $^{\pm 18.2}$ & 2.6 $^{\pm 19.3}$ & -0.7 $^{\pm 20.4}$  \\  
46 & 37.7 $^{\pm 11.4}$ & 35.7 $^{\pm 12.2}$ & 32.1 $^{\pm 13.2}$ & 29.7 $^{\pm 12.7}$ & 18.4 $^{\pm 13.2}$  \\  
47 & -17.9 $^{\pm 6.9}$ & -24.7 $^{\pm 7.5}$ & -26.5 $^{\pm 7.5}$ & -28.3 $^{\pm 8.7}$ & -24.7 $^{\pm 9.3}$  \\  
48 & -25.9 $^{\pm 11.2}$ & -27.7 $^{\pm 11.5}$ & -28.5 $^{\pm 11.8}$ & -25.4 $^{\pm 14.2}$ & -23.5 $^{\pm 15.6}$  \\  

    \hline
  \end{tabular}\caption{Full fit temperature coefficients for free-free in all 48 pixel regions, with associated 1-$\sigma$ errors (in $\mu K  R^{-1}$). \label{tab:TabFreeT}} 
\end{table}

\begin{table}
 \begin{tabular}{cccccc}
    \hline
     {\bf Pixel} & {\bf K-band } & {\bf Ka-band}  & {\bf Q-band } & {\bf V-band}  & {\bf W-band } \\
    \hline
   & \multicolumn{5}{c} 
      {\textbf{Free-Free Polarization C-C coefficients}} \\
      \\\hline

1 & -3 $^{\pm 5.8}$ & 0.8 $^{\pm 5.9}$ & -5.4 $^{\pm 5.6}$ & -0.7 $^{\pm 6.8}$ & 1.5 $^{\pm 8.3}$ \\ 
2 & -2.6 $^{\pm 5.7}$ & 0.3 $^{\pm 5.7}$ & 1.7 $^{\pm 5.4}$ & 1.8 $^{\pm 6.6}$ & 7.9 $^{\pm 7.9}$ \\ 
3 & -10.6 $^{\pm 3.5}$ & -0.3 $^{\pm 3.7}$ & -0.1 $^{\pm 3.6}$ & 2.3 $^{\pm 4.5}$ & 1.6 $^{\pm 4.9}$ \\ 
4 & -1.5 $^{\pm 1.2}$ & -0.4 $^{\pm 1.3}$ & -0.4 $^{\pm 1.2}$ & 0 $^{\pm 1.5}$ & -1.9 $^{\pm 1.8}$ \\ 
5 & -11.2 $^{\pm 2.8}$ & -0.7 $^{\pm 2.8}$ & -2.3 $^{\pm 2.8}$ & -1.4 $^{\pm 3.4}$ & 4.5 $^{\pm 4.1}$ \\ 
6 & 3.9 $^{\pm 2}$ & 0.2 $^{\pm 2}$ & -0.3 $^{\pm 1.9}$ & 0.1 $^{\pm 2.3}$ & 0.8 $^{\pm 2.7}$ \\ 
7 & -4.2 $^{\pm 1.6}$ & -1.7 $^{\pm 1.7}$ & -0.3 $^{\pm 1.5}$ & 0.2 $^{\pm 2}$ & -0.7 $^{\pm 2.4}$ \\ 
8 & 9 $^{\pm 3.5}$ & 1.4 $^{\pm 3.6}$ & -0.4 $^{\pm 3.4}$ & -0.5 $^{\pm 4.2}$ & -0.5 $^{\pm 5}$ \\ 
9 & 1.6 $^{\pm 1.6}$ & 0.4 $^{\pm 1.7}$ & 0.4 $^{\pm 1.6}$ & 0.7 $^{\pm 1.9}$ & 1.3 $^{\pm 2.3}$ \\ 
10 & -0.3 $^{\pm 1.2}$ & 1.3 $^{\pm 1.2}$ & 0.1 $^{\pm 1.2}$ & -0.4 $^{\pm 1.4}$ & -1.3 $^{\pm 1.7}$ \\ 
11 & 0.5 $^{\pm 1.1}$ & 0.2 $^{\pm 1.1}$ & 1 $^{\pm 1.1}$ & 0.2 $^{\pm 1.4}$ & -0.3 $^{\pm 1.6}$ \\ 
12 & 0.8 $^{\pm 1.5}$ & 0 $^{\pm 1.5}$ & 0.7 $^{\pm 1.6}$ & -0.8 $^{\pm 1.9}$ & -1.9 $^{\pm 2.3}$ \\ 
13 & -1.1 $^{\pm 0.8}$ & 0 $^{\pm 0.8}$ & -0.5 $^{\pm 0.8}$ & -0.7 $^{\pm 0.9}$ & -0.3 $^{\pm 1.1}$ \\ 
14 & -4.2 $^{\pm 0.8}$ & -0.8 $^{\pm 0.8}$ & -1.2 $^{\pm 0.8}$ & -0.1 $^{\pm 0.9}$ & -1.1 $^{\pm 1.1}$ \\ 
15 & -1.5 $^{\pm 0.4}$ & -0.4 $^{\pm 0.3}$ & -0.2 $^{\pm 0.3}$ & 0 $^{\pm 0.4}$ & 0 $^{\pm 0.5}$ \\ 
16 & 0.5 $^{\pm 0.3}$ & 0.1 $^{\pm 0.3}$ & 0.4 $^{\pm 0.4}$ & 0 $^{\pm 0.4}$ & -0.1 $^{\pm 0.5}$ \\ 
17 & -1 $^{\pm 0.5}$ & 0.3 $^{\pm 0.5}$ & 0.2 $^{\pm 0.5}$ & 0.1 $^{\pm 0.6}$ & 0.2 $^{\pm 0.7}$ \\ 
18 & 0.1 $^{\pm 0.9}$ & 0.7 $^{\pm 1}$ & -0.3 $^{\pm 0.9}$ & -0.2 $^{\pm 1.1}$ & -0.2 $^{\pm 1.4}$ \\ 
19 & 0 $^{\pm 0.2}$ & -0.1 $^{\pm 0.2}$ & 0 $^{\pm 0.2}$ & 0 $^{\pm 0.2}$ & -0.2 $^{\pm 0.3}$ \\ 
20 & -0.6 $^{\pm 0.4}$ & 0.2 $^{\pm 0.5}$ & 0 $^{\pm 0.4}$ & 0 $^{\pm 0.5}$ & 0.1 $^{\pm 0.7}$ \\ 
21 & -2.1 $^{\pm 1.9}$ & 1.2 $^{\pm 2}$ & -1.1 $^{\pm 1.9}$ & -1 $^{\pm 2.3}$ & -1.1 $^{\pm 2.9}$ \\ 
22 & -0.8 $^{\pm 0.3}$ & -0.2 $^{\pm 0.3}$ & -0.3 $^{\pm 0.3}$ & 0 $^{\pm 0.4}$ & 0.4 $^{\pm 0.5}$ \\ 
23 & 0.2 $^{\pm 0.5}$ & 0 $^{\pm 0.5}$ & -0.1 $^{\pm 0.5}$ & -0.2 $^{\pm 0.6}$ & -0.2 $^{\pm 0.8}$ \\ 
24 & -3.5 $^{\pm 0.8}$ & -1.1 $^{\pm 0.8}$ & -1 $^{\pm 0.8}$ & -1 $^{\pm 1}$ & -0.2 $^{\pm 1.2}$ \\ 
25 & 0.9 $^{\pm 0.5}$ & 0.5 $^{\pm 0.5}$ & 0.2 $^{\pm 0.5}$ & 0 $^{\pm 0.6}$ & 0 $^{\pm 0.7}$ \\ 
26 & 0.2 $^{\pm 0.2}$ & 0.1 $^{\pm 0.2}$ & -0.1 $^{\pm 0.2}$ & -0.1 $^{\pm 0.2}$ & 0.1 $^{\pm 0.2}$ \\ 
27 & -0.4 $^{\pm 0.2}$ & -0.1 $^{\pm 0.2}$ & 0 $^{\pm 0.2}$ & 0 $^{\pm 0.3}$ & -0.1 $^{\pm 0.3}$ \\ 
28 & 2.6 $^{\pm 1.7}$ & 1.3 $^{\pm 1.7}$ & -0.1 $^{\pm 1.6}$ & -0.6 $^{\pm 2.1}$ & 0.4 $^{\pm 2.3}$ \\ 
29 & -3.6 $^{\pm 0.7}$ & -1.8 $^{\pm 0.7}$ & -1.4 $^{\pm 0.7}$ & 0.6 $^{\pm 0.9}$ & 0.2 $^{\pm 1}$ \\ 
30 & -4.1 $^{\pm 1.7}$ & -0.4 $^{\pm 1.7}$ & -1.7 $^{\pm 1.7}$ & -0.5 $^{\pm 2.1}$ & -0.5 $^{\pm 2.5}$ \\ 
31 & -1.4 $^{\pm 0.3}$ & 0 $^{\pm 0.3}$ & -0.2 $^{\pm 0.3}$ & 0.1 $^{\pm 0.3}$ & -0.2 $^{\pm 0.4}$ \\ 
32 & -0.5 $^{\pm 0.4}$ & 0.2 $^{\pm 0.4}$ & -0.2 $^{\pm 0.4}$ & 0 $^{\pm 0.4}$ & 0.3 $^{\pm 0.5}$ \\ 
33 & -0.3 $^{\pm 0.2}$ & 0.1 $^{\pm 0.2}$ & 0.1 $^{\pm 0.2}$ & 0.2 $^{\pm 0.3}$ & 0.3 $^{\pm 0.3}$ \\ 
34 & -0.1 $^{\pm 0.1}$ & -0.1 $^{\pm 0.1}$ & -0.1 $^{\pm 0.1}$ & 0 $^{\pm 0.2}$ & 0.2 $^{\pm 0.2}$ \\ 
35 & -0.2 $^{\pm 0.1}$ & -0.1 $^{\pm 0.1}$ & -0.1 $^{\pm 0.1}$ & 0 $^{\pm 0.1}$ & 0 $^{\pm 0.1}$ \\ 
36 & -1 $^{\pm 0.4}$ & -0.3 $^{\pm 0.4}$ & -0.3 $^{\pm 0.3}$ & -0.2 $^{\pm 0.4}$ & 0.5 $^{\pm 0.5}$ \\ 
37 & -1 $^{\pm 3.4}$ & 0.9 $^{\pm 3.4}$ & 1.5 $^{\pm 3.3}$ & 0.2 $^{\pm 3.8}$ & 0.6 $^{\pm 4.8}$ \\ 
38 & -7.3 $^{\pm 1.8}$ & -4.1 $^{\pm 1.9}$ & -0.8 $^{\pm 1.8}$ & 0.3 $^{\pm 2.2}$ & -0.4 $^{\pm 2.7}$ \\ 
39 & -7.1 $^{\pm 2.7}$ & -1.3 $^{\pm 2.8}$ & -2.5 $^{\pm 2.7}$ & 0.2 $^{\pm 3.3}$ & -0.3 $^{\pm 4}$ \\ 
40 & 3.8 $^{\pm 1.1}$ & 0.3 $^{\pm 1.2}$ & 0.4 $^{\pm 1}$ & 0.3 $^{\pm 1.4}$ & 0.3 $^{\pm 1.5}$ \\ 
41 & 0 $^{\pm 0.2}$ & 0 $^{\pm 0.2}$ & 0.1 $^{\pm 0.2}$ & -0.1 $^{\pm 0.2}$ & 0 $^{\pm 0.3}$ \\ 
42 & 0.1 $^{\pm 0.5}$ & 0 $^{\pm 0.6}$ & -0.2 $^{\pm 0.5}$ & 0 $^{\pm 0.6}$ & 0 $^{\pm 0.8}$ \\ 
43 & -0.9 $^{\pm 1.2}$ & 0.5 $^{\pm 1.1}$ & -1 $^{\pm 1}$ & 0 $^{\pm 1.3}$ & 0.2 $^{\pm 1.5}$ \\ 
44 & -7.3 $^{\pm 2.2}$ & -2.6 $^{\pm 2.3}$ & 0.1 $^{\pm 2.2}$ & 0 $^{\pm 2.7}$ & 0.5 $^{\pm 3.2}$ \\ 
45 & 5 $^{\pm 5.9}$ & 7.3 $^{\pm 6.1}$ & 1 $^{\pm 6}$ & -0.7 $^{\pm 7.3}$ & -1.2 $^{\pm 8.5}$ \\ 
46 & 1.2 $^{\pm 4.5}$ & 1.2 $^{\pm 4.8}$ & -2.4 $^{\pm 4.6}$ & -1.3 $^{\pm 5.6}$ & 7.2 $^{\pm 6.7}$ \\ 
47 & 2.2 $^{\pm 1.6}$ & 1.6 $^{\pm 1.7}$ & 1.8 $^{\pm 1.5}$ & -0.5 $^{\pm 1.8}$ & 1 $^{\pm 2.2}$ \\ 
48 & -0.4 $^{\pm 3.4}$ & 2.1 $^{\pm 3.6}$ & -0.3 $^{\pm 3.3}$ & -2.8 $^{\pm 4}$ & 2 $^{\pm 5}$ \\

    \hline
  \end{tabular}\caption{Full fit coefficients for polarized free-free in all 48 pixel regions, with associated 1-$\sigma$ errors (in $\mu K  R^{-1}$). \label{tab:TabFreeP}} 
\end{table}

\end{document}